\documentclass[11pt]{article}
\usepackage{cite}
\usepackage{scalerel}
\usepackage{amsmath,amsfonts,amssymb}
\usepackage[small,bf,hang]{caption}
\usepackage{slashed}

\usepackage{latexsym,epsfig}
\usepackage{arydshln}
\usepackage{extarrows}

\usepackage{graphicx}
\usepackage{xcolor}
\usepackage{float}
\usepackage{tikz}
\usetikzlibrary{positioning,arrows,calc}
\usetikzlibrary{arrows.meta}
\usetikzlibrary{decorations.pathmorphing}
\usetikzlibrary{decorations.markings}
\usetikzlibrary{shapes.geometric}
\usetikzlibrary{patterns}

\newif\ifstartcompletesineup
\newif\ifendcompletesineup
\pgfkeys{
    /pgf/decoration/.cd,
    start up/.is if=startcompletesineup,
    start up=true,
    start up/.default=true,
    start down/.style={/pgf/decoration/start up=false},
    end up/.is if=endcompletesineup,
    end up=true,
    end up/.default=true,
    end down/.style={/pgf/decoration/end up=false}
}
\pgfdeclaredecoration{complete sines}{initial}
{
    \state{initial}[
        width=+0pt,
        next state=upsine,
        persistent precomputation={
            \ifstartcompletesineup
                \pgfkeys{/pgf/decoration automaton/next state=upsine}
                \ifendcompletesineup
                    \pgfmathsetmacro\matchinglength{
                        0.5*\pgfdecoratedinputsegmentlength / (ceil(0.5* \pgfdecoratedinputsegmentlength / \pgfdecorationsegmentlength) )
                    }
                \else
                    \pgfmathsetmacro\matchinglength{
                        0.5 * \pgfdecoratedinputsegmentlength / (ceil(0.5 * \pgfdecoratedinputsegmentlength / \pgfdecorationsegmentlength ) - 0.499)
                    }
                \fi
            \else
                \pgfkeys{/pgf/decoration automaton/next state=downsine}
                \ifendcompletesineup
                    \pgfmathsetmacro\matchinglength{
                        0.5* \pgfdecoratedinputsegmentlength / (ceil(0.5 * \pgfdecoratedinputsegmentlength / \pgfdecorationsegmentlength ) - 0.4999)
                    }
                \else
                    \pgfmathsetmacro\matchinglength{
                        0.5 * \pgfdecoratedinputsegmentlength / (ceil(0.5 * \pgfdecoratedinputsegmentlength / \pgfdecorationsegmentlength ) )
                    }
                \fi
            \fi
            \setlength{\pgfdecorationsegmentlength}{\matchinglength pt}
        }] {}
    \state{downsine}[width=\pgfdecorationsegmentlength,next state=upsine]{
        \pgfpathsine{\pgfpoint{0.5\pgfdecorationsegmentlength}{0.5\pgfdecorationsegmentamplitude}}
        \pgfpathcosine{\pgfpoint{0.5\pgfdecorationsegmentlength}{-0.5\pgfdecorationsegmentamplitude}}
    }
    \state{upsine}[width=\pgfdecorationsegmentlength,next state=downsine]{
        \pgfpathsine{\pgfpoint{0.5\pgfdecorationsegmentlength}{-0.5\pgfdecorationsegmentamplitude}}
        \pgfpathcosine{\pgfpoint{0.5\pgfdecorationsegmentlength}{0.5\pgfdecorationsegmentamplitude}}
}
    \state{final}{}
}

\tikzset{
corner/.style={line width=1pt,dashed,draw=black,dash pattern=on 6pt off 4pt},
scalar/.style={line width=1pt,draw=black},
photon/.style={line width=1pt,decorate, draw=black,
    decoration={complete sines,aspect=0,amplitude=1.25mm,segment length=1.5mm,start up,end up}},
}

\def\hybrid{
        \topmargin -20pt
        \oddsidemargin 0pt
        \headheight 0pt \headsep 0pt
        \textwidth 6.25in 
        \textheight 9.5in 
        \marginparwidth .875in
        \parskip 5pt plus 1pt \jot = 1.5ex}

\hybrid

 \csname
@addtoreset\endcsname{equation}{section}

\def\moth{\mathsurround=0pt}
\newdimen\zo \zo=0pt

\def\tick{\leaders\hrule height 0.5ex depth 0pt \hskip 0.5pt}
\def\upboxfill{$\moth \setbox\zo\hbox{\tick}%
  \hskip 3pt\hbox to 0pt{$\tick$\hss}\hrulefill \hbox to 7.5pt{$\tick$\hss}$}

\def\dtick{\leaders\hrule height .34pt depth 0.5ex \hskip 0.5pt}
\def\downboxfill{$\moth \setbox\zo\hbox{\dtick}%
  \hskip 2pt\hbox to 0pt{$\dtick$\hss}\hrulefill \hbox to 2pt{$\dtick$\hss}$}

\def\bec{\begin{center}}
\def\ec{\end{center}}

\def\cL{{\cal L}}
\def\cD{\mathfrak{D}}
\def\calD{\mathcal{D}}
\def\cF{{\cal F}}
\def\cO{{\cal O}}
\def\cA{{\cal A}}

\def\cS{{\cal S}}

\def\cH{{\cal H}}

\def\cT{{\cal T}}

\def\cA{{\cal A}}

\def\Tr{{\rm Tr}}

\def\cO{{\cal O}}

\def\del{\partial}

\def\Tr{{\rm Tr}}

 \def\det{{\rm det\,}}
\def\be{\begin{equation}}
\def\ee{\end{equation}}
\def\bea{\begin{eqnarray}}
\def\eea{\end{eqnarray}}
\def\ba{\begin{array}}
\def\ea{\end{array}}

\def\Dt{\frac{D}{Dt}}

\begin{document}

\begin{titlepage}
\rightline{}
\rightline{March  2021}
\rightline{HU-EP-21/07-RTG}  
\begin{center}
\vskip 1.5cm
 {\Large \bf{Beta functions for the duality-invariant sigma model 
 }}
\vskip 1.7cm

{\large\bf {Roberto Bonezzi, Tomas Codina and Olaf Hohm}}
\vskip 1.6cm

{\it  Institute for Physics, Humboldt University Berlin,\\
 Zum Gro\ss en Windkanal 6, D-12489 Berlin, Germany}\\[1.5ex] 
 ohohm@physik.hu-berlin.de, 
roberto.bonezzi@physik.hu-berlin.de, 
tomas.codina@physik.hu-berlin.de
\vskip .1cm

\vskip .2cm

\end{center}

\bigskip\bigskip
\begin{center} 
\textbf{Abstract}

\end{center} 
\begin{quote}

The 
$O(d,d)$ invariant worldsheet theory for bosonic string theory with 
$d$ abelian isometries is employed to compute the beta functions and Weyl anomaly at one-loop. 
We show that vanishing of the Weyl anomaly coefficients implies the equations of motion of the 
 Maharana-Schwarz action. We give a self-contained introduction into the required techniques, 
 including  beta functions, the Weyl anomaly for two-dimensional sigma models 
 and the background field method. This sets  the stage for 
 a sequel to this paper on generalizations to higher loops 
 and $\alpha'$ corrections.

\end{quote} 
\vfill
\setcounter{footnote}{0}
\end{titlepage}

\tableofcontents

\section{Introduction}

String theory is a theory of quantum gravity that at low energies reduces to Einstein gravity coupled to matter fields. 
The physical spacetime in which this theory is defined is usually referred to as the target space, to distinguish it from the two-dimensional worldsheet 
of the fundamental string. Physical observables such as scattering amplitudes may be  computed using data of this worldsheet 
theory, and the interplay of worldsheet and target space techniques is one of the intriguing features of string theory. 
For instance, demanding that the classical scale invariance of the worldsheet theory is preserved  quantum mechanically, 
i.e, that the Weyl anomaly vanishes, yields as a consistency condition target space equations that, to lowest order in 
the (inverse) string tension $\alpha'$, are equivalent to the Einstein equations. Higher loop corrections then imply that 
Einstein's theory receives (infinitely many) higher-derivative $\alpha'$ corrections. Notably, $\alpha'$ corrections in the target space are a feature 
already of \textit{classical} string theory that can be derived from \textit{quantum} considerations of the worldsheet theory.

In this paper we compute the trace anomaly for a  duality invariant worldsheet theory at one-loop level. 
By duality we refer to the phenomenon that on the space of string backgrounds (solutions of the stringy Einstein equations) with $d$ abelian isometries 
there is a global $O(d,d,\mathbb{R})$ `T-duality' invariance \cite{Meissner:1991zj}. Specifically, the effective action obtained by dimensionally reducing along 
these $d$ dimensions (i.e.~taking the fields to be independent of $d$ coordinates) is $O(d,d,\mathbb{R})$ invariant to all orders in $\alpha'$ \cite{Sen:1991zi}. 
This in turn poses strong constraints on the possible higher-derivative corrections, which are only known to a few orders in $\alpha'$. 
For instance, the most general $O(d,d,\mathbb{R})$ invariant equations for purely time-dependent backgrounds 
(as relevant for cosmological FLRW backgrounds with flat spatial metric)   were recently determined 
to all orders in $\alpha'$ \cite{Hohm:2019jgu}, but there is a finite number of free parameters at each order in $\alpha'$ that are not constrained by 
duality and hence must be determined by other methods \cite{Codina:2020kvj}. 

Traditionally, one would first determine the higher-derivative corrections for the full target space 
string theory (say by requiring vanishing of the Weyl anomaly) and, second, dimensionally reduce along $d$ directions. 
Both steps are technically challenging (to put it mildly). Here we aim to circumvent the need for a two-step procedure by employing a duality invariant worldsheet 
theory that is directly adapted to the dimensional reduction and is manifestly $O(d,d,\mathbb{R})$ invariant. This worldsheet action, which generalizes 
a formulation due to Tseytlin \cite{Tseytlin:1990nb,Tseytlin:1990va}, was originally given  by Schwarz and Sen \cite{Schwarz:1993mg} and more recently 
re-derived  in \cite{Blair:2016xnn,Bonezzi:2020ryb}. In particular in \cite{Bonezzi:2020ryb} it was shown that this model  is a consistent 
truncation of  the standard worldsheet theory in which internal momentum and winding modes are set to zero, being  the worldsheet counterpart of 
 taking the target space fields to be independent of $d$ coordinates. Moreover, it was shown that the duality symmetries become anomalous 
 quantum mechanically due to the presence of chiral bosons. The anomaly can then be cancelled by assigning a non-trivial duality transformation 
 to the 2-form B-field according to a Green-Schwarz mechanism. This result  gives a worldsheet interpretation for the observation that 
 at order $\alpha'$ the target space equations are only $O(d,d,\mathbb{R})$ invariant provided the  singlet B-field 
 transforms non-trivially \cite{Eloy:2019hnl,Eloy:2020dko}, in analogy to the heterotic string \cite{Hull:1985jv,Sen:1986nm}. 
 While this is a one-loop effect, the anomaly itself is finite and hence does not enter the one-loop beta function computation performed here, 
 in agreement with the fact that the Green-Schwarz deformation is invisible to lowest order in $\alpha'$. The anomaly will contribute, however,  to any 
 two-loop beta function computation.

 In the remainder of this paper we show at one loop that the duality invariant worldsheet theory can be used to derive directly the dimensionally reduced
 $O(d,d,\mathbb{R})$ invariant  target space equations by demanding vanishing of the Weyl anomaly. 
 Specifically, 
 we derive the $O(d,d,\mathbb{R})$ invariant theory to zeroth order in $\alpha'$ (second order in derivatives) 
 and show that it coincides with the theory derived by Maharana and Schwarz through dimensional reduction. 
 In this we generalize previous work in \cite{Berman:2007xn,Berman:2007yf} by including the (doubled) Kaluza-Klein vector fields and the external B-field, 
 which leads to significant technical complications that we deal with in due course. 
  We also use the opportunity to give a self-contained introduction into the general methods   needed for this computation. 
 While to some extent this is textbook material, we found that at the level of detail needed for our applications 
  the required technology is somewhat scattered through the literature of the 1980s, 
 and we hope that our exposition will be helpful to the community. 
 This sets the stage for a sequel  to this paper, in which we extend  the analysis to two loops and hence to first order in $\alpha'$.  
  This will be an important additional  test for the quantum consistency of this non-standard duality invariant sigma model (whose worldsheet diffeomorphism invariance, in particular, 
  is realized in a non-manifest fashion). Indeed, unexpected features may emerge given that the target space equations cannot 
  be written entirely in terms of the generalized metric that enters the duality invariant sigma model \cite{Eloy:2019hnl,Eloy:2020dko}.

 The rest of this paper is organized as follows. In sec.~2 we introduce the core notions needed to compute the Weyl anomaly, in particular 
 the beta functions whose vanishing is related to (but not equivalent to) the vanishing of the Weyl anomaly. 
 In sec.~3 we explain the background field method and clarify various issues that arise when computing the beta functions 
 for the conventional string worldsheet theory. These two sections are essentially  review but hopefully condense the needed techniques in a useful manner. 
 Our main new results are presented in sec.~4 where we display the one-loop computation for the duality invariant worldsheet that yields the 
 Maharana-Schwarz theory. Some technical details are collected in an appendix.

\section{The Trace Anomaly 
}\label{sec: Weyl anomaly}

Our goal is to extract target space field equations from a string sigma model  by requiring  quantum consistency \cite{Friedan:1980jf,Friedan:1980jm,Callan:1985ia,Tseytlin:1988rr,Callan:1989nz}. One ensures  that the non-linear sigma model is conformally invariant quantum mechanically  by imposing the vanishing of the Weyl anomaly. 
This condition is often stated to be equivalent 
to the vanishing of the renormalization-group (RG) beta functions $\beta^i$, but this is not quite correct for the string sigma model: ${\beta^i=0}$ ensures only \emph{rigid} scale invariance or, equivalently, the vanishing of the integrated trace $\int d^2\sigma\sqrt{\gamma}\,T^\alpha{}_\alpha$ of the energy-momentum tensor. The quantum string is consistent provided  the stronger condition $T^\alpha{}_{\alpha}=0$ holds locally. This leads to target space equations
for the true ``Weyl anomaly coefficients'' $\bar\beta^i$ of the schematic form 
\begin{equation}\label{betabarintro}
\bar\beta^i=\beta^i +
\Delta_{{\cal W}}\varphi^i=0\;,
\end{equation}
where $\varphi^i$ are the target space fields for which $\beta^i$ are the beta functions, 
and $\Delta$ is the operator implementing the gauge transformations of $\varphi^i$, with field-dependent effective parameter ${\cal W}$. 
In this section we will review the derivation of an operator expression for the Weyl anomaly \cite{Tseytlin:1986tt,Tseytlin:1986ws,Curci:1986hi}, which will elucidate the origin of the extra terms in \eqref{betabarintro}.
We shall follow closely the general discussion given by Tseytlin in \cite{Tseytlin:1986ws}.

\subsection{Generalities}

We consider   renormalizable field theories in 
two dimensions. 
The main example relevant for our subsequent applications is 
the string sigma model with target space metric $g_{\mu\nu}(X)$ 
\begin{equation}\label{sigmaexample}
S =\frac{1}{2\lambda}\int d^2\sigma \sqrt{\gamma}\, \gamma^{\alpha\beta}g_{\mu\nu}(X)\del_\alpha X^\mu\del_{\beta} X^\nu \;,  
\end{equation}
where $\gamma_{\alpha\beta}$ is the Euclidean worldsheet metric and $\lambda=2\pi\alpha'$. 
Recall that the $g_{\mu\nu}(X)$, which are functions on the $D-$dimensional target space, are viewed as a collection of infinitely many coupling constants.  
Upon quantization we have to distinguish between bare couplings and renormalized couplings. 
Using the  dot product 
\begin{equation}
f\cdot g=\int d^Dx\,f(x)\,g(x) \;, 
\end{equation}
we write the bare action as 
\begin{equation}\label{bare S}
S_0=\int d^n\sigma\,A_{i0}\cdot\varphi^i_0\;,    
\end{equation}
where $\varphi^i_0$ and $A_{i0}$ denote bare couplings and composite operators, respectively, 
and $n=2+\epsilon$, where $\epsilon$ is the dimensional  regularization parameter. 
For instance, for the  string sigma model (\ref{sigmaexample}) one has 
\begin{equation}\label{bareCouplings}
\varphi^i_0=g_{0\mu\nu}(x)\;,\quad A_{i0}=\frac{1}{2\lambda}\sqrt{\gamma}\gamma^{\alpha\beta}\del_\alpha X^\mu\del_\beta X^\nu\,\delta^D(x-X(\sigma))\;. 
\end{equation} 
We denote  the renormalized couplings by $\varphi^i$ and  choose them to be dimensionless.  The bare couplings $\varphi^i_0$ are taken to have mass dimension $\epsilon=n-2$, implying that the bare operators $A_{i0}$ have dimension 2.

In the following we perform renormalization via minimal subtraction, in which the counterterms are purely divergent. 
Expressing the counterterms as a Laurent series in the dimensional regularization parameter $\epsilon$ 
one writes the bare action \eqref{bare S} as a sum 
of the renormalized action  and the counterterms, 
 \be\label{useful bare}
 \begin{split}
  S_0 &= S_{\rm ren}+S_{\rm c.t.}\;, \\
  S_{\rm ren} &= \int d^n\sigma\,\mu^\epsilon A_{i0}\cdot \varphi^i \;, \qquad
  S_{\rm c.t.} = \int d^n\sigma\,\mu^\epsilon A_{i0}\cdot \sum_{n=1}^\infty\frac{1}{\epsilon^n}\,T^i_n(\varphi) \;, 
 \end{split}
 \ee 
where  $\mu$ is the renormalization scale that is introduced in order  to keep the renormalized couplings dimensionless. 
The action (\ref{useful bare}) is used to derive the Feynman rules, and the counterterms $T^i_n$ are fixed by demanding that they cancel the divergent contributions to the quantum effective action. 
Comparison of (\ref{useful bare}) with the canonical form (\ref{bare S}) of the bare action gives the bare couplings in terms of the renormalized ones: 
\begin{equation}\label{Phi0}
\varphi^i_0=\mu^\epsilon\Big[\varphi^i+\sum_{n=1}^\infty\frac{1}{\epsilon^n}\,T^i_n(\varphi)\Big]\;. 
\end{equation}
With this relation we may compute the beta functions associated to the couplings $\varphi^i$ in $n=2+\epsilon$ dimensions, which are defined by 
 \be
  \beta^i := \frac{d\varphi^i}{dt} + \epsilon\,\varphi^i\,, \qquad   t:= \log\mu \;, 
 \ee
where the derivative of $\varphi$ is   obtained by requiring that the bare couplings $\varphi^i_0$ do not depend on this  renormalization scale: 
\be
 \frac{d\varphi^i_0}{dt}=0   
 \,. 
\ee 
 Differentiating \eqref{Phi0} gives
\begin{equation}
0=\epsilon\varphi^i+T^i_1+\cO(\epsilon^{-1})+\frac{d\varphi^i}{dt}+\frac{1}{\epsilon}\,\frac{d\varphi^j}{dt}\frac{\del}{\del\varphi^j}T_1^i+\cO(\epsilon^{-2}) \;.  \end{equation}
Matching the $\cO(\epsilon)$ and $\cO(\epsilon^0)$ terms one obtains
\begin{equation}\label{tDerivative}
\hat\beta^i:= \frac{d\varphi^i}{dt}= 
-\epsilon\,\varphi^i+\beta^i\;,\quad \beta^i=-T^i_1+\varphi\cdot\frac{\del}{\del\varphi}T^i_1\;,   
\end{equation}
where $\hat\beta^i$ is a short-hand notation that shall be useful in what follows.  
Note that the expression for $\beta^i$ is exact in terms of the counterterm $T^i_1$ and the usual perturbative evaluation arises from the loop expansion of the latter.
The higher order terms in $\epsilon^{-n}$ provide the so-called pole relations between higher order counterterms \cite{tHooft:1973mfk,AlvarezGaume:1981hn}. 
Note that the above operator $\varphi\cdot\frac{\del}{\del\varphi}$ has to be understood in terms of the functional derivative as 
\begin{equation}\label{func der}
f\cdot\frac{\del F}{\del\varphi}=\int d^Dx\,f^i(x)\,\frac{\delta F}{\delta\varphi^i(x)}\;.    
\end{equation}

The partition function of the theory is given by the path integral
\begin{equation}
Z(\varphi, \mu)=e^{-W(\varphi, \mu)}=\int\calD X\,e^{-S_0}\;,
\end{equation}
and quantum expectation values are denoted by 
$\langle\cdots\rangle=\frac{1}{Z}\int\calD X\cdots\,e^{-S_0}$. 
Given  that  the theory is renormalizable, the bare action (\ref{useful bare}) contains all the required counterterms to render the above path integral finite. 
Since the bare action (\ref{bare S}) written in terms of the bare couplings does not depend on $\mu$
we can now derive the renormalization group equation
\begin{equation}
\frac{dW}{dt}=\frac{\del W}{\del t}+\hat\beta^i\,\frac{\del W}{\del\varphi^i}=0\;, 
\end{equation}
where we used (\ref{tDerivative}). 
The next ingredient we need is a prescription to define renormalized composite operators $[A_i]\,$ for the bare couplings in (\ref{bare S}), 
where by $[\cdots]$ we mean some normal-ordering prescription ensuring finiteness. 
Recalling (\ref{bare S}), the integral of the bare operator $A_{i0}$ is given by\footnote{ Notice that  for 
the worldsheet action (\ref{sigmaexample}) we have \begin{equation*}
\frac{\delta S_0}{\delta g_{0\mu\nu}(x)}=\frac{1}{2\lambda}\int d^n\sigma \,\delta^D(x-X(\sigma))\,\del_\alpha X^\mu\del^\alpha X^\nu\;,  \end{equation*}
i.e., the derivative  is still integrated over the worldsheet.}
\begin{equation}\label{Ai0 integrated}
\int d^n\sigma\, A_{i0}=\frac{\del S_0}{\del\varphi^i_0}\;.    
\end{equation}
Accordingly, we now define the renormalized composite operators $[A_i]$ by demanding 
\begin{equation}\label{Ai ren integrated}
\int d^n\sigma[A_i]=\frac{\del S_0}{\del\varphi^i}\;. 
\end{equation}
The quantum expectation value is then given by 
\begin{equation}
\Big\langle\int d^n\sigma[A_i]\Big\rangle  =\frac{\del W}{\del\varphi^i}  \,, 
\end{equation}
that is guaranteed to be finite, because it is  the derivative of a finite quantity by a finite parameter. 
Given \eqref{Ai ren integrated} and \eqref{Ai0 integrated} 
the relation between bare and renormalized operators  
is known up to possible total derivative terms:
\begin{equation}\label{Ai ren unintegrated}
[A_i]=A_{j0}\cdot\frac{\del\varphi^j_0}{\del\varphi^i}+\del_\alpha\Omega^\alpha_i\;.    
\end{equation}
Assuming that the set $\{A_{i0}\}$ is a complete basis of dimension 2 operators (modulo the bare equations of motions $\frac{\delta S_0}{\delta X^\mu}$ that we shall always discard, having zero expectation value), the total derivative part can also be expanded in terms of $A_{i0}$, namely
\begin{equation}\label{Qs defined}
\del_\alpha\Omega^\alpha_i=A_{j0}\cdot\Lambda^j_i\;,\quad\Lambda_i^j=\mu^\epsilon\sum_{n=1}^\infty\frac{1}{\epsilon^n}\,Q^j_{ni}(\varphi)  \;.  
\end{equation}
This allows us to define the renormalization matrix as
\begin{equation}\label{Ren matrix}
[A_i]=A_{j0}\cdot Z^j_i\;,\quad Z^j_i=\frac{\del\varphi^j_0}{\del\varphi^i}+\Lambda^j_i    \,. 
\end{equation}
Given the relation \eqref{Phi0} between bare and renormalized couplings, we infer 
\begin{equation}\label{Z factor}
Z^i_j=\mu^\epsilon\Big[\delta^i_j+\sum_{n=1}^\infty\frac{1}{\epsilon^n}\,X^i_{nj}(\varphi)\Big]\;,\quad X^i_{nj}=\frac{\del T^i_n}{\del\varphi^j}+Q^i_{nj}\;.
\end{equation}
It should be emphasized that while for ordinary field theories the $Z^i_j$ are finite-dimensional matrices, 
for the string sigma model they are actually differential operators. 
Specifically, one has $T^i_1=T^g_{1\mu\nu}=-\alpha'R_{\mu\nu}$ at one-loop order, with $R_{\mu\nu}$ the target space Ricci tensor, giving
\begin{equation}
\frac{\del T^i_1}{\del\varphi^j}\;\rightarrow\frac{\delta T_{1\mu\nu}(x)}{\delta g_{\lambda\rho}(y)}=\frac{\alpha'}{2}\,\Big[\delta^{(\lambda}_\mu\delta^{\rho)}_\nu\nabla^2-2\,\delta^{(\lambda}_{(\mu}\nabla^{\rho)}\nabla_{\nu)}+g^{\lambda\rho}\nabla_{(\mu}\nabla_{\nu)}\Big]_x\delta^D(x-y) \;, 
\end{equation}
which back in (\ref{Z factor}) implies that $Z^i_j$ is a differential operator. 
It is often more convenient to contract such operators with arbitrary functions, 
\begin{equation}\label{OperatorOnF}
\frac{\del T^i_1}{\del\varphi^j}\cdot F_j    \;\rightarrow\;\frac{\alpha'}{2}\,\Big[\nabla^2 F_{\mu\nu}-2\nabla^\lambda\nabla_{(\mu}F_{\nu)\lambda}+\nabla_\mu\nabla_\nu F\Big]\;. 
\end{equation}

At this point, the main goal is to find an operator expression for the trace of the energy-momentum tensor. 
At the quantum level one can choose 
\begin{equation}\label{trace T}
T^\alpha{}_\alpha=\frac{2}{\sqrt{\gamma}}\,\gamma^{\alpha\beta}\frac{\delta S_0}{\delta\gamma^{\alpha\beta}} \;,   
\end{equation}
which implies that the expectation value
\begin{equation}
\big\langle T^\alpha{}_\alpha\big\rangle=\frac{2}{\sqrt{\gamma}}\,\gamma^{\alpha\beta}\frac{\delta W}{\delta\gamma^{\alpha\beta}}  \end{equation}
is finite.
The Weyl variation of the bare operators is given by 
\begin{equation}\label{WeylVAR}
2\,\gamma^{\alpha\beta}\frac{\del}{\del\gamma^{\alpha\beta}}A_{i0}=-\epsilon\,A_{i0}+\del_\alpha\omega^\alpha_i \,, \end{equation}
where, again, we took into account possible total derivative terms. 
For the $A_{i0}$ defined in (\ref{bareCouplings}) this may be  verified by a direct computation (in which case there are no total derivative terms). 
It should be noted that since the bare operators $A_{i0}$ enter (\ref{WeylVAR})  this total derivative is given directly by the Weyl variation of the Lagrangian, implying  
that it requires no quantum computation.

In the remainder of this subsection we derive a general expression for $T^{\alpha}{}_{\alpha}$ and the relation between the RG beta functions $\beta^{i}$ and the Weyl anomaly coefficient $\bar{\beta}^i$. 
We can, as before, use the completeness of the basis $\{A_{i0}\}$ modulo equations of motion to expand
\begin{equation}\label{lambda defined}
\del_\alpha\omega^\alpha_i=A_{j0}\cdot\lambda^j_i(\varphi_0)\;,\quad\lambda^j_i(\varphi_0)=\lambda^j_i(\varphi)+\cO(\epsilon^{-1}) \;.  \end{equation}
The trace operator is then computed as
\begin{equation}
\begin{split}
\sqrt{\gamma}\,T^\alpha{}_\alpha&=2\,\gamma^{\alpha\beta}\frac{\delta}{\delta\gamma^{\alpha\beta}}S_0=2\,\gamma^{\alpha\beta}\frac{\delta}{\delta\gamma^{\alpha\beta}}\int d^n\sigma\,A_{i0}\cdot\varphi^i_0\\
&=(-\epsilon\,A_{i0}+\del_\alpha\omega^\alpha_i)\cdot\varphi^i_0=A_{i0}\cdot\psi^i\;,
\end{split}    
\end{equation}
where
\begin{equation}
\begin{split}
\psi^i&=-\epsilon\,\varphi^i_0+\lambda^i_j(\varphi_0)\cdot\varphi^j_0\\
&=\mu^{\epsilon}\Big[-\epsilon\,\varphi^i-T^i_1(\varphi)+\lambda^i_j(\varphi)\cdot\varphi^j+\cO(\epsilon^{-1})\Big]\;,
\end{split}    
\end{equation}
thanks to \eqref{Phi0}.
The last step consists in rewriting the bare operators in terms of the renormalized ones, \emph{i.e.} $A_{i0}=[A_j]\cdot(Z^{-1})^j_i$ where, by using \eqref{Z factor},
\begin{equation}
(Z^{-1})^j_i=\mu^{-\epsilon}\,\Big[\delta^j_i-\frac{1}{\epsilon}\,X^j_{1i}+\cO(\epsilon^{-2})\Big]\;. \end{equation}
Combining the last two expressions we find
\begin{equation}\label{combined derivatives}
(Z^{-1})^i_j\cdot\psi^j=-\epsilon\,\varphi^i-T^i_1+\varphi\cdot\frac{\del}{\del\varphi}T^i_1+\lambda^i_j\cdot\varphi^j+Q^i_{1j}\cdot\varphi^j\;.    
\end{equation}
In the above formula all the divergent terms must cancel out, since by definition both $T^\alpha{}_\alpha$ and $[A_i]$ are finite. Upon letting $\epsilon=0$ the Weyl anomaly operator can thus be written as
\begin{equation}\label{trace operator}
\sqrt{\gamma}\,T^\alpha{}_\alpha=[A_i]\cdot{\bar\beta}^i  \;,  
\end{equation}
where the Weyl anomaly coefficients are given by
\begin{equation}\label{betabar general}
\bar\beta^i=\beta^i+\big(\lambda^i_j+Q^i_{1j}\big)\cdot\varphi^j\;.    
\end{equation}
This formula is the precise form of (\ref{betabarintro}), for which we will see in the next subsection that the 
extra term encodes a target space gauge variation. 
One can see that the Weyl anomaly coefficients $\bar\beta^i$ differ from the RG $\beta-$functions by two total derivative terms: the $\lambda-$contribution can be found simply by varying the classical action, while  the $Q-$contribution, which is much harder to compute, can be found by direct renormalization of the dimension two operators $A_i\,$.

\subsection{String Sigma Model}

We can now specialize the general procedure outlined above to the case of the string sigma model coupled to all massless background fields  \cite{Tseytlin:1986ws}, 
and evaluate the right-hand side of (\ref{betabar general}). In particular, we analyze the dilaton contributions.  
We start from the Euclidean bare action
\begin{equation}\label{bare action sigma model}
\begin{split}
S_0&=\frac{1}{2\lambda}\int d^n\sigma\,\Big\{\sqrt{\gamma}\gamma^{\alpha\beta}\del_\alpha X^\mu\del_\beta X^\nu g_{0\mu\nu}(X)+i\,\varepsilon^{\alpha\beta}\del_\alpha X^\mu\del_\beta X^\nu B_{0\mu\nu}(X)+\alpha'\,\sqrt\gamma\bar R\,\Phi_0(X)\Big\}  \;,  
\end{split}    
\end{equation}
with the loop-counting parameter ${\lambda=2\pi\alpha'}\,$. The dimensionally extended worldsheet curvature $\bar R$  is defined with the unusual normalization $\bar R=\frac{1}{n-1}\,R^{(n)}$ \cite{Tseytlin:1986ws,Curci:1986hi}, which  is chosen purely for future convenience. The antisymmetric density $\varepsilon^{\alpha\beta}$ is the most problematic object to define in $n=2+\epsilon$ dimensions. Following \cite{Tseytlin:1986ws}, we choose
\begin{equation}\label{varepsilon}
\varepsilon^{\alpha\beta}=\sqrt\gamma\, e^\alpha{}_a\,e^\beta{}_b\,\epsilon^{ab}\;,  \end{equation}
where $e_\alpha{}^a(\sigma)$ are worldsheet (zwei$\,+\,\epsilon$)-beins, and $\epsilon^{ab}$ is a \emph{constant} antisymmetric Lorentz tensor obeying $\epsilon^{ac}\epsilon_{bc}=\delta^a_b$ in $n$ dimensions. 
Under a Weyl variation one has
\begin{equation}
2\,\gamma^{\alpha\beta} \frac{\delta}{\delta\gamma^{\alpha\beta}}\,\int d^n\sigma\,\varepsilon^{\gamma\delta}B_{\gamma\delta}=e^{\alpha}{}_{a}\,\frac{\delta}{\delta e^{\alpha}{}_{a}}\,\int d^n\sigma\,\varepsilon^{\gamma\delta}B_{\gamma\delta}=-\epsilon\,\varepsilon^{\alpha\beta}B_{\alpha\beta}\;,    
\end{equation}
where $B_{\alpha\beta}$ is the pullback of the target space $B-$field.
The above choice for the continuation to $n$ dimensions has an obvious problem: the action \eqref{bare action sigma model} is not gauge-invariant under transformations of the $B-$field $\delta B_{\mu\nu}=\del_{[\mu}\lambda_{\nu]}$, since the conformal factor of the metric does not drop out for $n=2+\epsilon$. For instance, in conformal gauge $e_\alpha{}^a=e^\rho\delta_\alpha^a$ one has $\varepsilon^{\alpha\beta}=e^{\epsilon\rho}\epsilon^{\alpha\beta}$ and $\del_\alpha(\varepsilon^{\alpha\beta})=\epsilon\,\del_\alpha\rho\,\varepsilon^{\alpha\beta}$. This problem ultimately leads to a trace operator $\sqrt{\gamma}T^\alpha{}_\alpha$ that violates the $B-$field gauge invariance \cite{Tseytlin:1986ws}. The proposal of \cite{Tseytlin:1986ws} was to modify the definition of the trace anomaly operator by further total derivative counterterms, whose role is precisely to cancel non gauge-invariant contributions. In the following we will ignore terms containing $\del_\alpha\varepsilon^{\alpha\beta}$ (they do not alter the general form of $T^\alpha{}_\alpha$ when $\epsilon\to0$) and comment on this issue at the end of this section.

Given the sigma model action \eqref{bare action sigma model}, our goal is to determine the general form of the total derivative terms $\lambda^i_j\cdot\varphi^j$ and $Q^i_{1j}\cdot\varphi^j$ appearing in the Weyl anomaly coefficients \eqref{betabar general}. In order to do so, it is useful to classify the possible total derivative terms. 
There are two total derivative operators of second order in  worldsheet derivatives that are worldsheet diffeomorphism invariant. 
We denote these operators by $d_{\gamma}$ and $d_{\varepsilon}$,  which act on target space vector fields $N_\mu$ and $K_\mu$ as 
\begin{equation}\label{N}
\begin{split}
d_{\gamma}^\mu\cdot N_\mu& :=  \frac{1}{2\lambda}\,\del_\alpha\,\big[\sqrt\gamma \gamma^{\alpha\beta}\del_\beta X^\mu\,N_\mu(X)\big]\;,\\ 
d^\mu_{\varepsilon}\cdot K_\mu& :=  \frac{i}{2\lambda}\,\del_\alpha\,\big[\varepsilon^{\alpha\beta}\del_\beta X^\mu\,K_\mu(X)\big] \;. 
\end{split}
\end{equation}
According to our previous discussion, it should be possible to expand these total derivatives (modulo the bare equations of motion) in terms of a basis of dimension 2 bare operators $\{A_{i0}\}$, which in the present case are given by
\begin{equation}\label{A basis}
\begin{split}
A_{g0}\cdot F^g&=\frac{1}{2\lambda}\,\sqrt{\gamma}\gamma^{\alpha\beta}\del_\alpha X^\mu\del_\beta X^\nu\,F^g_{\mu\nu}(X)\;,\quad A_{B0}\cdot F^B=\frac{i}{2\lambda}\,\varepsilon^{\alpha\beta}\del_\alpha X^\mu\del_\beta X^\nu\,F_{\mu\nu}^B(X)\;,\\
A_{\Phi0}\cdot F^\Phi&=\frac{\alpha'}{2\lambda}\,\sqrt{\gamma}\bar R\,F^\Phi(X)\;,  \end{split}    
\end{equation}
upon contraction with arbitrary functions. 
In order to show that any total derivative \eqref{N} can be expanded in terms of \eqref{A basis}, we shall need the bare equations of motion
 computed from \eqref{bare action sigma model}: \begin{equation}\label{bareom}
\frac{\lambda}{\sqrt{\gamma}}\,g^{\mu\nu}_0\frac{\delta S_0}{\delta X^\nu}=-\calD^\alpha_0\del_\alpha X^\mu+\frac{i}{2\sqrt\gamma}\,\varepsilon^{\alpha\beta}\del_\alpha X^\nu\del_\beta X^\lambda\,H^\mu_0{}_{\nu\lambda}+\frac{\alpha'}{2}\,\bar R\,\nabla^\mu_0\Phi_0=0\;,
\end{equation}
where $H_{\mu\nu\lambda}=3\,\del_{[\mu}B_{\nu\lambda]}$. All quantities with a subscript $0$ are constructed from the bare target space fields, and 
$\calD_{\alpha}$ is the covariant derivative w.r.t~the worldsheet and target space Levi-Civita connections, i.e., 
\begin{equation}
\calD_\alpha v_\beta^\mu=\del_\alpha v_\beta^\mu-\Gamma_{\alpha\beta}^\gamma(\gamma)\,v_\gamma^\mu+\del_\alpha X^\nu\,\Gamma^\mu_{\nu\lambda}(g)\,v_\beta^\lambda\;.    
\end{equation}
Expanding the derivatives in \eqref{N} we obtain, for instance,
\begin{equation}
\begin{split}
2\lambda\,d_{\gamma}^{\mu} \cdot N_{\mu} &=\sqrt\gamma \gamma^{\alpha\beta}\del_\alpha X^\mu\del_\beta X^\nu\,\del_\mu N_\nu(X)+N_\mu(X)\del_\alpha\big(\sqrt\gamma \gamma^{\alpha\beta}\del_\beta X^\mu\big)\\
&=\sqrt\gamma \gamma^{\alpha\beta}\del_\alpha X^\mu\del_\beta X^\nu\,\nabla_{0\mu} N_\nu(X)+\sqrt\gamma\,N_\mu(X)\,\calD_0^\alpha\del_\alpha X^\mu\;.
\end{split}    
\end{equation}
Using the bare equations of motion \eqref{bareom} one can  see that any total derivative term can be expanded in the basis \eqref{A basis} as:
\begin{equation}\label{expand on A basis}
\begin{split}
d_{\gamma}^{\mu} \cdot N_{\mu} &=A^{\mu\nu}_{g0}\cdot\Big(\nabla_{0(\mu}N_{\nu)}\Big)+A^{\mu\nu}_{B0}\cdot\Big(\tfrac12\,H^\lambda_{0\mu\nu}\,N_\lambda\Big)+A_{\Phi0}\cdot\Big(\tfrac12\,\nabla^\mu_0\Phi_0\,N_\mu\Big)\;,\\
d^\mu_{\varepsilon}\cdot K_\mu &=A^{\mu\nu}_{B0}\cdot\Big(\del_{[\mu}K_{\nu]}\Big)\;.
\end{split}    
\end{equation}

Let us emphasize that the above decomposition takes the form of operators $A_{i0}$ contracted with the target space gauge (or diffeomorphism) variation of the corresponding field, i.e. 
\begin{equation}\label{gauge form}
2\,d^\mu_\gamma\cdot N_\mu=A_{i0}\cdot\cL_{N}\varphi^i_0\;,\qquad 2\,d^\mu_\varepsilon\cdot K_\mu=A_{B0}\cdot\delta_K B_0\;,
\end{equation} 
where the Lie derivative on the $B-$field is meant to act in covariant form, meaning $\cL_N B_{\mu\nu}=N^\lambda H_{\lambda\mu\nu}$, modulo a gauge transformation.
The general form of \eqref{gauge form} is to be expected on general grounds: focusing on target space diffeomorphisms, for instance, we recall that the action \eqref{bare action sigma model} is invariant if one transforms \emph{both} the target space couplings $\varphi^i=(g_{\mu\nu}, B_{\mu\nu}, \Phi)$ and the worldsheet fields $X^\mu$. This yields the identity
\begin{equation}
\delta_\xi S=\int d^2\sigma\,\Big[-\xi^\mu(X)\,\frac{\delta S}{\delta X^\mu}+\cL_\xi\varphi^i\cdot\frac{\delta S}{\delta\varphi^i}\Big] =\int d^2\sigma\,\Big[-\xi^\mu(X)\,\frac{\delta S}{\delta X^\mu}+\cL_\xi\varphi^i\cdot A_i\Big]=0\;,   
\end{equation}
where we used that the integrated operators $A_i$ are obtained by differentiating the action w.r.t. the couplings, c.f.~(\ref{Ai0 integrated}). 
Since we are discarding the equations of motion operator, i.e.~$\frac{\delta S}{\delta X^\mu}=0$, the integral of the second term must vanish, implying that
$A_i\cdot\cL_\xi\varphi^i$ is a total worldsheet derivative, hence proportional to $d^\mu_\gamma\cdot\xi_\mu$. The argument for the $B-$field gauge transformation is completely analogous.

With the expansion \eqref{expand on A basis} we are ready to determine the general form of the $\lambda^i_j$ and $Q^i_{1j}$ terms in the Weyl anomaly. We start from the $\lambda^i_j$ contribution that, recalling \eqref{WeylVAR} and \eqref{lambda defined}, can be found by the Weyl variation of the Lagrangian. Upon using the general variation of the Ricci scalar
\begin{equation}
\delta R=\gamma_{\alpha\beta}\nabla^2\delta\gamma^{\alpha\beta}-\nabla_\alpha\nabla_\beta\,\delta\gamma^{\alpha\beta}+R_{\alpha\beta}\,\delta\gamma^{\alpha\beta}\;,
\end{equation}
the Weyl variation of \eqref{bare action sigma model} yields
\begin{equation}
\begin{split}
\sqrt{\gamma}\,T^\alpha{}_\alpha&=2\,\gamma^{\alpha\beta}\frac{\delta S_0}{\delta\gamma^{\alpha\beta}}=\frac{1}{2\lambda}\,\Big[-\epsilon\,\sqrt{\gamma}\gamma^{\alpha\beta}\del_\alpha X^\mu\del_\beta X^\nu g_{0\mu\nu}(X)-\epsilon\,i\,\varepsilon^{\alpha\beta}\del_\alpha X^\mu\del_\beta X^\nu B_{0\mu\nu}(X)\\
&\hspace{36mm}-\epsilon\,\alpha'\sqrt{\gamma}\bar R\,\Phi_0(X)\Big]+\frac{1}{2\pi}\,\sqrt{\gamma}\,\nabla^\alpha\del_\alpha\Phi_0(X)\\[2mm]
&=A_{g0}\cdot(-\epsilon\,g_0)+A_{B0}\cdot(-\epsilon\,B_0)+A_{\Phi0}\cdot(-\epsilon\,\Phi_0)+\frac{1}{2\pi}\,\del_\alpha\big[\sqrt{\gamma}\gamma^{\alpha\beta}\del_\beta\Phi_0\big]\;.
\end{split}    
\end{equation}
The total derivative term in here can be expressed in terms  of \eqref{N}: 
\begin{equation}\label{Nphi}
\frac{1}{2\pi}\del_\alpha\big[\sqrt{\gamma}\gamma^{\alpha\beta}\del_\beta\Phi_0\big]
= d_{\gamma}^\mu\cdot N_\mu\;, \qquad \text{where} \qquad N_\mu=2\alpha'\del_\mu\Phi_0 \;. 
\end{equation}
By using the decomposition \eqref{expand on A basis}, the trace operator can thus be put in the form $\sqrt{\gamma}\,T^\alpha{}_\alpha=A_{i0}\cdot\psi^i$,
with
\begin{equation}
\begin{split}
\psi^g_{\mu\nu}&=-\epsilon\,g_{0\mu\nu}+2\alpha'\,\nabla_{0\mu}\del_\nu\Phi_0\;,\quad \psi^B_{\mu\nu}=-\epsilon\,B_{0\mu\nu}+\alpha'\,H^\lambda_0{}_{\mu\nu}\,\del_\lambda\Phi_0\;,\\
\psi^\Phi&=-\epsilon\,\Phi_0+\alpha'\,\nabla_0^\mu\Phi_0\,\del_\mu\Phi_0\;.
\end{split}    
\end{equation}
Comparing the above result with the general formulas
\begin{equation}
\psi^i=-\epsilon\,\varphi^i_0+\lambda^i_j(\varphi_0)\cdot\varphi^j_0\;,\quad\bar\beta^i=\beta^i+\lambda^i_j(\varphi)\cdot\varphi^j+Q^i_{1j}(\varphi)\cdot\varphi^j\;,    
\end{equation}
we can read off the $\lambda^i_j$ from the first equation, and using this in the second equation one obtains 
the $\lambda-$contribution to the Weyl anomaly coefficients in terms of the familiar dilaton terms:
\begin{equation}\label{betabar partial}
\begin{split}
\bar\beta^g_{\mu\nu}&=\beta^g_{\mu\nu}+2\alpha'\,\nabla_\mu\nabla_\nu\Phi+\cdots \;,\quad \bar\beta^B_{\mu\nu}=\beta^B_{\mu\nu}+\alpha'\,H^\lambda{}_{\mu\nu}\nabla_\lambda\Phi+\cdots\;,\\[2mm]
\bar\beta^\Phi&=\beta^\Phi+\alpha'\,(\nabla\Phi)^2+\cdots\;,
\end{split}    
\end{equation}
where the dots stand for the contributions $Q^i_{1j}\cdot\varphi^j\,$.

Instead of trying to determine the $Q-$contributions to $\bar\beta^i$, we shall only fix their general form. Let us recall from \eqref{Ai ren unintegrated} and \eqref{Qs defined} that objects involving $Q^i_{1j}$ arise from the decomposition along the basis $\{A_{i0}\}$ of the total derivatives $\del_\alpha\Omega^\alpha_j$. Since any total worldsheet derivative can only depend on two distinct target space vectors (see \eqref{N}), the $Q-$contributions can be parametrized in terms of two undetermined field-dependent vectors: $W_\mu$, related to the $d_\gamma$ structure, and $L_\mu$, related to the $d_\varepsilon$ structure. Since the decomposition \eqref{expand on A basis} is valid for any pair  of vector fields, the $W-$contribution to the Weyl anomaly must be of the form of a target space diffeomorphism, c.f. \eqref{gauge form}, while the $L-$contribution can only take the form of a $B-$field gauge transformation. Including these terms in the expression for the Weyl anomaly coefficients, we obtain the final form
\begin{equation}\label{beta bars}
\begin{split}
\bar\beta^g_{\mu\nu}&=\beta^g_{\mu\nu}+2\alpha'\,\nabla_\mu\nabla_\nu\Phi+\nabla_{(\mu}W_{\nu)} \;,\quad \\[2mm]
\bar\beta^B_{\mu\nu}&=\beta^B_{\mu\nu}+\alpha'\,H^\lambda{}_{\mu\nu}\nabla_\lambda\Phi+\tfrac12\,H^\lambda{}_{\mu\nu}W_\lambda+\del_{[\mu}L_{\nu]}\;,\\[2mm]
\bar\beta^\Phi&=\beta^\Phi+\alpha'\,(\nabla\Phi)^2+\tfrac12\nabla^\mu\Phi\,W_\mu\;.
\end{split}    
\end{equation}
This gives the $Q$ contributions as the terms involving $W$ and $L$. 

Let us discuss this result in more detail. First of all, 
the renormalization of the dilaton operator $A_\Phi$ does not produce total derivatives. This can be understood by observing that the scalar curvature operator $\bar R\sim\del\del\gamma$ cannot mix with total derivative operators, which contain $\del X$, under renormalization.
A similar argument, which we will review in the next section, shows that the $\beta-$functions $\beta^g_{\mu\nu}$ and $\beta^B_{\mu\nu}$ do not depend on the dilaton, and $\beta^\Phi$ itself is only linear in $\Phi$, i.e.~$\beta^\Phi=\Delta(g,B)\,\Phi+\omega(g,B)$. We shall thus stress that the dilaton dependence displayed in \eqref{beta bars} is exact to all orders in $\alpha'$ and, in particular, $W_\mu(g,B)$ and $L_\mu(g,B)$ do not depend on $\Phi$.

We conclude this section by discussing the issue with the $B-$field gauge invariance. As we have pointed out at the beginning of this subsection, the definition of the $\varepsilon^{\alpha\beta}$ density \eqref{varepsilon} in $n$ dimensions breaks gauge invariance, in that the vectors $W_\mu$ and $L_\mu$ have in principle a general dependence on $B$, rather than on $H=dB$. 
Another way to circumvent this problem, compared to the proposal of \cite{Tseytlin:1986ws}, is to derive a manifestly gauge invariant perturbative expansion in two dimensions, and to extend the gauge invariant interactions to $n=2+\epsilon$ dimensions only afterwards. Finally, one can try to fix $W $ and $L$ by more indirect means (see \emph{e.g.} \cite{Metsaev:1987zx}). For instance, by expanding in powers of $\alpha'$ and imposing covariance, the most general form of $W$ is
$W_\mu(g,B)=\sum_{L=1}^\infty(\alpha')^LW_\mu^{(L)}(R,\nabla,H)$, where $W_\mu^{(L)}$ contains $2L-1$ derivatives of the metric and $B-$field. This already implies that at one-loop $W_\mu^{(1)}=0$ and
\begin{equation}
W_\mu(R,\nabla,H)=(\alpha')^2\,\Big(a_1\,\nabla_\mu R+a_2\,\nabla_\mu H^2+a_3\,H_{\mu\nu\lambda}\nabla_\rho H^{\rho\nu\lambda}\Big)+\cO(\alpha')^3\;, \end{equation}
and similarly for $L_{\mu}$, respectively. 
For our subsequent applications to the duality invariant sigma model  in sec.~4 we will also find that at one-loop level these vectors vanish.

\section{ 
Background Field Method}

In this section we revisit the techniques for computing the $\beta-$functionals of the string nonlinear sigma model defined by
\begin{equation}\label{sigma g+phi}
S=\frac{1}{2\lambda}\int d^2x\,\sqrt{\gamma}\Big[\gamma^{\alpha\beta} g_{\mu\nu}(X)\,\del_\alpha X^\mu\del_\beta X^\nu+\alpha'\,R^{(2)}\,\Phi(X)\Big] \;,\quad\lambda=2\pi\alpha'\;,
\end{equation}
on a curved \emph{Euclidean} worldsheet of spherical topology.
For now  we discard the coupling to the Kalb-Ramond field $B_{\mu\nu}$.  
We will  compute the quantum effective action $\Gamma$ using the background-field method 
and deduce the corresponding $\beta-$functions by renormalization of the couplings. 

\subsection{Effective Action}

We start by considering a generic field theory with Euclidean action $S[\phi]\,$. The generating functional of all correlators is given by
\begin{equation}\label{Z}
Z[J]=\int \calD\phi\,e^{-S[\phi]+J\cdot\phi}=Z[0]\,\Big(1+\sum_{n=1}^{\infty}\tfrac{1}{n!}\,J_1...J_n\,\langle\phi_1...\phi_n\rangle\Big)  \;, 
\end{equation}
where spacetime integrals and positions are condensed in the notation
\begin{equation}
\begin{split}
&J\cdot\phi:=\int d^nx\,J(x)\,\phi(x)\;,\\
&J_1...J_n\,\langle\phi_1...\phi_n\rangle:=\int d^nx_1...d^nx_n\,J(x_1)...J(x_n)\,\langle\phi(x_1)...\phi(x_n)\rangle\;, 
\end{split}    
\end{equation}
etc. The factor $Z:=Z[0]$
ensures that the correlators in \eqref{Z} are normalized, $\langle1\rangle=1\,$.
The generating functional of the connected correlators, $W[J]$, is defined by
\begin{equation}
W[J]:=\log\frac{Z[J]}{Z[0]}  \;, 
\end{equation}
which ensures that $W[0]=0$, meaning that all the vacuum bubbles are subtracted: 
\begin{equation}
W[J]=\sum_{n=1}^{\infty}\frac{1}{n!}\,J_1...J_n\,\langle\phi_1...\phi_n\rangle_{\rm connected}  \;.  
\end{equation}

Following standard terminology  we define the field $\varphi$, which is a function of the source $J$, 
as the quantum expectation value: 
\begin{equation}\label{varphi of J}
\varphi(J):=\frac{\delta W}{\delta J}=\langle\phi\rangle_J\;. 
\end{equation}
Assuming tadpole cancellation one has $\varphi(0)=\langle\phi\rangle=0\,$. If $\varphi(J)$ is invertible, the generating functional of one-particle-irreducible (1PI) diagrams, 
\emph{i.e.}~the quantum effective action $\Gamma[\varphi]$, is given by the Legendre transform of $W[J]\,$:
\begin{equation}
\Gamma[\varphi]=J\cdot\varphi-W[J]  \;.  
\end{equation}
In order to see that $\Gamma$ is genuinely a functional of $\varphi$ only, one can compute the variation
\begin{equation}
\delta\Gamma=\delta J\cdot\varphi+J\cdot\delta\varphi-\frac{\delta W}{\delta J}\cdot\delta J=J\cdot\delta\varphi\;,
\end{equation}
where we have used \eqref{varphi of J}. This also gives the quantum equation of motion in the form
\begin{equation}\label{J of varphi}
\frac{\delta\Gamma}{\delta\varphi}=J  \,,   
\end{equation}
that determines $J$ as a function of $\varphi\,$.

We can now use these definitions to manipulate the path integral:
\begin{equation}
\begin{split}
e^{W[J]}&=e^{-\Gamma[\varphi]+J\cdot\varphi}=\frac{1}{Z}\int\calD\phi\,e^{-S[\phi]+J\cdot\phi} \;\rightarrow\\
e^{-\Gamma[\varphi]}&=\frac{1}{Z}\int\calD\phi\,e^{-S[\phi]+J\cdot(\phi-\varphi)}\;.
\end{split}
\end{equation}
Using the definition \eqref{J of varphi} and shifting the integration variable by
$\phi=\varphi+\pi$ we obtain the most convenient form for the background field method:
\begin{equation}\label{Gamma in BG}
e^{-\frac{1}{\hbar}\Gamma[\varphi]}=\frac{1}{Z}\int\calD\pi\,\exp\Big\{-\frac{1}{\hbar}\,\Big(S[\varphi+\pi]-\frac{\delta\Gamma}{\delta\varphi}\cdot\pi\Big)\Big\}\;,   
\end{equation}
where we have reinstated Planck's constant $\hbar\,$. 
Here $\varphi$ is viewed as the classical background and $\pi$ as the quantum fluctuation. 

Let us mention that there is an all-order subtraction  in \eqref{Gamma in BG} 
given implicitly by the term $-\frac{\delta\Gamma}{\delta\varphi}\cdot\pi$, which removes non-(1PI)  contributions from $\Gamma\,$. 
In fact, $\Gamma[\varphi]$  appears on both sides of  \eqref{Gamma in BG}, 
which is therefore an integral equation. 
As usual, one can solve it perturbatively order by order in loops.
More precisely, one starts by writing $\Gamma$ as a power series in $\hbar\,$:
\begin{equation}
\Gamma[\varphi]=S[\varphi]+\hbar\,\Gamma_{ 1l}[\varphi]+{\cal O}(\hbar^2)  \;,  
\end{equation}
and expands the shifted action $S[\varphi+\pi]$ in powers of the fluctuation $\pi$ as
\begin{equation}
S[\varphi+\pi]=S[\varphi]+\frac{\delta S}{\delta\varphi}\cdot\pi+\frac12\,\pi\cdot\frac{\delta^2S}{\delta\varphi\delta\varphi}\cdot\pi+{\cal O}(\pi^3) \;. \end{equation}
The linear term in $\pi$ above is canceled by the subtraction in \eqref{Gamma in BG} to lowest order and no higher subtractions are needed at one-loop, as one can see by counting $\hbar$ powers. Similarly, one-loop diagrams require terms at most quadratic in $\pi$ from the above expansion. As the last ingredient, we shall extract the pure kinetic term for the fluctuations:
\begin{equation}
S_{2\pi}[\pi;\varphi]:=\frac12\,\pi\cdot\frac{\delta^2S}{\delta\varphi\delta\varphi}\cdot\pi=S_0[\pi]+S_{\rm int}[\pi;\varphi]=S_0[\pi]+{\cal O}(\varphi)\,, 
\end{equation}
and define the partition function of the free theory
\begin{equation}
Z_0:=\int\calD\pi\,e^{-\frac{1}{\hbar}S_0[\pi]}  \;.  
\end{equation}
This gives the expression for the one-loop contribution to the effective action:
\begin{equation}\label{gamma BG 1l}
e^{-\Gamma_{1l}[\varphi]}=\frac{1}{Z_0}\int\calD\pi\,e^{-\frac{1}{\hbar}\,S_{2\pi}[\pi;\varphi]} =\left\langle e^{-\frac{1}{\hbar}\,S_{\rm int}[\pi;\varphi]}\right\rangle_0  \;, 
\end{equation}
where the subscript $0$ denotes the normalized average with respect to the free action $S_0[\pi]$. 

\subsection{Nonlinear Sigma Model}

It is possible, in principle, to apply the background-field method in the form just described to the nonlinear sigma model
\begin{equation}\label{metric sigma}
S=\frac{1}{2\lambda}\int d^2x\,g_{\mu\nu}(X)\,\del_\alpha X^\mu\del^\alpha X^\nu \;,
\end{equation}
where, for the time being, we are considering a flat worldsheet. However, applying the linear background-quantum split by defining $X^\mu=\varphi^\mu+\pi^\mu$, leads to a perturbative expansion lacking manifest target space covariance, since the fluctuation field $\pi^\mu$ is a coordinate difference and thus has no geometric meaning. To remedy this 
we employ a field redefinition of $\pi^{\mu}$ as follows \cite{AlvarezGaume:1981hn,Mukhi:1985vy}: We consider a geodesic 
$X^\mu(t)$, where $t$ is the affine parameter (and we suppress  the dependence on the worldsheet coordinates $x^\alpha$), 
such that $X^\mu(0)=\varphi^\mu$ and $X^\mu(1)=\varphi^\mu+\pi^\mu$, with tangent vector $\xi^\mu(t)=\frac{dX^\mu}{dt}\,$. We then  use the tangent vector at $t=0$, $\xi^\mu:=\xi^\mu(0)$, as the quantum field for the expansion. Since it is a genuine vector, this ensures manifest target space covariance.

It is possible to derive the exact nonlinear relation $\pi^\mu(\xi)=\xi^\mu-\frac12\,\Gamma^\mu_{\nu\lambda}\,\xi^\nu\xi^\lambda+\cO(\xi^3)$ implementing the field redefinition and use it to write down the covariant expansion. This procedure, however, becomes very cumbersome after a few orders, even when resorting to a normal coordinate system. 
A considerable simplification was found in \cite{Mukhi:1985vy} by noting that one usually needs to expand only the Lagrangian, which is just a scalar. We thus consider a scalar field evaluated along the geodesic: $\Phi(t)\equiv\Phi(X(t))\,$. We can expand it around $t=0$, yielding
\begin{equation}
\Phi(t)=\sum_{n=0}^\infty\frac{t^n}{n!}\,\left.\frac{d^n\Phi}{dt^n}\right\rvert_{t=0} \;.   
\end{equation}
Since $\Phi$ is a scalar field, the derivative along the geodesic $\frac{d}{dt}\equiv\xi^\mu\del_\mu$ is already manifestly covariant, in that
\begin{equation}
\frac{D\Phi}{Dt}:=\xi^\mu\nabla_\mu\Phi=\xi^\mu\del_\mu\Phi=\frac{d\Phi}{dt}\;,    
\end{equation}
which defines the covariant derivative $\frac{D}{Dt}$ along the geodesic. One can of course also define $\frac{D}{Dt}=\xi^\mu\nabla_\mu$ on general 
spacetime tensors evaluated on the curve, and
\begin{equation}
\frac{D}{Dt}T^{\mu\dots}_{\nu\dots}(t)=\frac{d}{dt}T^{\mu\dots}_{\nu\dots}+\xi^\lambda\,\Gamma^\mu_{\lambda\rho}\,T^{\rho\dots}_{\nu\dots}+\cdots    
\end{equation}
for objects defined only on the curve.
Given that any application of $\frac{d}{dt}$ maps the scalar into a scalar, it immediately follows by induction that 
\begin{equation}
\frac{d^n\Phi}{dt^n}=\frac{d^{n-1}}{dt^{n-1}}\left(\xi\cdot\nabla\Phi\right) 
=(\xi\cdot\nabla)^n\Phi=\frac{D^n\Phi}{Dt^n}\;. 
\end{equation}
This yields the covariant expansion
\begin{equation}\label{covariant expansion}
\Phi(t)=\sum_{n=0}^\infty\frac{t^n}{n!}\,
\left.\frac{D^n\Phi}{Dt^n}\right\rvert_{t=0} 
\;. 
\end{equation}
Let us also mention that, thanks to the geodesic equation $\frac{D\xi^\mu}{Dt}=0$, one can freely rearrange the tangent vectors as
\begin{equation}
\frac{D^n\Phi}{Dt^n}=(\xi\cdot\nabla)^n\Phi=\xi^{\mu_1}\cdots \xi^{\mu_n}\,\nabla_{\mu_1}\cdots\nabla_{\mu_n}\Phi\;.    
\end{equation}

We now apply  the above result to the Lagrangian, for which we 
recall that $\varphi^{\mu}=X^{\mu}(0)$ and  $\varphi^\mu+\pi^\mu=X^\mu(1)$  so that (\ref{covariant expansion}) should be 
evaluated at $t=1$. This gives \cite{Howe:1986vm} \begin{equation}\label{LagrangianExp}
\cL(\varphi+\pi(\xi))=\exp\left(\frac{D}{Dt}\right)\cL(\varphi)\;.    
\end{equation}
In order to apply this to the sigma model we need to consider the pullback to the worldsheet. 
To this end we use  the worldsheet-dependent geodesic $X^\mu(x;t)$ and  $\xi^\mu(x;t)$ to compute 
\begin{equation}
\begin{split}
\frac{D}{Dt}\del_\alpha X^\mu &=\frac{d}{dt}\del_\alpha X^\mu+\xi^\nu\Gamma^\mu_{\nu\lambda}\del_\alpha X^\lambda
=\del_\alpha\xi^\mu+\del_\alpha X^\nu\Gamma^\mu_{\nu\lambda}\xi^\lambda
=: D_\alpha\xi^\mu\;, 
\end{split}    
\end{equation}
where we have defined the covariant derivative $D_\alpha$ on the worldsheet. 
This covariant derivative acts as $D_\alpha=\del_\alpha X^\mu\nabla_\mu$ on the pullback of target space tensors.
To act further with $\Dt$ we need  the commutator
\begin{equation}
\begin{split}
\Big[\Dt, D_\alpha\Big]&=\big[\xi^\mu\nabla_\mu, \del_\alpha X^\nu\nabla_\nu\big]\\
&=\xi^\mu\del_\alpha X^\nu\,\big[\nabla_\mu, \nabla_\nu\big]+\Big(\Dt\del_\alpha X^\nu\Big)\nabla_\nu-(D_\alpha\xi^\mu)\nabla_\mu\\
&=\xi^\mu\del_\alpha X^\nu\,R_{\mu\nu}^\#\;,
\end{split}    
\end{equation}
where $R_{\mu\nu}^\#$ is the Riemann tensor acting as an operator, \emph{e.g.}~$R_{\mu\nu}^\#V^\rho=R_{\mu\nu}{}^\rho{}_\lambda\,V^\lambda\,$.
One can thus determine, for instance,
\begin{equation}
\Big(\Dt\Big)^2\del_\alpha X^\mu=\Dt D_\alpha\xi^\mu=\Big[\Dt, D_\alpha\Big]\xi^\mu=\xi^\nu\del_\alpha X^\lambda\,R_{\nu\lambda}{}^\mu{}_\rho\,\xi^\rho  \;. 
\end{equation}
 
These tools allow one to systematically expand the sigma model action in a simple and recursive manner. Using (\ref{LagrangianExp}) the first few orders of the expansion of the worldsheet sigma model (\ref{metric sigma}) are easily obtained as
\begin{equation}\label{Mukhi metric}
\begin{split}
S_{0\xi}&=\frac{1}{2\lambda}\int d^2x\,\Big[g_{\mu\nu}(\varphi)\,\del_\alpha\varphi^\mu\del^\alpha\varphi^\nu\Big]\;,\\[2mm]
S_{1\xi}&=\frac{1}{2\lambda}\int d^2x\,\Dt\Big[g_{\mu\nu}(X)\,\del_\alpha X^\mu\del^\alpha X^\nu\Big]\Big\rvert_{t=0}=\frac{1}{\lambda}\int d^2x\,\Big[g_{\mu\nu}(\varphi)\,D_\alpha\xi^\mu\del^\alpha\varphi^\nu\Big]\;,\\[2mm]
S_{2\xi}&=\frac{1}{\lambda}\int d^2x\,\frac12\,\Dt\Big[g_{\mu\nu}(X)\,D_\alpha\xi^\mu\del^\alpha X^\nu\Big]\Big\rvert_{t=0}\\
&=\frac{1}{2\lambda}\int d^2x\,\Big[g_{\mu\nu}(\varphi)\,D^\alpha\xi^\mu D_\alpha\xi^\nu+R_{\mu\nu\lambda\rho}(\varphi)\,\del^\alpha\varphi^\mu\del_\alpha\varphi^\rho\,\xi^\nu\xi^\lambda\Big]\;,
\end{split}    
\end{equation}
where we recall that, after taking the derivatives, evaluating at $t=0$ amounts to replacing $X^{\mu}$ by $\varphi^{\mu}$. 
This already exhausts the terms needed at one-loop.

\subsection{One-loop Effective Action}

Having found the covariant background-field expansion $S_{2\xi}$ in \eqref{Mukhi metric}, we can write the one-loop effective action in terms of  the path integral
\begin{equation}\label{1loopPI}
e^{-\Gamma_{1l}[\varphi]}=\frac{1}{Z_0}\int\calD\xi\,e^{-S_{2\xi}[\xi;\varphi]} \,, \end{equation}
with
\begin{equation}\label{S2xi}
S_{2\xi}[\xi;\varphi]=\frac{1}{2\lambda}\int d^2x\,\Big[g_{\mu\nu}\,D^\alpha\xi^\mu D_\alpha\xi^\nu+R_{\mu\nu\lambda\rho}\,\del^\alpha\varphi^\mu\del_\alpha\varphi^\rho\,\xi^\nu\xi^\lambda\Big] \;, \end{equation}
where all target space tensors are evaluated at $\varphi^\mu\,$. Comparing \eqref{1loopPI} with the general formulas \eqref{Gamma in BG} and \eqref{gamma BG 1l}, one can see that the linear subtraction $S[\varphi+\pi]-\frac{\delta S}{\delta\varphi}\cdot\pi$ has not been performed in the usual way, since we rather subtracted $\frac{\delta S}{\delta\varphi}\cdot\xi$ from $S[\varphi+\pi(\xi)]\,$. This choice leads to a covariant one-loop effective action even for off-shell backgrounds $\varphi
^\mu$, while it coincides with any other choice for  on-shell backgrounds  \cite{Hull:1985rc}.
Let us note that the general form of the effective action is
\begin{equation}
\Gamma[\varphi]=\frac{1}{2\lambda}\int d^2x\,\Big[\cT_{\mu\nu}(g)\,\del^\alpha\varphi^\mu\del_\alpha\varphi^\nu+ \cdots \Big]   \;, 
\end{equation}
where $\cT_{\mu\nu}$ is some target space tensor, which prior to renormalization contains  divergent coefficients. 
The divergent part of $\cT_{\mu\nu}$ determines the renormalization of the metric and hence determines the $\beta-$function. 
For this reason, it is sufficient for our purposes to consider contributions to $\Gamma$ with only two factors of $\del_\alpha\varphi^\mu$, 
while the ellipsis denote terms with more than two derivatives of $\varphi$.

Let us now examine the perturbative expansion of the action. 
The kinetic term of \eqref{S2xi}, $g_{\mu\nu}(\varphi)\del^\alpha\xi^\mu\del_\alpha\xi^\nu$, has a non-standard form, since $g_{\mu\nu}(\varphi)$ is not constant on the worldsheet. To overcome this difficulty it is customary to introduce vielbeins $e_\mu^a(\varphi)$ and flatten the fluctuation by introducing $\xi^a=e_\mu^a\,\xi^\mu\,$. The covariant derivative $D_\alpha\xi^a=\del_\alpha\xi^a+\del_\alpha\varphi^\mu\,\omega_{\mu}{}^a{}_b\,\xi^b$ now involves the spin connection, and the action takes the form
\begin{equation}\label{S2xiflat}
\begin{split}
S_{2\xi}[\xi;\varphi]&=\frac{1}{2\lambda}\int d^2x\,\Big[D^\alpha\xi^a D_\alpha\xi_a+R_{\mu ab\nu}\,\del^\alpha\varphi^\mu\del_\alpha\varphi^\nu\,\xi^a\xi^b\Big]\\
&=\frac{1}{2\lambda}\int d^2x\,\del^\alpha\xi^a \del_\alpha\xi_a+S_{\rm int}[\xi;\varphi]\;,
\end{split}
\end{equation}
which has a standard kinetic term.
One can see that the worldsheet coupling constant $\lambda=2\pi\alpha'$ 
is the loop counting parameter, implying that the effective action at $L$ loops is of order $(\alpha')^{L-1}$. 
Writing out  the interaction part of \eqref{S2xiflat} we obtain
\begin{equation}\label{sint expansion}
\begin{split}
S_{\rm int}[\xi;\varphi]&= \frac{1}{2\lambda}\int d^2x\,\Big[2\,\del^\alpha\varphi^\mu\,\omega_{\mu\,ab}\,\xi^b\del_\alpha\xi^a+\del^\alpha\varphi^\mu\del_\alpha\varphi^\nu\,\omega_{\mu\,ca}\,\omega_\nu{}^c{}_b\,\xi^a\xi^b \\
&\qquad\qquad\qquad  + R_{\mu ab\nu}\,\del^\alpha\varphi^\mu\del_\alpha\varphi^\nu\,\xi^a\xi^b\Big] \\
&=: S_{\omega}+S_{\omega\omega}+S_R\;.
\end{split}    
\end{equation}

We are now ready to evaluate  the  one-loop effective action, which is given by
\begin{equation}
e^{-\Gamma_{1l}[\varphi]}=\left\langle e^{-S_{\rm int}[\xi;\varphi]}\right\rangle    \,, 
\end{equation}
where angle brackets denote  normalized free expectation values. The only terms contributing to the renormalization of the metric are
\begin{equation}\label{Gamma1lexpansion}
\Gamma_{1l}=\langle S_R\rangle+\langle S_{\omega\omega}\rangle-\tfrac12\,\langle S_{\omega}^2\rangle_{1{\rm PI}}+\cdots\;, \end{equation}
where dots stand for terms with more than two factors of $\del_\alpha\varphi^\mu$.
The propagator can be derived from the free part of \eqref{S2xiflat} and reads
\begin{equation}
\langle\xi^a(x)\,\xi^b(y)\rangle=\lambda\delta^{ab}G(x-y)\;,\quad G(x)=\int\frac{d^2p}{(2\pi)^2}\,\frac{e^{ip\cdot x}}{p^2}\;. \end{equation}
On dimensional grounds, and using gauge invariance, it follows  
that the terms $S_\omega$ and $S_{\omega\omega}$ involving the spin connection
cannot contribute to UV divergences. For the sake of completeness, we shall compute their contribution to \eqref{Gamma1lexpansion} nonetheless and show that it is UV finite. 
Taking the expectation values and performing the Wick contractions we have
\begin{equation}
\begin{split}
\langle S_{2\omega}\rangle&-\tfrac12\,\langle S_{1\omega}^2\rangle=\tfrac12 \int d^2x\,\omega_\alpha{}^{ab}(x)\,\omega^\alpha{}_{ab}(x)\,G(0)\\
&+\tfrac12\int d^2xd^2y\,\omega_\alpha{}^{ab}(x)\,\omega_{\beta\,ab}(y)\,\Big[G(x-y)\del^\alpha\del^\beta G(x-y)-\del^\alpha G(x-y)\del^\beta G(x-y)\Big]  \;, 
\end{split}
\end{equation}
where we introduced the pullback $\omega_\alpha{}^{ab}=\del_\alpha\varphi^\mu\,\omega_{\mu}{}^{ab}$, that behaves as an $SO(D)$ gauge field in two dimensions.
Going to momentum space by defining
\begin{equation}
\omega_{\alpha\,ab}(x)=\int\frac{d^2p}{(2\pi)^2}\,\omega_{\alpha\,ab}(p)\,e^{ip\cdot x}    
\end{equation}
one obtains 
\begin{equation}
\begin{split}
\langle S_{2\omega}\rangle-\tfrac12\,\langle S_{1\omega}^2\rangle&=\tfrac12\int\frac{d^2p}{(2\pi)^2}\,\omega_\alpha{}^{ab}(p)\,\Pi^{\alpha\beta}(p)\,\omega_{\beta\,ab}(-p) \;,\\
\Pi^{\alpha\beta}(p)&=-\tfrac12\int\frac{d^2k}{(2\pi)^2}\,\frac{(2k^\alpha+p^\alpha)(2k^\beta+p^\beta)-2\delta^{\alpha\beta}(p+k)^2}{k^2(p+k)^2}\;.
\end{split}    
\end{equation}
In order to regularize the polarization tensor $\Pi^{\alpha\beta}$
we continue to $n=2+\epsilon$ dimensions and introduce an infrared mass regulator $m^2$ by changing the propagator\footnote{This amounts to adding the mass term $\frac{m^2}{2\lambda}\int d^2x\,\xi^a\xi_a$ to the action, that suffices to regularize  infrared divergences at one-loop.} as $\frac{1}{p^2}\to\frac{1}{p^2+m^2}\,$. This yields
\begin{equation}
\Pi_{\rm reg}^{\alpha\beta}(p)=-\tfrac12\,\mu^{2-n}\int\frac{d^nk}{(2\pi)^n}\,\frac{(2k^\alpha+p^\alpha)(2k^\beta+p^\beta)-2\delta^{\alpha\beta}[(p+k)^2+m^2]}{(k^2+m^2)[(p+k)^2+m^2]}   \;, 
\end{equation}
where we introduced the arbitrary mass parameter $\mu$ to keep $\Pi^{\alpha\beta}_{\rm reg}$ dimensionless. After some standard manipulations for one-loop integrals (see appendix \ref{Feynman} for details) one obtains
\begin{equation}\label{omega bubble}
\begin{split}
\Pi_{\rm reg}^{\alpha\beta}(p)&=(\delta^{\alpha\beta}p^2-p^\alpha p^\beta)\,\Pi_{\rm reg}(p^2)\;,\\
\Pi_{\rm reg}(p^2)&=\frac{2\,\mu^{2-n}}{(4\pi)^{n/2}}\Gamma(2-\tfrac{n}{2})\int_{-1/2}^{+1/2}dy\,y^2\,\Big[(\tfrac14-y^2)p^2+m^2\Big]^{\frac{n}{2}-2}\;,
\end{split}    
\end{equation}
which is perfectly finite for $n=2$, provided one keeps $m^2$ fixed. This shows that the $\omega-$terms are UV finite and do not contribute to renormalization. 
Note that for this computation it was necessary to introduce the infrared regulator, because the integral is IR divergent in $n=2\,$:
\begin{equation}
\Pi_{\rm reg}(p^2)\;\stackrel{n=2}{\longrightarrow}\;-\frac{1}{2\pi p^2}\Big[\frac{R}{2}\log\frac{R-1}{R+1}+1\Big]\;,\quad R=\Big(1+\frac{4m^2}{p^2}\Big)^{1/2}\;.
\end{equation}
While at this point it is clear that the logarithmic divergence appearing as $m^2\to0$ is of infrared nature, it would have been dangerous to set $m^2=0$ beforehand, since the IR divergence would have reincarnated in a $\frac{1}{\epsilon}$ pole.

From the expansion \eqref{Gamma1lexpansion} we are left with the single divergent contribution
\begin{equation}
\Gamma_{1l}^{\rm div}=\langle S_R\rangle=-\frac12\,G(0)\int d^2x\,R_{\mu\nu}\,\del^\alpha\varphi^\mu\del_\alpha\varphi^\nu \;. \end{equation}
As before, we regularize the propagator at coinciding points $G(0)=\int\frac{d^2p}{(2\pi)^2}\frac{1}{p^2}$ by continuing to $n=2+\epsilon$ dimensions and introducing the IR mass regulator:
\begin{equation}\label{G(0)}
G(0)_{\rm reg}=\mu^{2-n}\int\frac{d^np}{(2\pi)^n}\,\frac{1}{p^2+m^2}=\frac{1}{4\pi}\,\left(\frac{m^2}{4\pi\mu^2}\right)^{\frac{\epsilon}{2}}\Gamma(-\epsilon/2) \;. \end{equation}
In order to extract the pole, we expand the gamma function $\Gamma(x)=\frac{1}{x}-\gamma+{\cal O}(x)$, where $\gamma$ is the Euler-Mascheroni constant, and obtain
\begin{equation}
\Gamma_{1l}^{\rm div}=\frac{1}{4\pi}\,\left(\frac{1}{\epsilon}+\log\frac{m}{\mu}\right)\,\int d^2x\,R_{\mu\nu}\,\del^\alpha\varphi^\mu\del_\alpha\varphi^\nu \;.  \end{equation}
Here we redefined the renormalization scale as $4\pi e^{-\gamma}\mu^2\rightarrow\mu^2$, as it is customary in the $\overline{\rm MS}$ scheme. At this point we can fix the one-loop counterterm by demanding that it cancels the divergence:
\begin{equation}\label{counter g}
S_{\rm c.t.}=-\frac{1}{4\pi\epsilon}\int d^2x\,R_{\mu\nu}\,\del^\alpha\varphi^\mu\del_\alpha\varphi^\nu\;, \end{equation}
which yields the renormalized coupling at one-loop order:
\begin{equation}
\Gamma_{\rm ren}=\frac{1}{2\lambda}\int d^2x\,\Big(g_{\mu\nu}+\frac{\lambda}{2\pi}\,\log\frac{m}{\mu}\,R_{\mu\nu}\Big)\,\del^\alpha\varphi^\mu\del_\alpha\varphi^\nu\;. \end{equation}

In order to extract the $\beta-$function, we apply the method outlined in the previous section. Following eq.~\eqref{useful bare} we write the bare action as 
\begin{equation}
S_0=S+S_{\rm c.t.}=\frac{1}{2\lambda}\int d^nx\,g_{\mu\nu}^0\,\del^\alpha\varphi^\mu\del_\alpha\varphi^\nu =\frac{1}{2\lambda}\int d^nx\,\mu^{\epsilon}\,\big(g_{\mu\nu}+T_{\mu\nu}(g)\big)\,\del^\alpha\varphi^\mu\del_\alpha\varphi^\nu  \;,  
\end{equation}
where, using (\ref{counter g}),  $T_{\mu\nu}=-\frac{\lambda}{2\pi\epsilon}R_{\mu\nu}$. 
This determines  the bare metric to be 
\begin{equation}
g_{\mu\nu}^0=\mu^\epsilon\,\Big(g_{\mu\nu}-\frac{\lambda}{2\pi\epsilon}\,R_{\mu\nu}\Big)\equiv\mu^\epsilon\,\Big(g_{\mu\nu}-\frac{\alpha'}{\epsilon}\,R_{\mu\nu}\Big)\;.    
\end{equation}
Defining $t:=\log\mu$ and requiring 
$g_{\mu\nu}^0$ to be independent of $t$  one obtains
\begin{equation}
0=\frac{dg_{\mu\nu}^0}{dt}=e^{\epsilon t}\Big(\epsilon\,g_{\mu\nu}-\alpha'\,R_{\mu\nu}+\frac{dg_{\mu\nu}}{dt}-\frac{\alpha'}{\epsilon}\,\frac{dg_{\lambda\rho}}{dt}\cdot \frac{\del}{\del g_{\lambda\rho}}\,R_{\mu\nu}\Big) \;.   
\end{equation}
Matching the order $\epsilon$ and $\epsilon^0$ terms yields
\begin{equation}\label{DerivativeSTEPP}
\frac{dg_{\mu\nu}}{dt}=-\epsilon\,g_{\mu\nu}+\beta_{\mu\nu}(g)\;, \quad  \beta_{\mu\nu}(g)=\alpha'\,\Big(1-g\cdot\frac{\del}{\del g}\Big)\,R_{\mu\nu}\;.
\end{equation}
We recall that the operator $g\cdot\frac{\del}{\del g}$ should be regarded as the integrated functional derivative as in \eqref{func der}, 
but here it can be reduced to an ordinary parametric derivative as follows: 
\begin{equation}
g_{\lambda\rho}\cdot \frac{\del}{\del g_{\lambda\rho}}\,T_{\mu\nu}(g)=\Lambda\frac{\del}{\del\Lambda}T_{\mu\nu}(\Lambda g)\Big\rvert_{\Lambda=1}\;. 
\end{equation}
This can be verified by computing the right-hand side, viewing  $T_{\mu\nu}$ as a function of $\tilde g_{\lambda\rho}=\Lambda\,g_{\lambda\rho}$ 
and applying the (functional) chain rule.
Note that the operator $\Lambda\frac{\del}{\del\Lambda}$ counts the number of $g_{\mu\nu}$ minus the number of $g^{\mu\nu}$. 
For the Ricci tensor this operator has zero eigenvalue, because the Christoffel symbols contain one $g$ and one $g^{-1}$ 
and there is no further metric needed in defining the Ricci tensor. 
Using this back in (\ref{DerivativeSTEPP}) we can finally read off  the  one-loop beta function \cite{Friedan:1980jf}
\begin{equation}\label{beta g}
\beta_{\mu\nu}(g)=\alpha'\,R_{\mu\nu}\;.
\end{equation}

\subsection{The Dilaton}

In this section we include the dilaton coupling in the sigma model and compute its one-loop $\beta-$function. Treating the dilaton term is technically more involved, since it requires computations on a curved worldsheet. Moreover, the resulting $\beta-$functions of the metric-dilaton system are such that $\beta^g_{\mu\nu}=0$ and $\beta^\Phi=0$ are \emph{not} the correct target space equations, which are instead given by  $\bar{\beta}^g_{\mu\nu}=0$ and $\bar{\beta}^\Phi=0$,   as we have discussed in section \ref{sec: Weyl anomaly}. 

We now compute  the background field expansion for  the sigma model action including the dilaton: 
\begin{equation}\label{sigma complete}
S=\frac{1}{2\lambda}\int d^2x\,\sqrt{\gamma}\Big[\gamma^{\alpha\beta}\,g_{\mu\nu}(X)\,\del_\alpha X^\mu\del_\beta X^\nu+\alpha'\,R^{(2)}\,\Phi(X)\Big] \;. 
\end{equation}
Here we need to consider an arbitrary curved worldsheet with metric $\gamma_{\alpha\beta}$ and spherical topology. The curved two-metric does not affect our discussion of the background-field expansion, since it does not involve the field $X^\mu=\varphi^\mu+\pi^\mu(\xi)\,$. This leads immediately to the quadratic action
\begin{equation}\label{S2xi with everything}
\begin{split}
S_{2\xi}&=\frac{1}{2\lambda}\int d^2x\,\sqrt{\gamma}\Big[\gamma^{\alpha\beta} D_\alpha\xi^a D_\beta\xi_a+\gamma^{\alpha\beta}R_{\mu ab\nu}\,\del_\alpha\varphi^\mu\del_\beta\varphi^\nu\,\xi^a\xi^b\Big]\\
&+\frac{1}{8\pi}\int d^2x\sqrt{\gamma}R^{(2)}\,\xi^a\xi^b\nabla_a\nabla_b\Phi+\frac{m^2}{2\lambda}\int d^2x\sqrt{\gamma}\,\xi^a\xi_a\;,
\end{split}    
\end{equation}
where we have already included the mass term to regulate infrared divergences, and we denoted $\nabla_a\nabla_b\Phi:=e_a^\mu e_b^\nu\nabla_\mu\nabla_\nu\Phi\,$. 
Following conventional  perturbation theory, we shall expand $\gamma_{\alpha\beta}$ around flat space:
\begin{equation}
\gamma_{\alpha\beta}=\delta_{\alpha\beta}+h_{\alpha\beta}\;, 
\end{equation}
and consider the one-loop effective action perturbatively in powers of $h_{\alpha\beta}\,$. In particular, the propagators are still extracted from the flat-space free theory
\begin{equation}
S_0=\frac{1}{2\lambda}\int d^2x\,\Big[\del^\alpha\xi^a\del_\alpha\xi_a+m^2\,\xi^a\xi_a\Big]    \;, 
\end{equation}
and terms with any powers of $h_{\alpha\beta}$ are treated as interactions.

Before starting any computation, there is one immediate consequence that can be derived from the structure of the action: thanks to the coupling with the scalar curvature $R^{(2)}$, every term involving the dilaton appears with at least one factor of $h_{\alpha\beta}\,$. Since the lowest order coupling of the metric $g_{\mu\nu}$ (and $B-$field if present) only involves the flat background $\delta_{\alpha\beta}$ one can immediately show that the dilaton \emph{cannot} renormalize the metric nor the $B-$field at any order in perturbation theory. This implies that the $\beta-$functions $\beta_{\mu\nu}^g$ and $\beta_{\mu\nu}^B$ do not depend on the dilaton at any order in $\alpha'$, as anticipated in section \ref{sec: Weyl anomaly}. This fact already  shows that the $\beta-$functions alone cannot provide the correct field equations.

Having shown that the metric $\beta-$function is not affected by the dilaton, we are left to determine the $\beta-$function of the dilaton itself. To this end, we have to extract from $S_{\rm int}=S_{2\xi}-S_0$ the terms that can renormalize the coupling $\int d^2x\sqrt{\gamma}R^{(2)}\Phi$ in $e^{-\Gamma_{1l}}=\left\langle e^{-S_{\rm int}}\right\rangle$, 
which in particular do not contain $\partial_{\alpha}\varphi^{\mu}$ factors.  From \eqref{S2xi with everything} one can split the interacting action into several terms, according to the background fields involved. Since we will work to quadratic order in $h_{\alpha\beta}$, we have
\begin{equation}\label{sint split}
\begin{split}
S_{\rm int}&=S_{\rm int}^{h=0}+\big(S_{h\omega}+S_{hh\omega}\big)+\big(S_{h\omega\omega }+S_{hh\omega\omega }\big)+\big(S_{hR}+S_{hhR}\big)\\
&+\big(S_{h\del\xi\del\xi}+S_{hh\del\xi\del\xi}\big)+m^2\,\big(S_{h\xi\xi}+S_{hh\xi\xi}\big)+S_\Phi+{\cal O}(h^3)\;,    
\end{split}
\end{equation}
where the subscripts denote the schematic form of the vertices.
 
In order to perform the perturbative expansion in $h_{\alpha\beta}$ we shall need \cite{Hull:1985rc}
\begin{equation}
\sqrt{\gamma}\,\gamma^{\alpha\beta}=\delta^{\alpha\beta}-\bar h^{\alpha\beta}
+{\cal O}(h^2)  \;, 
\end{equation}
where we defined the trace-adjusted perturbation in arbitrary dimensions
\begin{equation}
\bar h_{\alpha\beta}=h_{\alpha\beta}-\tfrac12\,\delta_{\alpha\beta} h\;,\quad \bar h^\alpha{}_\alpha=-\tfrac{1}{2}\,(n-2)\,h\;.
\end{equation}
In order to recognize the Ricci scalar we shall also need
\begin{equation}\label{Ricci n}
\sqrt{\gamma}R^{(n)}=\del^\alpha\del^\beta\bar h_{\alpha\beta}-\tfrac12\,\Box h+\tfrac14\,\bar h^{\alpha\beta}\Box\bar h_{\alpha\beta}+\tfrac12\,(\del\cdot\bar h_\alpha)^2-\tfrac{1}{16}\,(n-2)\,h\Box h+\del_\alpha v^\alpha+{\cal O}(h^3)\;,    
\end{equation}
modulo total derivatives.

We are now ready to examine the various contributions from the interacting action. First of all, the entire first line in \eqref{sint split} cannot contribute to the renormalization of the dilaton coupling, since every term contains at least one factor of $\del_\alpha\varphi^\mu\,$. 
Taking into account the terms in the second line we find
\begin{equation}\label{gamma dilaton}
\begin{split}
\Gamma^{\Phi}_{1l}&=\langle S_{h\del\xi\del\xi}+m^2\,S_{h\xi\xi}\rangle+\langle S_{hh\del\xi\del\xi}+m^2\,S_{hh\xi\xi}\rangle\\
&\hspace{5mm}-\tfrac12\,\langle (S_{h\del\xi\del\xi}+m^2\,S_{h\xi\xi})^2\rangle_{1\rm PI}+\langle S_\Phi\rangle+{\cal O}(h^3) \;,   
\end{split}    
\end{equation}
where by the superscript $\Phi$ we mean possible contributions proportional to $\int d^2x\sqrt{\gamma} R^{(2)}\,$.

At this point we notice that the ``tadpole'' contributions from the first line have no external momentum flowing in the loop, meaning that they give (possibly UV divergent) contributions of the form $C\int d^2x\,F(h)$, with no derivatives acting on $h_{\alpha\beta}\,$. These do not contribute to the Ricci scalar, but rather to the cosmological constant
\begin{equation}
\Lambda \int d^2x\sqrt{\gamma}\;.   
\end{equation}
On dimensional grounds $\Lambda$ must scale as $m^2$ and thus vanish upon removing the IR regulator.
Similarly, the contributions from the $m^2S_{h\xi\xi}$ term in 
$\langle (S_{h\del\xi\del\xi}+m^2\,S_{h\xi\xi})^2\rangle_{1\rm PI}$ vanish in the massless limit. We are thus left with
\begin{equation}
\begin{split}
\Gamma_{1l}^\Phi&=\langle S_\Phi\rangle-\tfrac12\,\langle S_{h\del\xi\del\xi}^2\rangle_{1\rm PI} +{\cal O}(h^3)+{\cal O}(m)\;,\\
S_{h\del\xi\del\xi}&=-\frac{1}{2\lambda}\int d^2x\,\bar h^{\alpha\beta}\del_\alpha\xi^a\del_\beta\xi_a\;,\quad   S_\Phi=\frac{1}{8\pi}\int d^2x\sqrt{\gamma}R^{(2)}\,\xi^a\xi^b\nabla_a\nabla_b\Phi\;.
\end{split}    
\end{equation}
We start from the bubble diagram given by the double contraction term
\begin{equation}
\begin{split}
-\tfrac12\,\langle S_{h\del\xi\del\xi}^2\rangle_{\rm 1\rm PI}&=   -\frac{1}{8\lambda^2}\int d^2xd^2y\,\bar h^{\alpha\beta}(x)\,\bar h^{\gamma\delta}(y)\,\langle\del_\alpha\xi^a(x)\del_\beta\xi_a(x)\,\del_\gamma\xi^b(y)\del_\delta\xi_b(y)\rangle_{1\rm PI}\\
&=-\frac{D}{4}\int d^2xd^2y\,\bar h^{\alpha\beta}(x)\,\bar h^{\gamma\delta}(y)\,\del_\alpha\del_\gamma G(x-y)\,\del_\beta\del_\delta G(x-y)\\
&=-\frac{D}{4}\int\frac{d^2p}{(2\pi)^2}\,\bar h^{\alpha\beta}(p)\,\Pi_{\alpha\beta,\gamma\delta}(p)\,\bar h^{\gamma\delta}(-p)\;,
\end{split}    
\end{equation}
where the regularized polarization tensor is given by
\begin{equation}
\Pi^{\rm reg.}_{\alpha\beta,\gamma\delta}(p)=\mu^{2-n}\int\frac{d^nk}{(2\pi)^n}\,\frac{k_{(\alpha}(p+k)_{\beta)}k_{(\gamma}(p+k)_{\delta)}}{(k^2+m^2)((p+k)^2+m^2)}  \;.  
\end{equation}
The diagram is IR finite in the massless limit, allowing one to compute $\Pi^{\rm reg.}_{\alpha\beta,\gamma\delta}(p)$ at $m^2=0\,$. This gives \cite{Hull:1985rc}
\begin{equation}
\begin{split}
-\tfrac12\,\langle & S_{h\del\xi\del\xi}^2  \rangle_{1\rm PI}= -\frac{D}{16\pi}\,\Gamma(1-\tfrac{\epsilon}{2})\,B(2+\tfrac{\epsilon}{2}, 2+\tfrac{\epsilon}{2})\,\int\frac{d^2p}{(2\pi)^2}\,\left(\frac{p^2}{4\pi\mu^2}\right)^{\frac{\epsilon}{2}}\\
&\times\Big\{\Big(p_\alpha p_\beta\bar h^{\alpha\beta}(p)-\tfrac12\,p^2h(p)\Big)\,\frac{1}{p^2}\,\Big(p_\gamma p_\delta\bar h^{\gamma\delta}(-p)-\tfrac12\,p^2h(-p)\Big)\\
&\hspace{5mm}-\frac{2}{\epsilon(1+\tfrac12\epsilon)}\,\Big(p_\alpha\bar h^{\alpha\beta}(p)\,p_\gamma\bar h^\gamma{}_\beta(-p)-\tfrac12\,\bar h^{\alpha\beta}(p)\,p^2\bar h_{\alpha\beta}(-p)+\tfrac18\,\epsilon\,h(p)\,p^2h(-p)\Big)\Big\}   \,, 
\end{split}    
\end{equation}
where $B(x,y)=\frac{\Gamma(x)\Gamma(y)}{\Gamma(x+y)}$ is the Euler beta function. The first term above is UV finite and non-local, corresponding to the covariant term
\begin{equation}
\int d^2x\sqrt{\gamma}\,R^{(2)}\frac{1}{\Box}\,R^{(2)}  \;.   
\end{equation}
The second term, which is the one we are after, is divergent, yielding
\begin{equation}
\begin{split}
\Gamma^\Phi_{\rm div.}&\supset  \frac{D}{48\pi}\,\frac{1}{\epsilon}\int\frac{d^2p}{(2\pi)^2}\,\Big(p_\alpha\bar h^{\alpha\beta}(p)\,p_\gamma\bar h^\gamma{}_\beta(-p)-\tfrac12\,\bar h^{\alpha\beta}(p)\,p^2\bar h_{\alpha\beta}(-p)+\tfrac18\,\epsilon\,h(p)\,p^2h(-p)\Big) \\
&=\frac{D}{48\pi}\,\frac{1}{\epsilon}\int d^2x\,\Big[(\del\cdot\bar h_\alpha)^2+\tfrac12\,\bar h^{\alpha\beta}\Box\bar h_{\alpha\beta}-\tfrac18\,\epsilon\,h\Box h\Big]\\
&=\frac{D}{24\pi}\,\frac{1}{\epsilon}\int d^2x\sqrt{\gamma}R^{(2)}\;,
\end{split}    
\end{equation}
where in the last line we recognized the Ricci scalar to order $h^2\,$. The only other contribution to the divergent part of $\Gamma^\Phi_{1l}$ is $\langle S_\Phi\rangle$, which is straightforward to evaluate: 
\begin{equation}
\begin{split}
\langle S_\Phi\rangle&=\frac{1}{8\pi}\int d^2x\sqrt{\gamma}R^{(2)}\,\langle\xi^a\xi^b\rangle\nabla_a\nabla_b\Phi=\frac{\lambda}{8\pi}\,G(0)\,\int d^2x\sqrt{\gamma}R^{(2)}\,\nabla^2\Phi\\
&=-\frac{\lambda}{16\pi^2}\frac{1}{\epsilon}\int d^2x\sqrt{\gamma}R^{(2)}\,\nabla^2\Phi+{\cal O}(\epsilon^0)\;.
\end{split}
\end{equation}
In order to cancel the divergences one needs the counterterm
\begin{equation}
S_{\rm c.t.}=\frac{1}{\epsilon}\int d^2x\sqrt{\gamma}R^{(2)}\,\Big[-\frac{D}{24\pi}+\frac{\alpha'}{8\pi}\,\nabla^2\Phi\Big]  \;,  
\end{equation}
where we have used  $\lambda=2\pi\alpha'$. As in (\ref{Phi0}) this allows us to determine the bare dilaton $\Phi_0$ as
\begin{equation}
\Phi_0=\mu^\epsilon\Big[\Phi-\frac{1}{\epsilon}\,\frac{D}{6}+\frac{1}{\epsilon}\,\frac{\alpha'}{2}\,\nabla^2\Phi\Big]\;.    
\end{equation}
Finally, using $\frac{d\Phi_0}{d t}=0$ we compute for the one-loop $\beta-$function $\beta^{\Phi}=\frac{d\Phi}{dt}+\epsilon \Phi$:
\begin{equation}\label{beta Phi}
\beta^\Phi=\frac{D}{6}-\frac{\alpha'}{2}\,\nabla^2\Phi\;.    
\end{equation}

Let us make a few comments on this result. Collecting what we have reviewed so far, we have determined the one-loop $\beta-$functions for the metric-dilaton system:
\begin{equation}\label{naive g+phi}
\beta_{\mu\nu}^g=\alpha'\,R_{\mu\nu}\;,\quad 
\beta^\Phi=\frac{D}{6}-\frac{\alpha'}{2}\,\nabla^2\Phi\;.
\end{equation}
One needs to include the reparametrization ghost system corresponding to the worldsheet metric being  a dynamical field that needs to be integrated over in the path integral. 
As is well-known \cite{Polyakov:1981rd}, this shifts the constant term $D\rightarrow D-26 $. 
Moreover, it is apparent from \eqref{naive g+phi} that $\beta^i=0$ are not the correct field equations. Including the extra terms as dictated by \eqref{beta bars} one rather obtains
\begin{equation}\label{g+phi eom}
\bar\beta^g_{\mu\nu}=\alpha'\,\Big(R_{\mu\nu}+2\,\nabla_\mu\nabla_\nu\Phi\Big)\;,\quad\bar\beta^\Phi=\frac{D-26}{6}-\frac{\alpha'}{2}\,\Big(\nabla^2\Phi-2\,\nabla^\mu\Phi\nabla_\mu\Phi\Big) \;. 
\end{equation}
Setting these functions to zero provides the correct field equations associated to the target space effective action
\begin{equation}
S[g,\Phi]=\frac{1}{2\kappa_0^2}\int d^{D}x\sqrt{-g}\,e^{-2\Phi}\Big[-\frac{2(D-26)}{3\,\alpha'}+R+4\,\nabla^\mu\Phi\nabla_\mu\Phi\Big] \;.   
\end{equation}
One should note, however, that the one-loop result \eqref{naive g+phi} does \emph{not} give, in principle, the full $\cO(\alpha')$ contribution to $\beta^\Phi\,$. This is because the dilaton coupling in the sigma model \eqref{sigma g+phi} appears with one extra order of $\alpha'$ as compared to the other couplings. This implies that the full $\cO({\alpha'}^{\,L})$ dilaton $\beta-$function requires an $(L+1)-$loop computation. For instance, while the pure metric-dilaton
system does not produce any further $\cO(\alpha')$ term in \eqref{beta Phi}, including the $B-$field would result in a further contribution $-\frac{\alpha'}{24}\,H^2$ arising at two loops.

For this reason, together with the complication of computing on a curved worldsheet, it is often preferable to fix the dilaton equation from consistency \cite{Curci:1986hi,Tseytlin:1986ws,Callan:1986jb,Hull:1987yi, Bonezzi:2020jjq}. Rather than discussing the general procedure, the idea is easily understood by giving the details in the simplest case at hand. Let us suppose that we do not know the dilaton $\beta-$function. The metric $\beta-$function, together with the general relation \eqref{beta bars} fixes the metric field equation to
\begin{equation}\label{metric eom}
R_{\mu\nu}+2\,\nabla_\mu\nabla_\nu\Phi=0\;\longrightarrow\;R+2\,\nabla^2\Phi=0\;.    
\end{equation}
Taking the divergence of the first equation  one obtains
\begin{equation}
\begin{split}
\nabla^\nu R_{\mu\nu}&=-2\,\nabla^2\nabla_\mu\Phi=-2\,\big(\nabla_\mu\nabla^2\Phi+R_{\mu\nu}\,\nabla^\nu\Phi\big)  \\
&=-2\,\nabla_\mu\nabla^2\Phi+4\,\big(\nabla_\mu\nabla_\nu\Phi\big)\nabla^\nu\Phi=-2\,\nabla_\mu\big(\nabla^2\Phi-(\nabla\Phi)^2\big)\;.
\end{split}    
\end{equation}
Consistency with the Bianchi identity $\nabla^\nu R_{\mu\nu}\equiv\frac12\,\nabla_\mu R$ gives the constraint
\begin{equation}
0=\nabla_\mu\big(\nabla^2\Phi-(\nabla\Phi)^2+\tfrac14\,R\big)=\tfrac12\,\nabla_\mu\big(\nabla^2\Phi-2\,(\nabla\Phi)^2\big) \;,   
\end{equation}
that can be integrated to 
\begin{equation}\label{consistent phi}
\nabla^2\Phi-2\,(\nabla\Phi)^2=C\;,    
\end{equation}
for an undetermined  constant $C\,$. This shows that the dilaton equation in \eqref{g+phi eom} is correctly reproduced by \eqref{consistent phi}, apart from the constant $C$ that can be easily fixed by matching with the one-loop result (\ref{g+phi eom}) to be $C=\frac{D-26}{3\alpha'}\,$. Here we use that, on dimensional grounds,  
the constant term in (\ref{g+phi eom}) cannot receive $\alpha'$ corrections.

\section{Duality-invariant Sigma Model}

Having reviewed the necessary tools to extract target space equations from a given string sigma model, we turn in this section to the main subject of this paper. We shall compute the one-loop $\beta-$functions of the $O(d,d)-$invariant sigma model recently discussed  in \cite{Bonezzi:2020ryb}. This, together with a general analysis of the Weyl anomaly, gives the complete $O(d,d)-$invariant target space equations to lowest order in $\alpha'
$, which  agree with the low-energy effective action constructed by Maharana and Schwarz in \cite{Maharana:1992my}. This result generalizes the earlier works of \cite{Berman:2007xn,Berman:2007yf} by including the external $B-$field and Kaluza-Klein gauge vectors and provides a positive check on the viability of the sigma model \cite{Bonezzi:2020ryb} to determine higher order $\alpha'$ corrections. 

We shall start by briefly reviewing the features of the duality-invariant sigma model that will be used to compute the $\beta-$functions. In order to find the target space equations, we will derive an operator expression for the Weyl anomaly, following the general discussion of section \ref{sec: Weyl anomaly}. We will then turn to the actual computation of the $\beta-$functions and finally determine the target space field equations, concluding with an internal consistency check on the result.

\subsection{$O(d,d)$-invariant Worldsheet}

To set the stage, let us consider a $(D+d)-$dimensional target space manifold possessing $d$ abelian isometries. We shall choose coordinates split as $\hat x^{\hat\mu}=(x^\mu, y^i)$, with $\mu=0,\cdots,D-1$ denoting the ``external'' directions and $i=1,\cdots,d$ the ``internal'' directions along the isometries, such that the target space fields do not depend on the internal coordinates $y^i\,$. 

Upon dimensional reduction, the $(D+d)-$dimensional fields give rise to the $D-$dimensional metric $g_{\mu\nu}$, $B-$field $B_{\mu\nu}$ and dilaton $\phi$, together with an $O(d,d)$ multiplet of Kaluza-Klein gauge fields $\cA_\mu{}^M$ and an $O(d,d)-$valued symmetric matrix of scalar fields $\cH_{MN}$, also known as the generalized metric. From here on we shall denote by $M=1,\cdots,2d$ indices in the fundamental representation of $O(d,d)$, which will be raised and lowered by the $O(d,d)-$invariant metric $\eta_{MN}\,$. The scalar matrix $\cH_{MN}$, being $O(d,d)-$valued, has to obey the constraint $\cH_{MP}\eta^{PQ}\cH_{QN}=\eta_{MN}\,$. 

The construction of \cite{Bonezzi:2020ryb}, which is based on the dimensional reduction of the Polyakov sigma model, together with the early proposals of \cite{Tseytlin:1990nb,Gasperini:1991ak} and \cite{Blair:2013noa}, leads to a manifestly $O(d,d)-$invariant worldsheet action that generalizes the one of \cite{Tseytlin:1990nb} and agrees with \cite{Schwarz:1993mg,Blair:2016xnn}:
\begin{equation}\label{SclassicO(d,d) no phi}
\begin{split}
S=&-\frac{1}{2\lambda}\int d^2\sigma\,\Big[\sqrt{-h}h^{\alpha\beta}\del_\alpha X^\mu\del_\beta X^\nu\,g_{\mu\nu}(X)+\epsilon^{\alpha\beta}\del_\alpha X^\mu\del_\beta X^\nu\,B_{\mu\nu}(X)\\
&-\epsilon^{\alpha\beta}\del_\alpha Y^M\del_\beta X^\mu\,\cA_{\mu M}(X)-D_0Y^MD_1Y_M+u\,D_1Y^MD_1Y_M+e\,\cH_{MN}\,D_1Y^MD_1Y^N\Big]\;,
\end{split}
\end{equation}
where we denote by $h_{\alpha\beta}$ the worldsheet metric of Lorentzian signature. As before, we have set $\lambda=2\pi\alpha'\,$. 
In the above action
the worldsheet scalars $X^\mu$ are the coordinate embeddings corresponding to the ``external'' directions $x^\mu$, while the $O(d,d)$ fields $Y^M=(Y^i,\widetilde Y_i)$ correspond to the doubled ``internal'' sector \cite{Maharana:1992my,Hull:2006va}.
The gauge-covariant derivatives are defined as \cite{Maharana:1992my}
\begin{equation}
D_\alpha Y^M:=\del_\alpha Y^M+\del_\alpha X^\mu\cA_\mu{}^M(X)\equiv\del_\alpha Y^M+\cA_\alpha{}^M\;,
\end{equation}
and the Weyl-invariant metric combinations $e$ and $u$ are given by
\begin{equation}\label{e and u}
e=\frac{\big((h_{01})^2-h_{00}h_{11}\big)^{1/2}}{h_{11}}\;,\quad u=\frac{h_{01}}{h_{11}}\;.
\end{equation}
It is clear from \eqref{SclassicO(d,d) no phi} that the price for having a manifestly $O(d,d)-$invariant action that does not need any subsidiary constraints\footnote{The proposal of \cite{Pasti:1996vs} does provide manifestly Lorentz-invariant actions for chiral forms, but they are either non-local or non-polynomial.} is the lack of manifest worldsheet diffeomorphism invariance. The action \eqref{SclassicO(d,d) no phi} is diffeomorphism invariant nonetheless, albeit with the non-standard transformation rules
\begin{equation}\label{diff Y}
\begin{split}
\delta_\xi X^\mu&=\xi^\alpha\del_\alpha X^\mu\;,\qquad\qquad\;\;\;\;\;\, \delta_\xi h_{\alpha\beta}=\nabla_\alpha\xi_\beta+\nabla_\beta\xi_\alpha\;,\\  
\delta_\xi Y^M&=\xi^\alpha\del_\alpha Y^M-\xi^0\,\cD^M\;,\qquad\cD^M=D_0Y^M-u\,D_1Y^M-e\,\cH^{MN}D_1Y_N \;. 
\end{split}    
\end{equation}
It should be emphasized that this diffeomorphism invariance requires $\cH_{MN}$ to obey the $O(d,d)$ constraint $\cH_{MP}\cH^P{}_N=\eta_{MN}\,$.
Let us also mention that the action \eqref{SclassicO(d,d) no phi} is invariant under the ``time-local'' shifts $\delta_\Xi Y^M=\Xi^M(\sigma^0)\,$. This symmetry allows one to integrate the second-order $Y-$equations, which are given by $\del_1\cD^M=0$, to the first-order self-duality relations $\cD^M=0\,$. These in turn render  the equations of motion equivalent to the ones of \cite{Maharana:1992my}.

As for the symmetry under gauge transformations of the background fields, the discussion of target space diffeomorphisms and $B-$field gauge transformations is standard. The $U(1)$ Kaluza-Klein transformation $\delta_\lambda\cA_\mu{}^M=\del_\mu\lambda^M$ instead has to be accompanied by 
\begin{equation}
\delta_\lambda Y^M=-\lambda^M(X)\;,\quad \delta_\lambda B_{\mu\nu}=\frac12\,\cF_{\mu\nu}{}^M \lambda_M\;,    
\end{equation}
where $\cF_{\mu\nu}{}^M=2\,\del_{[\mu}\cA_{\nu]}{}^M$ is the abelian curvature two-form. The non-standard transformation of the $B-$field under the vector gauge symmetries \cite{Maharana:1992my} requires a modification of the naive field strength by an abelian Chern-Simons term
\begin{equation}\label{H with CS}
H_{\mu\nu\lambda}=3\,\del_{[\mu}B_{\nu\lambda]}-3\,\cA_{[\mu}{}^M\del_\nu\cA_{\lambda] M}   \;, 
\end{equation}
to ensure gauge invariance of $H_{\mu\nu\lambda}\,$. Correspondingly, the standard Bianchi identity $dH=0$ is replaced by
\begin{equation}
\del_{[\mu}H_{\nu\lambda\rho]}+\frac34\,\cF_{[\mu\nu}{}^M\cF_{\lambda\rho] M}=0\;.    
\end{equation}

In \cite{Bonezzi:2020ryb}, the worldsheet action \eqref{SclassicO(d,d) no phi} was derived without coupling to the dilaton. We now include the dilaton coupling by a standard Fradkin-Tseytlin term: \begin{equation}\label{SclassicO(d,d)}
\begin{split}
S=&-\frac{1}{2\lambda}\int d^2\sigma\,\Big[\sqrt{-h}h^{\alpha\beta}\del_\alpha X^\mu\del_\beta X^\nu\,g_{\mu\nu}(X)+\epsilon^{\alpha\beta}\del_\alpha X^\mu\del_\beta X^\nu\,B_{\mu\nu}(X)\\
&\hspace{23mm}-\epsilon^{\alpha\beta}\del_\alpha Y^M\del_\beta X^\mu\,\cA_{\mu\, M}(X)+\alpha'\sqrt{-h}R^{(2)}\phi(X)\Big]\\
&+\frac{1}{2\lambda}\int d^2\sigma\,\Big[D_0Y^MD_1Y_M-u\,D_1Y^MD_1Y_M-e\,\cH_{MN}\,D_1Y^MD_1Y^N\Big]\;,
\end{split}
\end{equation}
where $\phi$ is the $D-$dimensional $O(d,d)-$invariant dilaton \cite{Berman:2007yf}, which is related to the higher-dimensional standard dilaton $\Phi$ by
\begin{equation}\label{Oddphi}
\phi=\Phi-\frac14\,\log\det G_{ij}\;,    
\end{equation}
where $G_{ij}(X)$ is the internal metric.

We can give a heuristic justification for the appearance of the $O(d,d)-$invariant dilaton in the sigma model \eqref{SclassicO(d,d)}. In the Buscher procedure  \cite{Buscher:1987qj} to derive the action of $T-$duality at the worldsheet level, one trades the derivatives of the original coordinates $\del_\alpha Y^i$ for a curl-free vector field $V_\alpha^i\,$. The curl-free condition $\epsilon^{\alpha\beta}\del_\alpha V_\beta^i=0$ is imposed via Lagrange multipliers $\widetilde Y_i$, that play the role of dual coordinates. Integrating out the vector fields $V_\alpha^i$ gives rise to the $T-$dual action for the $\widetilde Y_i\,$. It is well-known that integrating out the auxiliary fields $V_\alpha^i$ at the quantum level produces a one-loop shift of the action proportional to $\int d^2\sigma\sqrt{-h}R^{(2)}\,\log\det G_{ij}\,$ \cite{Rocek:1991ps,Tseytlin:1991wr,Schwarz:1992te}. This results in the original dilaton $\Phi$ being shifted to the T-dual dilaton 
\begin{equation}\label{TdualPhi}
\widetilde\Phi=\Phi-\frac12\,\log\det G_{ij} \;. \end{equation}
Now consider the above action and integrate out the dual coordinates $\widetilde Y_i$ in the $O(d,d)$ multiplet $Y^M=(Y^i,\widetilde Y_i)\,$. Similarly to the Buscher procedure, this leads back to the original Polyakov sigma model at the classical level. Including the one-loop effect should shift the dilaton coupling $\phi$ (that we now suppose to be unknown) by 
\begin{equation}
\phi\rightarrow\phi+k\,\log\det G_{ij} \;,   
\end{equation}
for some constant $k$. If this is to give the original theory, one must have $\phi+k\,\log\det G_{ij}=\Phi$.
If one integrates out the original coordinates $Y^i$ instead, the shift has to be
\begin{equation}
\phi\rightarrow\phi-k\,\log\det G_{ij} \;,   
\end{equation}
since $Y^i$ couples to the metric $G_{ij}$ as $\widetilde Y_i$ does to the inverse $G^{ij}\,$. This time, integrating out the original coordinates produces the T-dual sigma model at tree-level, but to match at one-loop the shift has to produce the T-dual dilaton, \emph{i.e.}~$\phi-k\,\log\det G_{ij}  =\widetilde\Phi$. 
Using \eqref{TdualPhi} together with $\phi+k\,\log\det G_{ij}=\Phi$ established above then fixes $k=\frac14$. This reproduces \eqref{Oddphi}.

\subsection{Weyl Anomaly Equations}

Before attempting to compute the $\beta-$functions of the sigma model \eqref{SclassicO(d,d)}, we shall derive the analogue of the Weyl anomaly equations \eqref{beta bars} reviewed in section \ref{sec: Weyl anomaly}. This is necessary in order to establish the correct field equations $\bar \beta^i=0$ in terms of the RG $\beta-$functions.

We begin by  classifying  the possible diffeomorphism invariant dimension two operators. It is important to note that, besides being background field dependent, the transformations \eqref{diff Y} do not leave individually invariant any terms in the action (\ref{SclassicO(d,d)}) that contain $Y^M$.  This implies that the only classically diffeomorphism invariant  operator involving the $Y^M$ fields
is the $Y-$action itself. 

We now come to the equations of motion, that have to be used to derive the Weyl anomaly, as we have seen in section \ref{sec: Weyl anomaly}.
The $Y-$equation simply reads
\begin{equation}
\del_1\cD^M=0\;. 
\end{equation}
Upon exploiting the time-dependent  $\Xi^M$ shift symmetry this is equivalent to the twisted self-duality relation $\cD^M=0\,$.
The $X-$equation is considerably more involved and is given by
\begin{equation}\label{X eom}
\begin{split}
\sqrt{-h}\,g_{\mu\nu}\,\calD^\alpha\del_\alpha X^\nu&=\frac12\,\epsilon^{\alpha\beta}\del_\alpha X^\nu\del_\beta X^\lambda\,H_{\mu\nu\lambda}+\frac{\alpha'}{2}\,\sqrt{-h}\,R^{(2)}\,\del_\mu\phi+\epsilon^{\alpha\beta}\del_\alpha X^\nu D_\beta Y_M\,\cF_{\mu\nu}{}^M\\
&\;\; +\frac{e}{2}\,\del_\mu\cH_{MN}\,D_1Y^MD_1Y^N-\del_1X^\nu\,\cF_{\mu\nu}{}^M\,\cD_M\;,
\end{split}    
\end{equation}
where we recall  that the field strength $H_{\mu\nu\lambda}$ includes the Chern-Simons correction \eqref{H with CS}.

In order to write down the action as a sum of dimension two operators associated to different couplings as in \eqref{bare S}, one has to write out the covariant derivatives since the Kaluza-Klein vectors $\cA_\mu{}^M$ give rise to quadratic and cubic composite couplings:
\begin{equation}\label{coupling split}
\begin{split}
S&
=\int d^2\sigma\,\Big[A^{\mu\nu}_g\cdot g_{\mu\nu}+A^{\mu\nu}_{\tilde g}\cdot\big(g_{\mu\nu}+\cA_\mu{}^M\cH_{MN}\cA_\nu{}^N\big)+A^{\mu\nu}_{\cA^2}\cdot \big(\cA_\mu{}^M\cA_{\nu M}\big)\\
&+A^{\mu\nu}_B\cdot B_{\mu\nu}+A^{\mu M}_\cA\cdot\cA_{\mu M}+A^{\mu M}_{\cA\cH}\cdot\big(\cA_\mu{}^N\cH_{MN}\big)+A^{MN}_\eta\cdot\eta_{MN}+A^{MN}_\cH\cdot\cH_{MN}+A_\phi\cdot\phi\Big]\;,\end{split}    
\end{equation}
where we defined 
\begin{equation}\label{operators split}
\begin{split}
A^{\mu\nu}_g&=\frac{1}{2\lambda e}\,d_0 X^\mu d_0X^\nu \;,\qquad\hspace{10mm} A^{\mu\nu}_{\tilde g}=-\frac{e}{2\lambda}\,\del_1X^\mu\del_1X^\nu\;,\quad\hspace{1mm}
A^{\mu\nu}_{\cA^2}=\frac{1}{2\lambda}\,\del_1X^{(\mu}d_0X^{\nu)} \;,\\[2mm] A^{\mu\nu}_B&=-\frac{1}{2\lambda}\,\epsilon^{\alpha\beta}\del_\alpha X^\mu\del_\beta X^\nu\;,\qquad
A^{\mu\,M}_\cA=\frac{1}{\lambda}\,d_0X^\mu\del_1Y^M\;,\qquad A^{\mu\,M}_{\cA\cH}=-\frac{e}{\lambda}\,\del_1X^\mu\del_1Y^M\;,\\[2mm]
A^{MN}_\eta&=\frac{1}{2\lambda}\,\del_1Y^{(M}d_0Y^{N)}\;,\qquad\hspace{5mm} A^{MN}_\cH=-\frac{e}{2\lambda}\,\del_1Y^M\del_1Y^N\;,\quad\hspace{2mm} A_\phi=-\frac{\alpha'}{2\lambda}\,\sqrt{-h}\,R^{(2)}\;,
\end{split}    
\end{equation}
with $d_0:=\del_0-u\del_1$, and we suppressed  the factor $\delta^D(x-X(\sigma))\,$ in every $A_i$.

Since the linear couplings $\varphi^i_{\rm lin}=(g_{\mu\nu}, B_{\mu\nu}, \cA_{\mu\,M}, \cH_{MN}, \phi)$ all multiply different operators, their $\beta-$functions are extracted as usual from the corresponding expressions for the bare couplings:
\begin{equation}
\begin{split}
g_{0\mu\nu}&=\mu^\epsilon\Big[g_{\mu\nu}+\sum_{n=1}^\infty\frac{1}{\epsilon^n}\,T_{n\mu\nu}^g\Big]\;,\quad B_{0\mu\nu}=\mu^\epsilon\Big[B_{\mu\nu}+\sum_{n=1}^\infty\frac{1}{\epsilon^n}\,T_{n\mu\nu}^B\Big]\;,\quad \phi_0=\mu^\epsilon\Big[\phi+\sum_{n=1}^\infty\frac{1}{\epsilon^n}\,T_{n}^\phi\Big]\;,\\
\cA_{0\mu\,M}&=\mu^\epsilon\Big[\cA_{\mu\,M}+\sum_{n=1}^\infty\frac{1}{\epsilon^n}\,T_{n\mu\,M}^\cA\Big]\;,\quad\cH_{0MN}=\mu^\epsilon\Big[\cH_{MN}+\sum_{n=1}^\infty\frac{1}{\epsilon^n}\,T_{nMN}^\cH\Big]\;,\quad\eta_{0MN}=\mu^\epsilon\,\eta_{MN}\;.
\end{split}    
\end{equation}
Notice that, in order to keep both $X^\mu$ and $Y^M$ dimensionless in $n=2+\epsilon$ dimensions, one has to define the bare $O(d,d)$ metric as\footnote{We shall see in the following that $\eta_{MN}$ does not renormalize.} $\eta_{0MN}=\mu^\epsilon\,\eta_{MN}\,$. Due to this  one has to be careful with the position of the $O(d,d)$ indices. 
We take $A_{0\mu M}$ to have dimension $\epsilon$, which implies that 
$\cA_{0\mu}{}^M=\eta_0^{MN}\cA_{0\mu\,N}=\cA_\mu{}^M+\cdots $ is dimensionless and 
that $\cA_{0\alpha}{}^M$ has dimension one (as it should in order to define the bare covariant derivative $D^0_{\alpha}Y^M$).
Knowing the $\beta-$functions for the linear couplings, one can easily derive the ones for the composite couplings. For instance,
\begin{equation}\label{Leibniz example}
\begin{split}
&\frac{d}{dt}\,\big(\cA_\mu{}^N\,\cH_{MN}\big)=\frac{d\cA_{\mu\,P}}{dt}\,\eta^{PN}\,\cH_{MN}+\cA_{\mu\,P}\,\frac{d\eta^{PN}}{dt}\,\cH_{MN}+\cA_{\mu\,P}\,\eta^{PN}\,\frac{d\cH_{MN}}{dt}\\
&=\big(-\epsilon\,\cA_{\mu\,P}+\beta^\cA_{\mu\,P}\big)\,\eta^{PN}\,\cH_{MN}+\cA_{\mu\,P}\,\big(\epsilon\,\eta^{PN}\big)\,\cH_{MN}+\cA_{\mu\,P}\,\eta^{PN}\,\big(-\epsilon\,\cH_{MN}+\beta^\cH_{MN}\big)\\
&=-\epsilon\,\big(\cA_\mu{}^N\cH_{MN}\big)+\beta^\cA_{\mu\,N}\cH_M{}^N+\cA_\mu{}^N\beta^\cH_{MN}\;.
\end{split}    
\end{equation}

In order to compute  the Weyl anomaly as in \eqref{trace T} (albeit now in Lorentzian signature),
we need to define  the bare action in $n=2+\epsilon$ dimensions. Similar to the problem of defining the antisymmetric density $\varepsilon^{\alpha\beta}$ in $n$ dimensions, we now encounter the difficulty of defining $e$ and $u$ in arbitrary dimensions. If we keep the definitions \eqref{e and u} in arbitrary dimensions, $e$ and $u$ remain Weyl invariant, but in order to also keep $Y^M$ invariant one has to define the bare action for $Y$ as\footnote{To see this one can repeat the steps \cite{Bonezzi:2020ryb} that lead to the action \eqref{SclassicO(d,d)} starting from the Polyakov sigma model directly in $n=2+\epsilon$ dimensions.}
\begin{equation}
S_{0Y}=\frac{1}{2\lambda}\int d^n\sigma\,e^{\epsilon\rho}\,\big(D_1Y_M\big)_0\,\big(D_0Y^M-u\,D_1Y^M-e\,\cH^{MN}D_1Y_N\big)_0 \;, \end{equation}
where the subscript $0$ denotes that all the couplings, including the vectors $\cA_\mu{}^M$ hidden in $D_\alpha Y^M$, are the bare ones, and $\rho$ is defined by $e^{(2+\epsilon)\rho}=\sqrt{-h}\,$.

Taking the Weyl variation now leads to
\begin{equation}
\sqrt{-h}\,T^\alpha{}_\alpha=A_{i0}\cdot(-\epsilon\,\varphi^i_0)-\frac{1}{2\pi}\,\del_\alpha\big[\sqrt{-h}h^{\alpha\beta}\del_\beta X^\mu\del_\mu\phi_0\big]\;,    
\end{equation}
where all couplings $\varphi^i$ and operators $A_i$ are listed in \eqref{coupling split} and \eqref{operators split}.
By repeating the general procedure reviewed in section \ref{sec: Weyl anomaly} and using the equations of motion \eqref{X eom} to decompose the total derivative one finds
\begin{equation}\label{dilaton mess}
\begin{split}
\sqrt{-h}\,T^\alpha{}_\alpha&=[A_i]\cdot\hat\beta^i-\frac{1}{2\lambda}\,\Big\{\sqrt{-h}h^{\alpha\beta}\del_\alpha X^\mu\del_\beta X^\nu\,\big(2\alpha'\,\nabla_{0\mu}\del_\nu\phi_0\big)+\varepsilon^{\alpha\beta}\del_\alpha X^\mu\del_\beta X^\nu\,\big(\alpha'\,\del_\lambda\phi_0\,H_0^\lambda{}_{\mu\nu}\big)\\
&+\alpha'\,\sqrt{-h}\,\bar R\,\big(\alpha'\,\del_\mu\phi_0\nabla_0^\mu\phi_0\big)+e\,e^{\epsilon\rho}\,D_1^0Y^MD_1^0Y^N\,\big(\alpha'\,\del_\mu\phi_0\nabla_0^\mu\cH_{0MN}\big)\\
&+2\,\varepsilon^{\alpha\beta}\del_\alpha X^\mu D_\beta^0Y^M\,\big(\alpha'\,\nabla_0^\lambda\phi_0\,\cF_{0\lambda\mu\,M}\big)-2\,e^{\epsilon\rho}\,\del_1X^\mu\,\cD_0^M\,\big(\alpha'\,\nabla_0^\lambda\phi_0\,\cF_{0\lambda\mu\,M}\big)\Big\}\\
&=[A_i]\cdot\hat\beta^i+\alpha'\,A_{i0}\cdot\big(\cL_{\nabla_0\phi_0}\varphi^i_0\big)\;,    
\end{split}    
\end{equation}
where, for the moment, we have discarded the $Q-$type total derivative contributions coming from $[A_i]$ (see \eqref{combined derivatives}).
The Lie derivative displayed in the last line acts on $g_{\mu\nu}$, $\phi$ and $\cH_{MN}$ as usual, while it differs in the case of $\cA_\mu{}^M$ and $B_{\mu\nu}$ by gauge transformations:
\begin{equation}
\cL_\xi\cA_\mu{}^M:=\xi^\lambda\cF_{\lambda\mu}{}^M\;,\quad \cL_\xi B_{\mu\nu}:=\xi^\lambda\,\big(H_{\lambda\mu\nu}+\cA_{[\mu}{}^M\cF_{\nu]\lambda\,M}\big)\;.
\end{equation}
The main difference from the standard theory is in the transformation $\cL_\xi B_{\mu\nu}$, that is \emph{not} covariant, due to the mixing with the Kaluza-Klein vector fields. Such non-covariant term, however, is canceled in the sum $A_{i0}\cdot\big(\cL_\xi\varphi^i_0\big)$, as can be seen directly from the expanded form of \eqref{dilaton mess}, which is manifestly gauge-invariant.
Since both the $\beta-$functions of composite couplings (see e.g. \eqref{Leibniz example}) and the Lie derivative obey the Leibniz rule, one can see that the vanishing of the Weyl anomaly \eqref{dilaton mess}, which involves \emph{all} the couplings \eqref{coupling split}, is ensured by the vanishing of the $\bar\beta^i$ functions associated to the independent linear couplings, namely
\begin{equation}\label{Odd beta bar without Q}
\begin{split}
\bar\beta^i&=\beta^i+\alpha'\,\cL_{\nabla\phi}\varphi^i+\cdots=0   \quad{\rm for}\quad i=g,\phi,\cA,\cH \;,\\
\bar\beta_{\mu\nu}^B&=\beta_{\mu\nu}^B+\alpha'\,\nabla^\lambda\phi\,H_{\lambda\mu\nu}+\cdots=0\;,
\end{split}
\end{equation}
where the dots stand for the possible total derivative terms arising from $[A_i]$. 

Let us now study the total derivative terms related to $[A_i]$. 
These terms come from renormalization of the composite operators $A_i$ in flat space, and so we can take $h_{\alpha\beta}=\eta_{\alpha\beta}$ and thus $\rho=u=0$ and $e=1\,$. The possible dimension two total derivatives arising from $\del_\alpha X^\mu$ are  $d^\mu_h\cdot N_\mu$ and $d^\mu_\varepsilon\cdot K_\mu$ as in \eqref{N}.  
Coming now to possible dimension two total derivatives constructed from $\del_\alpha Y^M$, the situation is much more subtle. In flat space, the Lorentz transformations of $Y^M$ (that can be found by specializing the diffeomorphism transformations \eqref{diff Y}), are highly non-linear and background field dependent.
In order to determine possible total derivatives including $\del_\alpha Y^M$, we notice that time derivatives $\del_0 Y^M$ appear in the action \eqref{SclassicO(d,d)} only as $\cD^M$ or $\epsilon^{\alpha\beta}D_\beta Y^M$. Moreover, double time derivatives never appear acting on $Y^M$. For these reasons, one can naively construct two independent operators acting on different fields: $d^M_\epsilon\cdot V_M:=-\frac{1}{\lambda}\del_\alpha\big[\epsilon^{\alpha\beta}D_\beta Y^M\,V_M(X)\big]$ and $d^M_\cD\cdot W_M:=-\frac{1}{\lambda}\del_1\big[\cD^M\,W_M(X)\big]$, where we recall that $\cD^M$ is defined in \eqref{diff Y}. These two operators, however, cannot appear in an arbitrary way. One can see that the only (on-shell) Lorentz-invariant combination is given by
\begin{equation}
(d_\epsilon^M-d_\cD^M)\cdot V_M=:d^M\cdot V_M=-\frac{1}{\lambda}\Big[\del_\alpha\big(\epsilon^{\alpha\beta}D_\beta Y^M\,V_M(X)\big)-\del_1\big(\cD^M\,V_M(X)\big)\Big]\;,
\end{equation}
and depends on a single target space field $V_M$.
Decomposing the above expression gives 
\begin{equation}\label{V decomp}
\begin{split}
d^M\cdot V_M&=A_i\cdot\big(\delta_V\varphi^i\big)-\frac{1}{2\lambda}\,\del_\alpha\big[\epsilon^{\alpha\beta}\del_\beta X^\mu\,K_\mu'(X)\big]   \;,\\
\delta_V\cA_\mu{}^M&=\del_\mu V^M\;,\quad\delta_V B_{\mu\nu}=\frac12\,\cF_{\mu\nu}{}^M V_M\;, 
\end{split}
\end{equation}
where $K'_\mu:=V_M\cA_\mu{}^M$. 
This latter  contribution 
is of the $d^\mu_\varepsilon$ type and thus does not introduce new independent structures.

We have thus shown that in the $O(d,d)-$invariant sigma model any total derivative term giving rise to a $Q-$contribution to \eqref{Odd beta bar without Q} can be parametrized by two field-dependent vector fields $W_\mu(g,B,\cA,\cH)$ and $L_\mu(g,B,\cA,\cH)$ as in section 2, together with an $O(d,d)$ vector-valued scalar field $V_M(g,B,\cA,\cH)$. The contribution of $W_\mu$ and $L_\mu$ to the Weyl anomaly coefficients \eqref{Odd beta bar without Q} is the same as in \eqref{beta bars}, including the covariant Lie derivative of $\cA_\mu{}^M$, while the one from $V_M$ can be inferred from the decomposition \eqref{V decomp}. This finally gives the relation 
between $\beta$ and $\bar{\beta}$: 
\begin{equation}\label{eoms beta bar}
\begin{split}
\bar\beta_{\mu\nu}^g&=\beta_{\mu\nu}^g+2\alpha'\,\nabla_\mu\nabla_\nu\phi+\nabla_{(\mu}W_{\nu)} \;, \\ 
\bar\beta^\phi&=\beta^\phi+\alpha'\,(\nabla\phi)^2+\tfrac12\nabla^\mu\phi\,W_\mu\;,\\ \bar\beta^B_{\mu\nu}&=\beta^B_{\mu\nu}+\alpha'\,H^\lambda{}_{\mu\nu}\nabla_\lambda\phi+\tfrac12\,H^\lambda{}_{\mu\nu}W_\lambda+\del_{[\mu}L_{\nu]}+\tfrac12\,\cF_{\mu\nu}{}^M V_M \;,\\
\bar\beta_{\mu\,M}^\cA&=\beta_{\mu\,M}^\cA+\alpha'\,\nabla^\lambda\phi\,\cF_{\lambda\mu\,M}+\tfrac12\,W^\lambda\,\cF_{\lambda\mu\,M}+\nabla_\mu V_M\;,\\
\bar\beta_{MN}^\cH&=\beta_{MN}^\cH+\alpha'\,\nabla^\mu\phi\,\nabla_\mu\cH_{MN}+\tfrac12\,W^\mu\,\nabla_\mu\cH_{MN}\;.
\end{split}    
\end{equation}

As we have previously discussed, if one tries to compute the vectors $W_\mu$, $L_\mu$ and $V_M$ directly from the bare action, they will in general break gauge invariance in both the $B-$field and Kaluza-Klein gauge sectors, in agreement with the fact that our prescription for extending the action to $n=2+\epsilon$ dimensions violates said symmetries. One can employ a manifestly gauge invariant perturbative scheme and try to fix $W$, $L$ and $V$ by other means. 
Based on covariance, they can only depend on the Riemann tensor, covariant derivatives and the gauge-invariant field strengths $H_{\mu\nu\rho}$ and $\cF_{\mu\nu}{}^M\,$. 
One can then see that they must vanish at one-loop order, since there is no way of constructing a gauge-invariant vector, nor an $O(d,d)-$vector valued scalar with one spacetime derivative. We shall thus set all of them to zero and proceed in the following with the actual computation of the $\beta-$functions.

\subsection{Beta Functions of Duality Invariant Sigma Model}

Having found the dilaton contributions to the field equations, we can now consider the sigma model on a flat worldsheet, thereby discarding the dilaton coupling. After deriving the equations of motion for the other fields, the dilaton equation will be fixed by consistency.

In flat two-dimensional space one can safely perform the Wick rotation. We will thus study the Euclidean action
\begin{equation}\label{sigma Odd euclidean}
\begin{split}
S&=\frac{1}{2\lambda}\int d^2x\,\Big\{g_{\mu\nu}(X)\,\del^\alpha X^\mu\del_\alpha X^\nu+i\,\epsilon^{\alpha\beta}\,\Big[B_{\mu\nu}(X)\,\del_\alpha X^\mu\del_\beta X^\nu+\cA_\mu{}^M(X)\,\del_\alpha X^\mu D_\beta Y_M\Big]\\
&\hspace{23mm}-i\,D_1Y^MD_2Y_M+\cH_{MN}(X)\,D_1Y^MD_1Y^N\Big\}\;,    
\end{split}    
\end{equation}
where $x^2$ is the Euclidean time, $\epsilon^{12}=+1$ and we recall $D_\alpha Y^M=\del_\alpha Y^M+\cA_\mu{}^M(X)\,\del_\alpha X^\mu$.

\paragraph{Gauge-invariant perturbative expansion}

In order to perform the background field expansion we shall employ the covariant fluctuation $\xi^\mu$ and split
\begin{equation}
X^\mu=\varphi^\mu+\pi^\mu(\xi)   \;, 
\end{equation}
as we have extensively discussed.
If one now performs a \emph{linear} split $Y^M=\phi^M+\pi^M$ in the internal sector, manifest gauge invariance under the $U(1)$ Kaluza-Klein symmetries will be lost. The reason 
is that  $Y^M$ transforms as $\delta Y^M=-\lambda^M(X)$. Combining this with the split of $X^\mu\, $, one finds that the background field $\phi^M$ transforms with respect to the background gauge parameter:
\begin{equation}\label{delta BG phi}
\delta\phi^M=-\lambda^M(\varphi) \;,   
\end{equation}
ensuring that the background covariant derivative
\begin{equation}
D_\alpha\phi^M:=\del_\alpha\phi^M+\cA_\mu{}^M(\varphi)\,\del_\alpha\varphi^\mu    
\end{equation}
is gauge-invariant. The linear fluctuation $\pi^M$, however, transforms to all orders, as follows from  \eqref{delta BG phi}:
\begin{equation}\label{delta varpi}
\delta\pi^M=-\big[\lambda^M(X)-\lambda^M(\varphi)\big]=-\sum_{n=1}^\infty\frac{1}{n!}\,(\xi\cdot\nabla)^n\lambda^M(\varphi)\;, 
\end{equation}
where we used the covariant Taylor expansion in terms of $\xi^\mu\,$.
We can rewrite the right-hand side of \eqref{delta varpi} in terms of $\delta \cA_{\mu}{}^{M}=\nabla_{\mu}\lambda^M$ as 
\begin{equation}
-\sum_{n=1}^\infty\frac{1}{n!}\,(\xi\cdot\nabla)^n\lambda^M(\varphi)=-\sum_{n=0}^\infty\frac{1}{(n+1)!}\,(\xi\cdot\nabla)^n\Big(\xi^\mu\delta\cA_\mu{}^M(\varphi)\Big)\;. \end{equation}
Since the right-hand side here is a total gauge variation we can define a gauge invariant fluctuation $\zeta^M$ as follows: 
\begin{equation}\label{zeta}
\pi^M(\xi, \zeta)=\zeta^M-\sum_{n=0}^\infty\frac{1}{(n+1)!}\,(\xi\cdot\nabla)^n\,\Big(\xi^\mu\cA_\mu{}^M(\varphi)\Big)\;.
\end{equation}
This ensures that the perturbation theory is manifestly gauge invariant.

Even though it is guaranteed that the expansion in powers of $\xi^\mu$ and $\zeta^M$ will be gauge-invariant, we will not use \eqref{zeta} directly, just as one does not use $\pi^\mu(\xi)$ to derive the usual covariant expansion. We shall rather use \eqref{zeta} in order to find a set of covariant rules to perform the expansion. Let us recall the strategy for the covariant background field method: given any target space scalar $\Phi(X)$ 
its covariant $\xi-$expansion is given by
\begin{equation}
\Phi(X)=\Phi(\varphi+\pi(\xi))=e^{\calD}\Phi(\varphi)\;.  \end{equation}
The operator $\calD$ is defined in terms of  an auxiliary geodesic $X^{\mu}(t)$ as discussed  in  sec.~3: 
\begin{equation}
\calD \Phi := \left(\frac{D\Phi(t)}{Dt}\right)_{t=0}\;. 
\end{equation} 
This  yields 
the effective rules
\begin{equation}\label{cov rules 1}
\begin{split}
\calD T&=\xi^\mu\,\nabla_\mu T\;,\quad D_\alpha T=\del_\alpha\varphi^\mu\,\nabla_\mu T
\quad\text{for target space tensors}\;,\\
\calD\xi^\mu&=0\;,\quad \calD\del_\alpha\varphi^\mu=\del_\alpha\xi^\mu+\del_\alpha\varphi^\nu\,\Gamma^\mu_{\nu\lambda}(\varphi)\,\xi^\lambda=D_\alpha\xi^\mu\;,\\
[\calD, D_\alpha]&=\xi^\mu\del_\alpha\varphi^\nu\,R_{\mu\nu}^\#\quad\text{acting on target space tensors and}\,\xi^\mu\;
\end{split}    
\end{equation}
that ensure manifest covariance under external target space diffeomorphisms. 
Note that one should distinguish $D_\alpha V^\mu=\del_\alpha\varphi^\nu\nabla_\nu V^\mu$ from $D_\alpha\phi^M=\del_\alpha\phi^M+\del_\alpha\varphi^\mu\cA_\mu{}^M\,$.
Next, we derive effective rules for fields transforming under $U(1)$.  
Using that $D_\alpha Y^M$ is a scalar with respect to external (target space) diffeomorphisms and using \eqref{zeta}
we may now  derive the gauge-invariant expansion of $D_\alpha Y^M$:
\begin{equation}\label{DY expansion}
\begin{split}
D_\alpha Y^M&=\del_\alpha\phi^M+\del_\alpha\zeta^M+\del_\alpha X^\mu\,\cA_\mu{}^M(X)-\del_\alpha \sum_{n=0}^\infty\frac{1}{(n+1)!}\,(\xi\cdot\nabla)^n\,\Big(\xi^\mu\cA_\mu{}^M(\varphi)\Big)\\
&=\del_\alpha\phi^M+\del_\alpha\zeta^M+e^{\calD}\left[\del_\alpha \varphi^\mu\,\cA_\mu{}^M(\varphi)\right]-\del_\alpha \sum_{n=0}^\infty\frac{1}{(n+1)!}\,(\xi\cdot\nabla)^n\,\Big(\xi^\mu\cA_\mu{}^M(\varphi)\Big)\\
&=D_\alpha\phi^M+\del_\alpha\zeta^M+\sum_{n=1}^\infty\frac{1}{n!}\,\calD^n\left[\del_\alpha \varphi^\mu\,\cA_\mu{}^M(\varphi)\right]-\del_\alpha \sum_{n=0}^\infty\frac{1}{(n+1)!}\,(\xi\cdot\nabla)^n\,\Big(\xi^\mu\cA_\mu{}^M(\varphi)\Big)\;,
\end{split}    
\end{equation}
where we have extracted the zeroth order part from $e^{\calD}$ in order to recover $D_\alpha\phi^M$.
From now on we will omit the explicit evaluation of spacetime fields at the base point $\varphi^\mu$, since it is always implied in the expansion.
The final step to prove gauge invariance of the above expansion requires to evaluate the first $\calD$ on $\del_\alpha\varphi^\mu\cA_\mu{}^M\,$:
\begin{equation}
\begin{split}
\sum_{n=1}^\infty\frac{1}{n!}\,\calD^n\left[\del_\alpha\varphi^\mu\cA_\mu{}^M\right]&=\sum_{n=1}^\infty\frac{1}{n!}\,\calD^{n-1}\left[D_\alpha\xi^\mu\cA_\mu{}^M+\del_\alpha\varphi^\mu\xi^\nu\nabla_\nu\cA_\mu{}^M\right]\\
&=\sum_{n=1}^\infty\frac{1}{n!}\,\calD^{n-1}\left[\xi^\mu\del_\alpha\varphi^\nu\cF_{\mu\nu}{}^M+\del_\alpha\big(\xi^\mu\cA_\mu{}^M\big)\right]\;, 
\end{split}    
\end{equation}
where we used (\ref{cov rules 1}). 
Given that  $[\calD, D_\alpha]=0$ when acting on scalars, the non-covariant term can be rewritten as 
\begin{equation}
\sum_{n=1}^\infty\frac{1}{n!}\,\calD^{n-1} \del_\alpha\big(\xi^\mu\cA_\mu{}^M\big)  = 
\del_\alpha\sum_{n=0}^\infty\frac{1}{(n+1)!}\,\calD^n\left(\xi^\mu\cA_\mu{}^M\right) \;,   
\end{equation}
which cancels the last term in the third line of \eqref{DY expansion}. 
We have thus found the manifestly gauge-invariant expansion
\begin{equation}
D_\alpha Y^M=D_\alpha\phi^M+\del_\alpha\zeta^M+\sum_{n=0}^\infty\frac{1}{(n+1)!}\,\calD^n\left[\xi^\mu\del_\alpha\varphi^\nu\cF_{\mu\nu}{}^M\right]\;.   \end{equation}
Finally, by formally declaring 
\begin{equation}\label{cov rules 2}
\calD D_\alpha\phi^M:=\xi^\mu\del_\alpha\varphi^\nu\cF_{\mu\nu}{}^M\;,\quad \calD \del_\alpha\zeta^M:=0\;,    
\end{equation}
the entire expansion can be recast in the exponential form
\begin{equation}
D_\alpha Y^M=e^{\calD}\big(D_\alpha\phi^M+\del_\alpha\zeta^M\big) \;. \end{equation}
The set of covariant rules \eqref{cov rules 1} and \eqref{cov rules 2} allows one to write down the $\cO(\xi^n)$ term in the action recursively. 
By expanding in powers of $\xi$ one obtains the recursive relations 
\begin{equation}
\begin{split}
S[\varphi+\pi(\xi)]&=e^\calD S[\varphi]=\sum_{n=0}^\infty S_{n\xi}\;,\\
S_{n\xi}&=\frac{1}{n!}\,\calD^nS_{0\xi}=\frac1n\,\calD S_{(n-1)\xi}\;, 
\end{split}
\end{equation}
starting from
\begin{equation}\label{S0xi}
\begin{split}
S_{0\xi}&=\frac{1}{2\lambda}\int d^2x\,\Big\{g_{\mu\nu}\,\del^\alpha \varphi^\mu\del_\alpha \varphi^\nu+i\,\epsilon^{\alpha\beta}\,\Big[B_{\mu\nu}\,\del_\alpha \varphi^\mu\del_\beta \varphi^\nu+\cA_\mu{}^M\,\del_\alpha \varphi^\mu (D_\beta \phi_M+\del_\beta\zeta_M)\Big]\\
&\hspace{13mm}-i\,(D_1\phi^M+\del_1\zeta^M)(D_2\phi_M+\del_2\zeta_M)+\cH_{MN}\,(D_1\phi^M+\del_1\zeta^M)(D_1\phi^N+\del_1\zeta^N)\Big\}\;.   
\end{split}    
\end{equation}
This action is not manifestly gauge invariant due to the topological terms involving  $B_{\mu\nu}$ and $\cA_\mu{}^M$, which transform non trivially, 
but the higher-order terms $S_{n\xi}$ will be manifestly gauge invariant. To see this we compute explicitly $S_{1\xi}$:
\begin{equation}\label{S1xi preliminary}
\begin{split}
&S_{1\xi}=\calD S_{0\xi}=\frac{1}{2\lambda}\int d^2x\,\Big\{2\,g_{\mu\nu}\,D^\alpha\xi^\mu\del_\alpha\varphi^\nu+i\,\epsilon^{\alpha\beta}\,\xi^\lambda H^0_{\lambda\mu\nu}\,\del_\alpha\varphi^\mu\del_\beta\varphi^\nu\\
&+i\,\epsilon^{\alpha\beta}\Big[\xi^\nu\,\nabla_\nu\cA_\mu^M\del_\alpha\varphi^\mu(D_\beta\phi_M+\del_\beta\zeta_M)+\cA_\mu^MD_\alpha\xi^\mu\,(D_\beta\phi_M+\del_\beta\zeta_M)+\cA_\mu^M\del_\alpha\varphi^\mu\,\xi^\nu\del_\beta\varphi^\lambda\cF_{\nu\lambda\,M}\Big]\\
&-i\,\Big[(D_1\phi_M+\del_1\zeta_M)\xi^\mu\del_2\varphi^\nu\cF_{\mu\nu}^M+(D_2\phi_M+\del_2\zeta_M)\xi^\mu\del_1\varphi^\nu\cF_{\mu\nu}^M\Big]\\
&+\xi^\mu\,\nabla_\mu\cH_{MN}\,(D_1\phi^M+\del_1\zeta^M)(D_1\phi^N+\del_1\zeta^N)+2\,\cH_{MN}\,(D_1\phi^M+\del_1\zeta^M)\,\xi^\mu\del_1\varphi^\nu\cF_{\mu\nu}^N\Big\}\;,    
\end{split}    
\end{equation}
where we have already integrated by parts the term $2i\,\epsilon^{\alpha\beta}B_{\mu\nu}D_\alpha\xi^\mu\del_\beta\varphi^\nu$ in order to generate $H^0_{\lambda\mu\nu}=3\,\del_{[\lambda}B_{\mu\nu]}\,$. Integrating by parts $D_\alpha\xi^\mu$ in the term $\cA_\mu{}^MD_\alpha\xi^\mu\,(D_\beta\phi_M+\del_\beta\zeta_M)$ we finally obtain
\begin{equation}\label{S1xi}
\begin{split}
S_{1\xi}&=\frac{1}{2\lambda}\int d^2x\,\Big\{2\,g_{\mu\nu}\,D^\alpha\xi^\mu\del_\alpha\varphi^\nu+i\,\epsilon^{\alpha\beta}\,\xi^\lambda H_{\lambda\mu\nu}\,\del_\alpha\varphi^\mu\del_\beta\varphi^\nu-2i\,\xi^\mu\del_2\varphi^\nu\cF_{\mu\nu}{}^M\,(D_1\phi_M+\del_1\zeta_M)\\
&+\xi^\mu\,\nabla_\mu\cH_{MN}\,(D_1\phi^M+\del_1\zeta^M)(D_1\phi^N+\del_1\zeta^N)+2\,\cH_{MN}\,(D_1\phi^M+\del_1\zeta^M)\,\xi^\mu\del_1\varphi^\nu\cF_{\mu\nu}{}^N\Big\} \;.   
\end{split}    
\end{equation}
This is manifestly gauge-invariant under all symmetries, since it displays the invariant field strength $H_{\mu\nu\rho}=H^0_{\mu\nu\rho}-\frac32\,\cA_{[\mu}{}^M\cF_{\nu\rho]\,M}$, 
which ensures that the perturbative expansion will be manifestly gauge-invariant at all orders.

For the one-loop computation we only need the action to quadratic order in all fluctuations, which 
 is schematically given by $S_2=S_{\zeta\zeta}+S_{\zeta\xi}+S_{\xi\xi}\,$. The term $S_{\zeta\zeta}$ is easily obtained from \eqref{S0xi}:
\begin{equation}\label{Szetazeta}
S_{\zeta\zeta}=\frac{1}{2\lambda}\int d^2x\,\Big[-i\,\del_1\zeta^M\del_2\zeta_M+\cH_{MN}\,\del_1\zeta^M\del_1\zeta^N\Big] \;,   
\end{equation}
while $S_{\zeta\xi}$ is obtained by keeping the $\cO(\zeta)$ terms of \eqref{S1xi}:
\begin{equation}\label{Szetaxi}
S_{\zeta\xi}=\frac{1}{\lambda}\int d^2x\,\Big[\xi^\mu\,\nabla_\mu\cH_{MN}\,D_1\phi^M\del_1\zeta^N+\xi^\mu\del_1\zeta^M\cF_{\mu\nu}^N\,\big(\del_1\varphi^\nu\cH_{MN}-i\,\del_2\varphi^\nu\eta_{MN}\big)\Big]  \;.  
\end{equation}
$S_{\xi\xi}$ on the other hand requires one  to compute $\frac12\,\calD S_{1\xi}$ and set $\zeta^M=0$ in \eqref{S1xi}, finally yielding
\begin{equation}\label{Sxixi}
\begin{split}
S_{\xi\xi}&=\frac{1}{2\lambda}\int d^2x\,\Big\{g_{\mu\nu}\,D^\alpha\xi^\mu D_\alpha\xi^\nu+R_{\mu\nu\lambda\rho}\,\xi^\nu\xi^\lambda\,\del^\alpha\varphi^\mu\del_\alpha\varphi^\rho+\tfrac{i}{2}\,\epsilon^{\alpha\beta}\,\xi^\lambda\xi^\rho\,\nabla_\lambda H_{\rho\mu\nu}\,\del_\alpha\varphi^\mu\del_\beta\varphi^\nu\\
&+i\,\epsilon^{\alpha\beta}\,\xi^\lambda D_\alpha\xi^\mu H_{\lambda\mu\nu}\,\del_\beta\varphi^\nu+\tfrac12\,\xi^\mu\xi^\nu\,\nabla_\mu\nabla_\nu\cH_{MN}\,D_1\phi^M D_1\phi^N\\
&+D_1\phi^M\cF_{\mu\nu}^N\,\big(\xi^\mu D_1\xi^\nu\,\cH_{MN}-i\,\xi^\mu D_2\xi^\nu\,\eta_{MN}\big)+\xi^\lambda \xi^\mu\,D_1\phi^M\nabla_\lambda\cF_{\mu\nu}^N\,\big( \del_1\varphi^\nu\,\cH_{MN}-i\, \del_2\varphi^\nu\,\eta_{MN}\big)\\
&+\xi^\lambda\xi^\mu\del_1\varphi^\rho\cF_{\mu\nu}^M\cF_{\lambda\rho}^N\,\big( \del_1\varphi^\nu\,\cH_{MN}-i\, \del_2\varphi^\nu\,\eta_{MN}\big)+2\,\xi^\mu\xi^\nu\del_1\varphi^\lambda\,\nabla_\mu\cH_{MN}\,\cF_{\nu\lambda}^M\,D_1\phi^N\Big\}  \;.  
\end{split}    
\end{equation}

In order to have standard kinetic terms we introduce background vielbeins $e_\mu^a(\varphi)$ and consider the flattened fluctuation $\xi^a=e_\mu^a\,\xi^\mu\,$. This transforms the $\xi-$kinetic term into $D^\alpha\xi^a D_\alpha\xi_a$, with 
\begin{equation}
D_\alpha\xi^a=\del_\alpha\xi^a+\del_\alpha\varphi^\mu\,\omega_\mu{}^a{}_b\,\xi^b=\del_\alpha\xi^a+\omega_\alpha{}^a{}_b\,\xi^b  \;,  
\end{equation}
that allows us  to extract a standard propagator from $\frac{1}{2\lambda}\int d^2x\,\del^\alpha\xi^a\del_\alpha\xi_a\,$. 
As recalled in sec.~3, 
dimensional arguments using  local Lorentz symmetry ensure that the spin connection cannot contribute to UV divergences at one loop. This allows one to effectively treat all $D_\alpha\xi^a$
as $\del_\alpha\xi^a$ in the interaction vertices. 

A similar treatment is needed for the $\zeta-$kinetic term $S_{\zeta\zeta}$, given the non-constant background $\cH_{MN}(\varphi)\,$. We shall thus introduce $O(d,d)-$valued frames $E_M{}^A(\varphi)$, such that
\begin{equation}
\cH_{MN}=E_M{}^A\,h_{AB}\,E_N{}^B\;,\quad \eta_{MN}=E_M{}^A\,\eta_{AB}\,E_N{}^B\;, \end{equation}
where $\eta_{AB}$ has the same form as $\eta_{MN}$, while $h_{AB}$ is a \emph{constant} matrix that we choose to be a metric of $SO(d)\times SO(d)$ embedded in $O(d,d)\,$. The frames are then subject to \emph{local} $SO(d)\times SO(d)$ transformations that rotate the flat index $A\,$. We then flatten the fluctuations by defining $\zeta^A=E_M{}^A(\varphi)\,\zeta^M\,$. The derivatives, accordingly, are modified as
\begin{equation}
\del_\alpha\zeta^M=E^M{}_A\,D_\alpha\zeta^A  \;,\quad D_\alpha\zeta^A=\del_\alpha\zeta^A+\del_\alpha\varphi^\mu\,W_\mu{}^A{}_B\,\zeta^B = \del_\alpha\zeta^A+W_\alpha{}^A{}_B\,\zeta^B\;,
\end{equation}
where $W_\mu{}^{AB}$ is the Maurer-Cartan flat connection
\begin{equation}
W_\mu{}^{AB}=E^{MA}\del_\mu E_M{}^B=Q_\mu{}^{AB}+P_\mu{}^{AB}    
\end{equation}
that splits into an $SO(d)\times SO(d)$ connection $Q$ and an $SO(d)\times SO(d)$ tensor $P$, as familiar in coset constructions.
The kinetic term $S_{\zeta\zeta}$ is rewritten as
\begin{equation}
\begin{split}
S_{\zeta\zeta}&=\frac{1}{2\lambda}\int d^2x\,\Big[-i\,D_1\zeta^A D_2\zeta_A+h_{AB}\,D_1\zeta^AD_1\zeta^B\Big]\\
&=\frac{1}{2\lambda}\int d^2x\,\Big[-i\,\del_1\zeta^A \del_2\zeta_A+h_{AB}\,\del_1\zeta^A\del_1\zeta^B\Big]+S_W+S_{WW}  \;,  
\end{split}    
\end{equation}
where the free part defines the propagator, while the interaction terms involving the Maurer-Cartan form are given by
\begin{equation}
S_W=\frac{1}{\lambda}\int d^2x\,T^{\alpha\beta}_{AB}\,W_\alpha{}^A{}_C\,\del_\beta\zeta^B\zeta^C\;,\quad S_{WW}=\frac{1}{2\lambda}\int d^2x\,T^{\alpha\beta}_{AB}\,W_\alpha{}^A{}_C\,W_\beta{}^B{}_D\,\zeta^C\zeta^D\;,
\end{equation}
where $T^{\alpha\beta}_{AB}$ is defined by
\begin{equation}\label{T tensor}
T^{11}_{AB}=h_{AB}\;,\quad T^{12}_{AB}=T^{21}_{AB}=-\frac{i}{2}\,\eta_{AB}\;,\quad T^{22}_{AB}=0\;. \end{equation}
Despite the similarities with the Lorentz connection $\omega_\alpha{}^{ab}$, the Maurer-Cartan form has two important differences: first of all, it contains an $SO(d)\times SO(d)$ tensor, $P_\alpha{}^{AB}$, that \emph{does} contribute to divergences. Moreover, as it has been recently shown \cite{Bonezzi:2020ryb}, the local $SO(d)\times SO(d)$ symmetry is anomalous, due to the chiral nature of the $Y-$sector. The anomaly, that is canceled by a suitable $\cO(\alpha')$ transformation of $B_{\mu\nu}\,$ \cite{Eloy:2019hnl,Bonezzi:2020ryb}, is a one-loop effect but is \emph{finite}. Therefore, it does not affect our computation of the $\beta-$functions at this order, but it should intervene at higher loops, thereby triggering the Green-Schwarz type deformation of the field strength $H_{\mu\nu\rho}$ at order $\alpha'\,$.
For these reasons, the connection terms arising from both $S_{\zeta\zeta}$ and $S_{\zeta\xi}$ \emph{cannot} be ignored.

\paragraph{One-loop $\beta-$functions}

From the free kinetic terms we can derive the propagators, that read
\begin{equation}
\begin{split}
\langle \xi^a(x)\,\xi^b(y)\rangle&=\lambda\,\delta^{ab}\,G(x-y)\;,\quad G(x)=\int\frac{d^2p}{(2\pi)^2}\frac{e^{ip\cdot x}}{p^2}\;,\\
\langle \zeta^A(x)\,\zeta^B(y)\rangle&=\lambda\,G^{AB}(x-y)\;,\quad G^{AB}(x)=\int\frac{d^2p}{(2\pi)^2}\frac{e^{ip\cdot x}}{p^2}\,\Big[h^{AB}+i\,\frac{p_2}{p_1}\,\eta^{AB}\Big]\;.
\end{split}    
\end{equation}
The above massless propagators suffer from infrared divergences. The $\xi-$propagator $G(x)$ can be regularized in the IR by adding a mass term $\frac{m^2}{2\lambda}\int d^2x\,\xi^a\xi_a $ to the action, thereby modifying $\frac{1}{p^2}\to\frac{1}{p^2+m^2}$ as usual. The IR regularization of the $\zeta-$propagator is more subtle and is discussed in Appendix \ref{Appendix IR reg}.
The interaction vertices extracted from \eqref{Szetazeta}, \eqref{Szetaxi} and \eqref{Sxixi} can be summarized as
\begin{equation}
\begin{split}
S_{\zeta\zeta}&\;\longrightarrow\;S_W+S_{WW}\;,\\
S_{\zeta\xi}&\;\longrightarrow\;S_{\nabla\cH}+S_{W\nabla\cH}+S_\cF+S_{W\cF}\;,\\
S_{\xi\xi}&\;\longrightarrow\;S_R+S_{\nabla H}+S_H+S_{\nabla\nabla\cH}+S_{\cF'}+S_{\nabla\cF}+S_{\cF\cF}+S_{\nabla\cH\cF}\;,
\end{split}    
\end{equation}
where we schematically distinguished them according to the target space fields entering the interaction. Let us notice that the new terms $S_{W\nabla\cH}$ and $S_{W\cF}$ arise from the replacement $\del_\alpha\zeta^M=E^M{}_A\,(\del_\alpha\zeta^A+W_\alpha{}^{AB}\zeta_B)\,$. Another way to classify the vertices is according to the number of background fields $\del_\alpha\varphi^\mu$ and $D_\alpha\phi^M$ involved:
\begin{equation}
\begin{split}
\del\varphi\del\varphi&\;\longrightarrow\;S_{WW}+S_{W\cF}+S_R+S_{\nabla H}+S_{\cF\cF}\;,\\
\del\varphi D\phi&\;\longrightarrow\;S_{W\nabla\cH}+S_{\nabla\cF}+S_{\nabla\cH\cF}\;,\\
D\phi D\phi&\;\longrightarrow\;S_{\nabla\nabla\cH}\;,\\
\del\varphi&\;\longrightarrow\;S_W+S_\cF+S_H\;,\\
D\phi&\;\longrightarrow\;S_{\nabla\cH}+S_{\cF'}\;.
\end{split}    
\end{equation}
Since the divergences in $e^{-\Gamma_{1l}}=\left\langle e^{-S_{\rm int}}\right\rangle$ can only renormalize the dimension two operators $\del\varphi\del\varphi$, $\del\varphi D\phi$ and $D\phi D\phi$, they can only arise from the divergent part of
\begin{equation}\label{1loop divergent parts}
\begin{split}
\Gamma_{1l\,\del\varphi\del\varphi}&=\langle S_R\rangle+\langle S_{\nabla H}\rangle+\langle S_{WW}\rangle+\langle S_{\cF\cF}\rangle-\tfrac12\,\langle S_W^2+S_\cF^2+S_H^2\rangle_{1{\rm PI}}\;,\\
\Gamma_{1l\,D\phi D\phi}&=\langle S_{\nabla\nabla\cH}\rangle-\tfrac12\,\langle S_{\nabla\cH}^2+S_{\cF'}^2\rangle_{1{\rm PI}}\;,\\
\Gamma_{1l\,\del\varphi D\phi}&=\langle S_{\nabla\cF}\rangle+\langle S_{\nabla\cH\cF}\rangle-\langle S_\cF S_{\nabla\cH}\rangle_{1\rm{PI}}-\langle S_H S_{\cF'}\rangle_{1\rm{PI}}\;,
\end{split}    
\end{equation}
where we have set to zero all the contributions that have trivially vanishing contractions.

We are now ready to compute the divergent parts of \eqref{1loop divergent parts} starting from $\Gamma_{1l\,\del\varphi\del\varphi}$, that gives the $\beta-$functions of the metric $g_{\mu\nu}$ and $B-$field. The diagrams with a single contraction, \emph{e.g.} $\langle S_R\rangle$, are tadpole diagrams that give a purely local contribution. For instance, 
\begin{equation}\label{Ricci}
\begin{split}
\langle S_R\rangle&=\frac{1}{2\lambda}\int d^2x\,\langle\xi^a\xi^b\rangle\,R_{\mu ab\nu}\,\del^\alpha\varphi^\mu\del_\alpha\varphi^\nu=-\frac12\,G(0)\int d^2x\,R_{\mu\nu}\,\del^\alpha\varphi^\mu\del_\alpha\varphi^\nu\\
&=\frac{1}{4\pi\epsilon}\,\int d^2x\,R_{\mu\nu}\,\del^\alpha\varphi^\mu\del_\alpha\varphi^\nu +\cO(\epsilon^0)   \;, 
\end{split}    
\end{equation}
and
\begin{equation}\label{divH}
\begin{split}
\langle S_{\nabla H}\rangle&=\frac{i}{4\lambda}\int d^2x\,\epsilon^{\alpha\beta}\,\langle\xi^a\xi^b\rangle\,e^\lambda_ae^\rho_b\,\nabla_\lambda H_{\rho\mu\nu}\,\del_\alpha\varphi^\mu\del_\beta\varphi^\nu   \\
&=\frac{i}{4}\,G(0)\int d^2x\,\nabla^\rho H_{\rho\mu\nu}\,\epsilon^{\alpha\beta}\,\del_\alpha\varphi^\mu\del_\beta\varphi^\nu \\
&=-\frac{i}{8\pi\epsilon}\int d^2x\,\nabla^\rho H_{\rho\mu\nu}\,\epsilon^{\alpha\beta}\,\del_\alpha\varphi^\mu\del_\beta\varphi^\nu+\cO(\epsilon^0)\;, 
\end{split}    
\end{equation}
where we used the  $\frac{1}{\epsilon}$ contribution to $G(0)$ given in (\ref{G(0)}) (c.f.~(\ref{G0Appendix})). 
The bubble diagrams that arise from double contractions, such as $\langle S_H^2\rangle_{1{\rm PI}}$, are non-local, but their divergent contribution only comes from the local leading order, when expanding in external momenta.
As an example, let us consider $\langle S_H^2\rangle_{1{\rm PI}}\,$. Its contribution to $\Gamma_{1l\,\del\varphi\del\varphi}$ reads
\begin{equation}
\begin{split}
-\frac12\,\langle S_H^2\rangle_{1{\rm PI}}&=\frac{1}{8\lambda^2}\int d^2xd^2y\,\epsilon^{\alpha\beta}\epsilon^{\gamma\delta}\,\big(H_{ab\mu}\,\del_\beta\varphi^\mu\big)(x)\,\big(H_{cd\nu}\,\del_\delta\varphi^\nu\big)(y)\,\big\langle\xi^a(x)\,\del_\alpha\xi^b(x)\,\xi^c(y)\,\del_\gamma\xi^d(y)\big\rangle\\
&= \frac{1}{8}\int\frac{d^2p}{(2\pi)^2}\,\epsilon^{\alpha\beta}\epsilon^{\gamma\delta}\,\big(H_{\mu ab}\,\del_\beta\varphi^\mu\big)(p)\,\Pi_{\alpha\gamma}(p)\,\big(H_\nu{}^{ab}\,\del_\delta\varphi^\nu\big)(-p)\;,
\end{split}    
\end{equation}
where the polarization tensor is given by
\begin{equation}
\Pi_{\alpha\gamma}(p)=\frac12 \int\frac{d^2k}{(2\pi)^2}\,\frac{(p+2k)_\alpha(p+2k)_\gamma}{(k^2+m^2)((p+k)^2+m^2)}\;.   
\end{equation}
The divergent part of $\Pi_{\alpha\gamma}(p)$ comes from $\Pi_{\alpha\gamma}(0)$, which  is the only contribution we need to compute. Upon extending the integral to $n=2+\epsilon$ dimensions we have
\begin{equation}
\begin{split}
\Pi^{\rm reg}_{\alpha\gamma}(0) &=\mu^{2-n}\int\frac{d^nk}{(2\pi)^n}\,\frac{2\,k_\alpha k_\gamma}{(k^2+m^2)^2} =\mu^{2-n}\,\frac{2}{n}\,\delta_{\alpha\gamma}\,\int\frac{d^nk}{(2\pi)^n}\,\frac{k^2}{(k^2+m^2)^2} \\
&=\frac{\delta_{\alpha\gamma}}{4\pi}\,\left(\frac{m^2}{4\pi\mu^2}\right)^{\frac{\epsilon}{2}}\Gamma(-\epsilon/2)=\delta_{\alpha\gamma}\,G(0)=-\frac{\delta_{\alpha\gamma}}{2\pi\epsilon}+\cO(\epsilon^0)\;,
\end{split}    
\end{equation}
yielding
\begin{equation}\label{H^2}
\begin{split}
-\frac12\,\langle S_H^2\rangle_{1{\rm PI}}&=
 -\frac{1}{16\pi\epsilon}\int\frac{d^2p}{(2\pi)^2}\,\big(H_{\mu ab}\,\del_\alpha\varphi^\mu\big)(p)\,\big(H_\nu{}^{ab}\,\del^\alpha\varphi^\nu\big)(-p)+\cO(\epsilon^0)\\
 &=-\frac{1}{16\pi\epsilon}\int d^2x\,H_{\mu \lambda\rho}H_\nu{}^{\lambda\rho}\,\del_\alpha\varphi^\mu\del^\alpha\varphi^\nu+\cO(\epsilon^0)\;.
\end{split}    
\end{equation}

All the remaining integrals in (\ref{1loop divergent parts}) can be computed in a similar way, upon using the results of Appendix \ref{Appendix IR reg} for the $\zeta-$sector, 
the results of which we summarize now. 
Proceeding with the terms involving the Maurer-Cartan form $W_\mu{}^{AB}$, it is interesting to notice that the separate diagrams $\langle S_{WW}\rangle$ and $\langle S^2_W\rangle_{1{\rm PI}}$ are neither 2D-Lorentz nor $SO(d)\times SO(d)$ gauge-invariant:
\begin{equation}
\begin{split}
\langle S_{WW}\rangle&=\frac{1}{2\lambda}\int d^2x\,T^{\alpha\beta}_{AB}\,\del_\alpha\varphi^\mu\del_\beta\varphi^\nu\,W_\mu{}^A{}_C\,W_\nu{}^B{}_D\,\langle\zeta^C\zeta^D\rangle\\
&=-\frac{1}{4\pi\epsilon}\int d^2x\,W_\mu{}^{AC}W_\nu{}^{BD}\,\Big[h_{AB}\,h_{CD}\,\del_1\varphi^\mu\del_1\varphi^\nu-i\,\eta_{AB}\,h_{CD}\,\del_1\varphi^\mu\del_2\varphi^\nu\Big]+\cO(\epsilon^0)
\end{split}    
\end{equation}
and
\begin{equation}
\begin{split}
-\frac12\,\langle S_W^2\rangle_{1{\rm PI}}&=-\frac{1}{2\lambda^2}\int d^2xd^2y\,T^{\alpha\beta}_{AB}\,\big(\del_\alpha\varphi^\mu\,W_\mu{}^A{}_C\big)(x)\,T^{\gamma\delta}_{DE}\,\big(\del_\gamma\varphi^\nu\,W_\nu{}^D{}_F\big)(y)\\
&\hspace{20mm}\times\big\langle\del_\beta\zeta^B(x)\,\zeta^C(x)\,\del_\delta\zeta^E(y)\,\zeta^F(y)\big\rangle_{1{\rm PI}}\\
&=\frac{1}{4\pi\epsilon}\int d^2x\,W_{\mu\,AB}\,W_{\nu\,CD}\,\Big[\del_1\varphi^\mu\del_1\varphi^\nu\,\big(\tfrac34\,h^{AC}\,h^{BD}+\tfrac14\,\eta^{AC}\,\eta^{BD}\big)\\
&\hspace{15mm}+\del_2\varphi^\mu\del_2\varphi^\nu\,\big(\tfrac14\,h^{AC}\,h^{BD}-\tfrac14\,\eta^{AC}\,\eta^{BD}\big)-i\,\del_1\varphi^\mu\del_2\varphi^\nu\,\eta^{AC}\,h^{BD}\Big]+\cO(\epsilon^0)\;,
\end{split}    
\end{equation}
where the objects $T^{\alpha\beta}_{AB}$ have been defined in \eqref{T tensor}.
However, the two diagrams combine in a non-trivial way to form the Lorentz and gauge-invariant term
\begin{equation}\label{HMN for g}
\begin{split}
\langle S_{WW}\rangle-\frac12\,\langle S_W^2\rangle_{1{\rm PI}}&=\frac{1}{16\pi\epsilon}\int d^2x\,\del^\alpha\varphi^\mu\del_\alpha\varphi^\nu\,W_{\mu\,AB}\,W_{\nu\,CD}\,\big(\eta^{AC}\,\eta^{BD}-h^{AC}\,h^{BD}\big) +\cO(\epsilon^0)\\
&=\frac{1}{8\pi\epsilon}\int d^2x\,\del^\alpha\varphi^\mu\del_\alpha\varphi^\nu\,P_{\mu\,AB}\,P_{\nu}{}^{AB} +\cO(\epsilon^0)\\
&=\frac{1}{32\pi\epsilon}\int d^2x\,\del^\alpha\varphi^\mu\del_\alpha\varphi^\nu\,\del_\mu\cH_{MN}\,\del_\nu\cH^{MN} +\cO(\epsilon^0)\;.
\end{split}    
\end{equation}
The final contributions to $\Gamma_{1l\,\del\varphi\del\varphi}$, namely $\langle S_{\cF\cF}\rangle$ and $\langle S^2_\cF\rangle_{1{\rm PI}}$, behave in a similar manner: the separate diagrams are not Lorentz invariant, being given by
\begin{equation}
\begin{split}
\langle S_{\cF\cF}\rangle&=\frac{1}{2\lambda}\int d^2x\,\langle\xi^a\xi^b\rangle\,\del_1\varphi^\mu\,\cF_{\mu a}^M\,\cF_{\nu b}^N\,\big(\del_1\varphi^\nu\,\cH_{MN}-i\,\del_2\varphi^\nu\,\eta_{MN}\big)\\
&=-\frac{1}{4\pi\epsilon}\int d^2x\,\cF_{\mu \lambda}^M\,\cF_{\nu }{}^{\lambda\,N}\,\big(\del_1\varphi^\mu\del_1\varphi^\nu\,\cH_{MN}-i\,\del_1\varphi^\mu\del_2\varphi^\nu\,\eta_{MN}\big)+\cO(\epsilon^0)
\end{split}    
\end{equation}
and
\begin{equation}
\begin{split}
 -\frac12\,\langle S^2_\cF\rangle_{1{\rm PI}}&=-\frac{1}{2\lambda^2}\int d^2xd^2y\,\big[\cF_{\mu a}^N\,(\del_1\varphi^\mu\,\cH_{MN}-i\,\del_2\varphi^\mu\,\eta_{MN})\,E^M{}_A\big](x)\\
 &\times\,\big[\cF_{\nu b}^Q\,(\del_1\varphi^\nu\,\cH_{PQ}-i\,\del_2\varphi^\nu\,\eta_{PQ})\,E^P{}_B\big](y)\,\big\langle\xi^a(x)\,\del_1\zeta^A(x)\,\xi^b(y)\,\del_1\zeta^B(y)\big\rangle\\
 &=\frac{1}{8\pi\epsilon}\int d^2x\,\cF_{\mu\lambda}^M\,\cF_\nu{}^{\lambda\,N}\,\Big[\big(\del_1\varphi^\mu\del_1\varphi^\nu-\del_2\varphi^\mu\del_2\varphi^\nu\big)\,\cH_{MN}-2i\,\del_1\varphi^\mu\del_2\varphi^\nu\,\eta_{MN}\Big]+\cO(\epsilon^0)\;,
\end{split}    
\end{equation}
but combine into the Lorentz invariant term
\begin{equation}\label{FF for g}
\langle S_{\cF\cF}\rangle-\frac12\,\langle S^2_\cF\rangle_{1{\rm PI}}=-\frac{1}{8\pi\epsilon}\int d^2x\,\del^\alpha\varphi^\mu\del_\alpha\varphi^\nu\,\cH_{MN}\,\cF_{\mu\lambda}^M\,\cF_\nu{}^{\lambda\,N}+\cO(\epsilon^0)\;.
\end{equation}
Implementing minimal subtraction to cancel the divergences arising from \eqref{Ricci}, \eqref{H^2}, \eqref{HMN for g}, \eqref{FF for g} and \eqref{divH} we fix the counterterm Lagrangian involving $\del\varphi\del\varphi\,$:
\begin{equation}\label{g+B counter}
\begin{split}
S_{\rm c.t.}^{g+B}&=-\frac{1}{4\pi\epsilon}\int d^2x\,\del^\alpha\varphi^\mu\del_\alpha\varphi^\nu\,\Big[R_{\mu\nu}-\tfrac14\,H^2_{\mu\nu}-\tfrac12\,\big(\cH\cF\cF\big)_{\mu\nu}+\tfrac18\,\del_\mu\cH_{MN}\del_\nu\cH^{MN}\Big]\\&+\frac{i}{8\pi\epsilon}\int d^2x\,\epsilon^{\alpha\beta}\del_\alpha\varphi^\mu\del_\beta\varphi^\nu\,\nabla^\rho H_{\rho\mu\nu}\;,    
\end{split}    
\end{equation}
where we have defined $H^2_{\mu\nu}=H_{\mu\lambda\rho}\,H_\nu{}^{\lambda\rho}$ and $\big(\cH\cF\cF\big)_{\mu\nu}=\cH_{MN}\,\cF_{\mu\lambda}{}^M\,\cF_\nu{}^{\lambda\,N}$.
Comparing \eqref{g+B counter} with the classical background action
\begin{equation}
\frac{1}{2\lambda}\int d^2x\,\Big[g_{\mu\nu}\,\del^\alpha\varphi^\mu\del_\alpha\varphi^\nu+i\,\epsilon^{\alpha\beta}B_{\mu\nu}\,\del_\alpha\varphi^\mu\del_\beta\varphi^\nu\Big]
\end{equation}
fixes the normalization of the tensors $T_{1\mu\nu}^g$ and $T_{1\mu\nu}^B$ entering $g_{0\mu\nu}=\mu^\epsilon\,\big[g_{\mu\nu}+\frac{1}{\epsilon}\,T_{1\mu\nu}^g+\cdots\big]$ and $B_{0\mu\nu}\,$. This finally fixes the normalization of the one-loop $\beta-$functions:
\begin{equation}\label{MS beta g+B}
\begin{split}
\beta_{\mu\nu}^g&=\alpha'\,\Big[R_{\mu\nu}-\tfrac14\,H^2_{\mu\nu}-\tfrac12\,\big(\cH\cF\cF\big)_{\mu\nu}+\tfrac18\,\nabla_\mu\cH_{MN}\nabla_\nu\cH^{MN}\Big]  \;,\\
\beta_{\mu\nu}^B&=-\frac{\alpha'}{2}\,\nabla^\lambda H_{\lambda\mu\nu}\;,
\end{split}    
\end{equation}
where we recall  that $\beta^i=(-1+\Delta)\,T^i_1$, with the Euler operator
\begin{equation}
\Delta=g_{\mu\nu}\cdot\frac{\del}{\del g_{\mu\nu}}+B_{\mu\nu}\cdot\frac{\del}{\del B_{\mu\nu}}+\cA_{\mu\,M}\cdot\frac{\del}{\del \cA_{\mu\,M}}+\cH_{MN}\cdot\frac{\del}{\del \cH_{MN}}+\eta_{MN}\cdot\frac{\del}{\del \eta_{MN}}    
\end{equation}
having zero eigenvalue on all one-loop terms.

We now continue by computing the divergent part of $\Gamma_{1l\,D\phi D\phi}$, that yields the $\beta-$function of the generalized metric $\cH_{MN}\,$: the three diagrams in \eqref{1loop divergent parts} yield
\begin{equation}
\begin{split}
\langle S_{\nabla\nabla\cH}\rangle&= -\frac{1}{8\pi\epsilon}\int d^2x\,\nabla^2\cH_{MN}\,D_1\phi^M D_1\phi^N+\cO(\epsilon^0)\;,\\    
-\frac12\,\langle S^2_{\nabla\cH}\rangle_{1{\rm PI}}&= -\frac{1}{8\pi\epsilon}\int d^2x\,\cH_{MP}\nabla^\mu\cH^{PQ}\nabla_\mu\cH_{QN}\,D_1\phi^M D_1\phi^N+\cO(\epsilon^0)\;,\\    
-\frac12\,\langle S^2_{\cF'}\rangle_{1{\rm PI}}&= -\frac{1}{16\pi\epsilon}\int d^2x\,\Big[\cF_{\mu\nu\,M}\,\cF_N^{\mu\nu}-\cH_{MP}\cH_{NQ}\,\cF_{\mu\nu}^P\,\cF^{\mu\nu\,Q}\Big]\,D_1\phi^M D_1\phi^N+\cO(\epsilon^0)\;.
\end{split}    
\end{equation}
This fixes the corresponding counterterm Lagrangian to
\begin{equation}
\begin{split}
S_{\rm c.t.}^\cH&=\frac{1}{8\pi\epsilon}\int d^2x\,D_1\phi^MD_1\phi^N\,\Big[\nabla^2\cH_{MN}+\cH_{MP}\nabla^\mu\cH^{PQ}\nabla_\mu\cH_{QN}\\
&\hspace{20mm}+\tfrac12\,\cF_{\mu\nu\,M}\,\cF_N^{\mu\nu}-\tfrac12\,\cH_{MP}\cH_{NQ}\,\cF_{\mu\nu}^P\,\cF^{\mu\nu\,Q}\Big] \;, \end{split}    
\end{equation}
and comparing with the classical action
\begin{equation}
\frac{1}{2\lambda}\int d^2x\,\cH_{MN}\,\del_1\phi^M\del_1\phi^N+\cdots    
\end{equation}
gives the $\beta-$function
\begin{equation}\label{MS beta cH}
\beta_{MN}=-\frac{\alpha'}{2}\,\Big[\nabla^2\cH_{MN}+\cH_{MP}\nabla^\mu\cH^{PQ}\nabla_\mu\cH_{QN}+\tfrac12\,\cF_{\mu\nu\,M}\,\cF_N^{\mu\nu}-\tfrac12\,\cH_{MP}\cH_{NQ}\,\cF_{\mu\nu}^P\,\cF^{\mu\nu\,Q}\Big] \;.   
\end{equation}
Finally, the remaining diagrams in \eqref{1loop divergent parts} give the divergent part of $\Gamma_{1l\,\del\varphi D\phi}\,$:
\begin{equation}
\begin{split}
\langle S_{\nabla\cF}\rangle&=-\frac{1}{4\pi\epsilon}\int d^2x\,\nabla^\rho\cF_{\rho\mu}{}^M\,D_1\phi^N\,\big(\del_1\varphi^\mu\,\cH_{MN}-i\,\del_2\varphi^\mu\,\eta_{MN}\big)+\cO(\epsilon^0)\;,\\    
\langle S_{\nabla\cH\cF}\rangle&=-\frac{1}{2\pi\epsilon}\int d^2x\,\nabla^\rho\cH_{MN}\cF_{\rho\mu}{}^M\,D_1\phi^N\del_1\varphi^\mu+\cO(\epsilon^0)\;,\\
-\langle S_{\cF}S_{\nabla\cH}\rangle_{1{\rm PI}}&=\frac{1}{4\pi\epsilon}\int d^2x\,\Big[\nabla^\mu\cH_{MN}\,\cF_{\mu\nu P}D_1\phi^M\,\Big(\eta^{NP}\del_1\varphi^\nu-i\,\cH^{NP}\del_2\varphi^\nu\Big)\Big]+\cO(\epsilon^0)\;,\\
-\langle S_{H}S_{\cF'}\rangle_{1{\rm PI}}&=-\frac{1}{8\pi\epsilon}\int d^2x\,H^{\mu\nu}{}_{\rho}\,D_1\phi^M\,\Big[\cF_{\mu\nu M}\,\del_1\varphi^\rho-i\,\cH_{MN}\,\cF_{\mu\nu}{}^N\,\del_2\varphi^\rho\Big]+\cO(\epsilon^0)\;. 
\end{split}    
\end{equation}
This requires the counterterm 
\begin{equation}
\begin{split}
S_{\rm c.t.}^\cA&=-\frac{i}{4\pi\epsilon}\int d^2x\,D_1\phi_M\del_2\varphi^\nu\,\Big[\nabla^\mu\cF_{\mu\nu}{}^M+\cH^{MN}\nabla^\mu\cH_{NP}\,\cF_{\mu\nu}{}^P+\tfrac12\,\cH^{MN}H_{\nu\lambda\rho}\,\cF^{\lambda\rho}{}_N\Big]\\
&+\frac{1}{4\pi\epsilon}\int d^2x\,D_1\phi^M\del_1\varphi^\nu\,\Big[\nabla^\mu\big(\cH_{MN}\,\cF_{\mu\nu}{}^N\big)+\tfrac12\,H_{\nu\lambda\rho}\,\cF^{\lambda\rho}{}_M\Big]\\
&=\frac{1}{4\pi\epsilon}\int d^2x\,D_1\phi^Q\big(\del_1\varphi^\nu\,\cH_{MQ}-i\,\del_2\varphi^\nu\,\eta_{MQ}\big)\\
&\hspace{35mm}\times\Big[\nabla^\mu\cF_{\mu\nu}{}^M+\cH^{MN}\nabla^\mu\cH_{NP}\,\cF_{\mu\nu}{}^P+\tfrac12\,\cH^{MN}H_{\nu\lambda\rho}\,\cF^{\lambda\rho}{}_N\Big]\;.
\end{split}    
\end{equation}
Comparison with the original Lagrangian
\begin{equation}
\frac{1}{2\lambda}\int d^2x\,\Big[-2i\,\del_1\phi^M\del_2\varphi^\mu\,\cA_{\mu\,M}+\cdots\Big]    
\end{equation}
allows us  to determine the final $\beta-$function for the gauge fields:
\begin{equation}\label{MS beta A}
\beta_\nu^M=-\frac{\alpha'}{2}\,\Big[\nabla^\mu\cF_{\mu\nu}{}^M+\cH^{MN}\nabla^\mu\cH_{NP}\,\cF_{\mu\nu}{}^P+\tfrac12\,\cH^{MN}H_{\nu\lambda\rho}\,\cF^{\lambda\rho}{}_N\Big]  \;. 
\end{equation}

\subsection{Field Equations}

Knowing the $\beta-$functions, the equations of motion of the background fields $g_{\mu\nu}$, $B_{\mu\nu}$, $\cH_{MN}$ and $\cA_\mu{}^M$ are given by the vanishing of the Weyl anomaly coefficients \eqref{eoms beta bar}. Having determined by covariance that the vectors $W_\mu$, $L_\mu$ and $V_M$ must vanish at one-loop order, we are finally able to determine the field equations:  
\begin{equation}\label{MS eoms}
\begin{split}
R_{\mu\nu}-\tfrac14\,H_{\mu\lambda\rho}H_\nu{}^{\lambda\rho}-\tfrac12\,\cH_{MN}\cF_{\mu\lambda}{}^M\cF_\nu{}^{\lambda\,N}+\tfrac18\,\nabla_\mu\cH_{MN}\nabla_\nu\cH^{MN}+2\,\nabla_\mu\nabla_\nu\phi&=0 \;,\\
\nabla^\lambda H_{\lambda\mu\nu}-2\,\nabla^\lambda\phi\,H_{\lambda\mu\nu}&=0\;,\\
\nabla^2\cH_{MN}+\cH_{MP}\nabla^\mu\cH^{PQ}\nabla_\mu\cH_{QN}+\tfrac12\,\cF_{\mu\nu\, M}\cF^{\mu\nu}{}_N
-\tfrac12\cH_{MP}\cH_{NQ}\,\cF_{\mu\nu}{}^P\,\cF^{\mu\nu\,Q}\\
-2\,\nabla^\mu\phi\,\nabla_\mu\cH_{MN}&=0\;,\\
\nabla^\mu\cF_{\mu\nu}{}^M+\cH^{MN}\nabla^\mu\cH_{NP}\,\cF_{\mu\nu}{}^P+\tfrac12\,\cH^{MN}H_{\nu\lambda\rho}\,\cF^{\lambda\rho}{}_N-2\,\nabla^\mu\phi\,\cF_{\mu\nu}{}^M&=0\;.
\end{split}
\end{equation}
These are all the field equations except for the dilaton equation. 
Indeed, we performed the entire computation on a flat worldsheet, thereby ignoring the dilaton coupling. Consistency of the field equations \eqref{MS eoms} with the Bianchi identities, however, is enough to fix the dilaton equation of motion up to a constant, that vanishes in the critical dimension $D+d=26\,$.

In order to deal more efficiently with contractions of $O(d,d)$ indices, we shall employ the matrix notation
\begin{equation}
\cS\equiv\cS^M{}_N=\eta^{MP}\cH_{PN}   \;,\quad 
\cF^2_{\mu\nu}\equiv\left(\cF^2_{\mu\nu}\right)^M{}_N=\cF_{\mu\lambda}{}^M\cF_{\nu}{}^\lambda{}_N\;,\quad\cF^2=g^{\mu\nu}\cF^2_{\mu\nu}\;, \end{equation}
in which the $O(d,d)$ constraint on the generalized metric $\cH_{MN}$ reads $\cS^2=1$, and the field equation for $\cH_{MN}$ can be rewritten in the compact form
\begin{equation}
\nabla^2\cS+\cS\nabla^\mu\cS\nabla_\mu\cS+\tfrac12\,\cF^2-\tfrac12\,\cS\cF^2\cS-2\,\nabla^\mu\phi\nabla_\mu\cS=0\;.    
\end{equation}
We begin the derivation of the dilaton equation by taking the divergence of the metric equation of motion:
\begin{equation}\label{div Ricci}
\nabla^\nu R_{\mu\nu}=\nabla^\nu\Big[\tfrac14\,H^2_{\mu\nu}+\tfrac12\,\Tr\big(\cS\cF^2_{\mu\nu}\big)-\tfrac18\,\Tr\big(\nabla_\mu\cS\nabla_\nu\cS\big)-2\,\nabla_\mu\nabla_\nu\phi\Big] \;.   
\end{equation}
Upon using the field equations \eqref{MS eoms} together with the Bianchi identities
\begin{equation}
\nabla_{[\mu}\cF_{\nu\lambda]}{}^M=0\;,\quad \nabla_{[\mu}H_{\nu\lambda\rho]}+\frac34\,\cF_{[\mu\nu}{}^M\cF_{\lambda\rho]\,M}=0 \;, \end{equation}
the Ricci identity
\begin{equation}
\nabla^2\nabla_\mu\phi=\nabla_\mu\nabla^2\phi+R_{\mu\nu}\,\nabla^\nu\phi    
\end{equation}
and the constraint
\begin{equation}
\cS^2=1\;\longrightarrow\;\big(\nabla_\mu\cS\big)\cS=-\cS\nabla_\mu\cS \;,   
\end{equation}
the divergence \eqref{div Ricci} can be written as
\begin{equation}
\begin{split}
\nabla^\nu R_{\mu\nu}&=\nabla_\mu\Big[\tfrac{1}{24}\,H^2-\tfrac{1}{16}\,\Tr\big(\nabla^\nu\cS\nabla_\nu\cS\big)+\tfrac18\,\Tr\big(\cS\cF^2\big)-2\,\nabla^2\phi\Big] \\
&-2\,\nabla^\nu\phi\Big[R_{\mu\nu}-\tfrac14\,H^2_{\mu\nu}+\tfrac18\,\Tr\big(\nabla_\mu\cS\nabla_\nu\cS\big)-\tfrac12\,\Tr\big(\cS\cF^2_{\mu\nu}\big)\Big]\;.
\end{split}    
\end{equation}
Using again the metric equation in the second line one obtains
\begin{equation}
\nabla^\nu R_{\mu\nu}=\nabla_\mu\Big[\tfrac{1}{24}\,H^2-\tfrac{1}{16}\,\Tr\big(\nabla^\nu\cS\nabla_\nu\cS\big)+\tfrac18\,\Tr\big(\cS\cF^2\big)-2\,\nabla^2\phi+2\,\big(\nabla\phi\big)^2\Big]\;.    
\end{equation}
Finally, the metric Bianchi identity $\nabla^\nu R_{\mu\nu}=\frac12\,\nabla_\mu R$
gives the dilaton equation:
\begin{equation}\label{MS phi}
\nabla^2\phi-\big(\nabla\phi\big)^2+\tfrac14\,\Big(R+\tfrac18\,\Tr\big(\nabla^\mu\cS\nabla_\mu\cS\big)-\tfrac14\,\Tr\big(\cS\cF^2\big)-\tfrac{1}{12}\,H^2\Big)=0\;,  
\end{equation}
where we set the integration constant to zero.\footnote{The undetermined constant term can be deduced from $\beta^\phi=\frac{D+d-26}{6}-\frac{\alpha'}{2}\,\nabla^2\phi+\cdots$ and indeed vanishes in the critical dimension.}
Eliminating the Ricci scalar by using the trace of the metric equation one can rewrite it in the more familiar form 
\begin{equation}\label{MS phi'}
-\tfrac12\,\nabla^2\phi+\big(\nabla\phi\big)^2-\tfrac{1}{24}\,H^2-\tfrac{1}{16}\,\Tr\big(\cS\cF^2\big) =0\;,   
\end{equation}
which  is the form one should  find by direct computation of the dilaton $\beta-$function at two loops.
As promised, the vanishing Weyl anomaly conditions  \eqref{MS eoms}, together with \eqref{MS phi} or \eqref{MS phi'}, coincide with
the field equations derived from the Maharana-Schwarz low-energy effective field theory \cite{Maharana:1992my}
\begin{equation}
S=\int d^Dx\,\sqrt{-g}\,e^{-2\phi}\Big[R+4\,\nabla^\mu\phi\nabla_\mu\phi+\tfrac18\,\Tr\big(\nabla^\mu\cS\nabla_\mu\cS\big)-\tfrac14\,\Tr\big(\cS\cF^2\big)-\tfrac{1}{12}\,H^2\Big]\;.
\end{equation}

We conclude our discussion with a final consistency check of our procedure. According to the general derivation of \eqref{eoms beta bar}, the target space equations \eqref{MS eoms} are equivalent to the vanishing of the Weyl anomaly at one-loop order. This, in turn, should imply that the sigma model \eqref{sigma Odd euclidean} be UV-finite at one-loop when \eqref{MS eoms} holds, which we confirm in the following.  Collecting the results of this section for the divergent part of the effective action we have
\begin{equation}
\Gamma[\varphi,\phi]=S[\varphi,\phi]+\Gamma_{1l}[\varphi,\phi]+\cdots   \;, 
\end{equation}
where the classical action is given by
\begin{equation}\label{classical S}
\begin{split}
S[\varphi,\phi]&=\frac{1}{2\lambda}\int d^2x\,\Big\{g_{\mu\nu}\,\del^\alpha \varphi^\mu\del_\alpha \varphi^\nu+i\,\epsilon^{\alpha\beta}\,\Big[B_{\mu\nu}\,\del_\alpha \varphi^\mu\del_\beta \varphi^\nu+\cA_\mu^M\,\del_\alpha \varphi^\mu D_\beta \phi_M\Big]\\
&\hspace{23mm}-i\,D_1\phi^MD_2\phi_M+\cH_{MN}\,D_1\phi^MD_1\phi^N\Big\}\;, 
\end{split}    
\end{equation}
while, using the fact that $\beta^i=-T^i_1$ at one loop, the divergent contribution reads
\begin{equation}\label{Gamma div}
\begin{split}
\Gamma_{1l}[\varphi,\phi]&=\frac{1}{2\lambda\epsilon}\int d^2x\,\Big\{\beta_{\mu\nu}^g\,\del^\alpha\varphi^\mu\del_\alpha\varphi^\nu+i\,\epsilon^{\alpha\beta}\beta_{\mu\nu}^B\,\del_\alpha\varphi^\mu\del_\beta\varphi^\nu+\beta_{MN}\,D_1\phi^MD_1\phi^N\\
&\hspace{25mm}+2\,\beta_\mu^M\,D_1\phi^N\,\big(\cH_{MN}\,\del_1\varphi^\mu-i\,\eta_{MN}\,\del_2\varphi^\mu\big)\Big\}  +\cO(\epsilon^0)\;, 
\end{split}    
\end{equation}
with the $\beta-$functions given by \eqref{MS beta g+B}, \eqref{MS beta cH} and \eqref{MS beta A}. UV finiteness of the sigma model is not immediately apparent, since the target space equations are \emph{not} $\beta^i=0$, but rather
\begin{equation}\label{on-shell betas}
\begin{split}
\beta_{\mu\nu}^g&=\nabla_\mu\xi_\nu+\nabla_\nu\xi_\mu\;,\qquad\beta_{\mu\nu}^B=\xi^\lambda\,H_{\lambda\mu\nu}\;,\\
\beta_{MN}&=\xi^\mu\nabla_\mu\cH_{MN}\;,\qquad\;\;\;\, \beta_\mu^M=\xi^\lambda\,\cF_{\lambda\mu}{}^M\;,
\end{split}    
\end{equation}
with $\xi^\mu=-\alpha'\,\nabla^\mu\phi\,$. Substituting the conditions \eqref{on-shell betas} into \eqref{Gamma div} we obtain
\begin{equation}
\begin{split}
\Gamma_{1l}^{\rm div}[\varphi,\phi]&=\frac{1}{2\lambda\epsilon}\int d^2x\,\Big\{2\,\nabla_\mu\xi_\nu\,\del^\alpha\varphi^\mu\del_\alpha\varphi^\nu+i\,\epsilon^{\alpha\beta}\xi^\lambda\,H_{\lambda\mu\nu}\,\del_\alpha\varphi^\mu\del_\beta\varphi^\nu+\xi^\mu\nabla_\mu\cH_{MN}\,D_1\phi^MD_1\phi^N\\
&\hspace{25mm}+2\,\xi^\lambda\,\cF_{\lambda\mu}{}^M\,D_1\phi^N\,\big(\cH_{MN}\,\del_1\varphi^\mu-i\,\eta_{MN}\,\del_2\varphi^\mu\big)\Big\}\\
&=\frac{1}{\lambda\epsilon}\int d^2x\,\xi^\mu\,\Big\{-g_{\mu\nu}\,D^\alpha\del_\alpha\varphi^\nu+\tfrac{i}{2}\,\epsilon^{\alpha\beta}H_{\mu\nu\lambda}\,\del_\alpha\varphi^\nu\del_\beta\varphi^\lambda+\tfrac12\,\nabla_\mu\cH_{MN}\,D_1\phi^MD_1\phi^N\\
&\hspace{25mm}+\cF_{\mu\nu}{}^M\,D_1\phi^N\,\big(\cH_{MN}\,\del_1\varphi^\nu-i\,\eta_{MN}\,\del_2\varphi^\nu\big)\Big\}\;. 
\end{split}    
\end{equation}
This vanishes whenever the background fields $\varphi^\mu$ and $\phi^M$ are on the mass-shell of the classical action \eqref{classical S}. This is sufficient, since the only meaningful statement about finiteness of the effective action is for on-shell classical fields.

An equivalent conclusion can be reached by noting that the divergent part of the effective action can be rewritten as
\begin{equation}
\Gamma_{1l}^{\rm div}=\frac{1}{\epsilon}\,\delta_\xi S\;,    
\end{equation}
where $\delta_\xi$ acts \emph{only} on the couplings as
\begin{equation}\label{Lie plus gauge}
\begin{split}
\delta_\xi g_{\mu\nu}&=\cL_\xi g_{\mu\nu}\;,\qquad\qquad\;\;\,  \delta_\xi\cH_{MN}=\cL_\xi\cH_{MN}\;,\\
\delta_\xi\cA_\mu{}^M&=\cL_\xi\cA_\mu{}^M+\del_\mu\lambda^M\;,\quad\delta_\xi B_{\mu\nu}=\cL_\xi B_{\mu\nu}+2\,\del_{[\mu}\lambda_{\nu]}+\tfrac12\,\lambda_M\,\cF_{\mu\nu}{}^M\;,
\end{split}    
\end{equation}
where here $\cL_\xi$ denotes the ordinary (non-gauge-invariant) Lie derivative and
\begin{equation}
\xi^\mu=-\alpha'\,\nabla^\mu\phi\;,\quad \lambda^M=-\xi^\mu\cA_\mu{}^M\;,\quad \lambda_\mu=-\xi^\lambda B_{\lambda\mu}\;.    
\end{equation}
This can be seen by inspection upon using (\ref{on-shell betas}) in (\ref{Gamma div}). 
The variations \eqref{Lie plus gauge} are given by a target space diffeomorphism plus a gauge transformation. Since $S$ is diffeomorphism and gauge-invariant when \emph{both} couplings and fields transform, \eqref{Lie plus gauge} can be compensated  by a field redefinition
\begin{equation}
\varphi'^\mu=\varphi^\mu-\frac{1}{\epsilon}\,\xi^\mu(\varphi)\;,\quad\phi'^M=\phi^M-\frac{1}{\epsilon}\,\lambda^M(\varphi)   \;, 
\end{equation}
that removes all one-loop divergences, as expected.

\section{Conclusions and Outlook}

In this paper we have introduced the background material needed to compute beta functions for the duality invariant sigma model 
of \cite{Schwarz:1993mg,Blair:2016xnn,Bonezzi:2020ryb}  and shown that at one loop vanishing of the Weyl anomaly implies the 
$O(d,d)$ invariant target space equations of Maharana-Schwarz. This is an instructive  test for the results in \cite{Bonezzi:2020ryb}
according to which the presence of chiral bosons implies the existence of an anomaly that, however, can be cancelled by a Green-Schwarz 
mechanism. It is then an important consistency test that this model can be employed for quantum computations, as displayed here. 
In a forthcoming publication we will explore the computation of beta functions to two and higher loops. Apart from the expected technical 
complications there are new conceptual issues that arise beyond one loop, and we hope to report on these soon. 

A major motivation for the investigation of duality invariant beta functions is the relatively recent insight 
that in one dimension (dimensional reduction to only cosmic time) all $\alpha'$ corrections that are compatible with duality  can be completely classified, 
leaving a surprisingly small number of free coefficients at each order in $\alpha'$  \cite{Hohm:2019jgu}. 
Fixing  these coefficients to all orders in $\alpha'$ would be a major breakthrough as it would give us unprecedented access to truly stringy physics. 
Unfortunately, the results presented here show little  indication that this could eventually be done to all orders, at least when restricting to the method of 
  beta functions, which does not show enough simplification even in one dimension. 
 However,   other methods may be helpful, the obvious example being supersymmetry in case of superstring theory, 
 which should at least constrain the free coefficients that are not fixed by duality. Although it is not immediately clear  how to incorporate fermions 
 into the duality invariant sigma model explored here, one may aim to implement supersymmetry directly in the duality invariant target space theory.

\subsection*{Acknowledgements} 

We would like to thank Felipe Diaz-Jaramillo and Arkady Tseytlin for useful discussions and correspondence.

This work is funded  by the ERC Consolidator Grant ``Symmetries \& Cosmology" 
and by the Deutsche Forschungsgemeinschaft (DFG, German Research Foundation) - Projektnummer 417533893/GRK2575 ``Rethinking Quantum Field Theory".

\appendix

\section{Technical details}

\subsection{One-loop Feynman integrals}\label{Feynman}

In order to keep our exposition self-contained, we will review the standard tools for evaluating one-loop Feynman integrals in dimensional regularization, with $n=2+\epsilon$. All the divergent integrals encountered in the main text can be recast in terms of the generalized tadpole integral:
\begin{equation}\label{general tadpole}
I(p,q;D)=\int\frac{d^nk}{(2\pi)^n}\frac{(k^2)^p}{(k^2+D)^q}=\frac{1}{(4\pi)^{n/2}}\,\frac{\Gamma(q-p-n/2)\,\Gamma(p+n/2)}{\Gamma(q)\,\Gamma(n/2)}\,D^{-(q-p-n/2)} \;,   
\end{equation}
where $D$ is an arbitrary scalar.
The simplest example is the regularized propagator at coincident points:
\begin{equation}\label{G0Appendix}
\begin{split}
G(0)_{\rm reg}&=\mu^{2-n}\int\frac{d^{n}k}{(2\pi)^{n}}\,\frac{1}{k^2+m^2}\equiv\mu^{2-n}I(0,1;m^2)=\frac{1}{(4\pi)^{n/2}}\,\left(\frac{m^2}{\mu^2}\right)^{\frac{n}{2}-1}\Gamma(1-n/2)\\
&=\frac{1}{4\pi}\,\left(\frac{m^2}{4\pi\mu^2}\right)^{\frac{\epsilon}{2}}\Gamma(-\epsilon/2)=-\frac{1}{2\pi\epsilon}-\frac{1}{4\pi}\,\left(\gamma+\log\frac{m^2}{4\pi\mu^2}\right)+\cO(\epsilon)\;.
\end{split}    
\end{equation}
Divergent integrals, however, usually appear in the tensor form
\begin{equation}\label{tensor integral}
I_{\alpha_1\cdots\alpha_{m}}(q;D)=\int\frac{d^nk}{(2\pi)^n}\,\frac{k_{\alpha_1}\cdots k_{\alpha_{m}}}{(k^2+D)^q}\;,   
\end{equation}
where we stress that $D$ contains only \emph{scalar} parameters. This allows to exploit the $SO(n)$ covariance of the integral and total symmetry in the vector indices to deduce
\begin{equation}
I_{\alpha_1\cdots\alpha_{2p}}(q;D)=\frac{1}{F_p}\,\delta_{(\alpha_1\alpha_2}\cdots\delta_{\alpha_{2p-1}\alpha_{2p})}\,I(p,q;D)\;,\quad I_{\alpha_1\cdots\alpha_{2p+1}}=0\;,
\end{equation}
where the normalization factor can be computed by induction upon taking traces:
\begin{equation}
F_p=\frac{(n+2p-2)(n+2p-4)\cdots n}{(2p-1)(2p-3)\cdots 3}=\frac{(n+2p-2)!!}{(n-2)!!\,(2p-1)!!}\;. \end{equation}

More generally, at one-loop one also encounters bubble integrals that depend on external momenta. All the one-loop bubbles encountered in the text take the generic form
\begin{equation}\label{general bubble}
I_{\alpha_1\cdots\alpha_m}(p)=\int\frac{d^nk}{(2\pi)^n}\frac{k_{\alpha_1}\cdots k_{\alpha_m}}{(k^2+m^2)\big[(p-k)^2+m^2\big]}\;,
\end{equation}
for $m\leq4$. In these cases, the first step consists in combining denominators via Feynman's formula
\begin{equation}
\frac{1}{AB}=\int_0^1 dx\,\frac{1}{[x\,A+(1-x)\,B]^2} \;.   
\end{equation}
Applying it to \eqref{general bubble} yields, upon completing squares and shifting to $q_\alpha=k_\alpha-x\,p_\alpha$, 
\begin{equation}
I_{\alpha_1\cdots\alpha_m}(p)=\int_0^1dx \int\frac{d^nq}{(2\pi)^n}\frac{(q_{\alpha_1}+x\,p_{\alpha_1})\cdots (q_{\alpha_m}+x\,p_{\alpha_m})}{(q^2+D)^2}\;,\quad D=x(1-x)\,p^2+m^2\;,
\end{equation}
which is (modulo the integration over the Feynman parameter $x$) a linear combination of integrals of the form \eqref{tensor integral}.

\subsection{Infrared regularization in the internal sector}\label{Appendix IR reg}

When dealing with the massless propagator
\begin{equation}
G^{AB}(x)=\int\frac{d^2p}{(2\pi)^2}\frac{e^{ip\cdot x}}{p^2}\,\Big[h^{AB}+i\,\frac{p_2}{p_1}\,\eta^{AB}\Big]    
\end{equation}
we encounter two difficulties: the lack of manifest Lorentz invariance and infrared divergences. While the non-manifest Lorentz invariance cannot be helped at this level, and needs to be checked at the end of the computations, the infrared divergences need to be regulated.

While the infrared pole at $p^2=0$ is the usual one associated to massless particles, the entire IR-divergent line $p_1=0$ is associated with the $x^2-$local symmetry $\delta\zeta^A=\Xi^A(x^2)$ of the action. In order to regulate the IR behavior we shall add mass terms in the chiral basis $\zeta_\pm^A\,$:
\begin{equation}
S_{\zeta\zeta}^{\rm kin}=\frac{1}{2\lambda}\int d^2x\,\Big[\del_1\zeta^A_+(\del_1-i\del_2)\zeta_{+A}+\tfrac{m^2}{2}\,\zeta_+^A\zeta_{+A}-\del_1\zeta^A_-(\del_1+i\del_2)\zeta_{-A}-\tfrac{m^2}{2}\,\zeta_-^A\zeta_{-A}\Big]  \;,  
\end{equation}
where $\zeta^A_\pm=\frac12\,\big(\delta^A{}_B\pm h^A{}_B\big)\zeta^B\,$. This modifies the two-point functions as
\begin{equation}
\langle\zeta^A(x)\zeta^B(y)\rangle  =\lambda\,\Big(h^{AB}\,G_0(x-y)+\eta^{AB}\,G_1(x-y)\Big) \;, 
\end{equation}
with
\begin{equation}
\begin{split}
G_0(x)&=\int\frac{d^2p}{(2\pi)^2}\,e^{ip\cdot x}\frac{p_1^2+m^2/4}{p_1^2(p^2+m^2)+m^4/4} =\int\frac{d^2p}{(2\pi)^2}\,e^{ip\cdot x}\frac{1}{p^2+m^2}\,\Big(1+\cO(m^2)\Big)\;,\\
G_1(x)&=\int\frac{d^2p}{(2\pi)^2}\,e^{ip\cdot x}\frac{i\,p_1p_2}{p_1^2(p^2+m^2)+m^4/4} =\int\frac{d^2p}{(2\pi)^2}\,e^{ip\cdot x}\frac{i\,p_1p_2}{(p_1^2+m^2)(p^2+m^2)}\,\Big(1+\cO(m^2)\Big)\;.
\end{split}    
\end{equation}
We shall thus use the infrared regulated propagator
\begin{equation}
G^{AB}(x)=\int\frac{d^2p}{(2\pi)^2}\frac{e^{ip\cdot x}}{p^2+m^2}\,\Big[h^{AB}+\frac{i\,p_1p_2}{p_1^2+m^2}\,\eta^{AB}\Big]\;.\end{equation}
This way of regulating the infrared divergences breaks the $\Xi^A-$symmetry. The chiral nature of the $\zeta^A$ bosons, as well as two-dimensional Lorentz invariance, are expected to emerge only in the massless limit.

A typical class of divergent one-loop integrals that we encounter is of the form
\begin{equation}
\begin{split}
I_{\alpha\beta}^{ABCD}&=\int\frac{d^2k}{(2\pi)^2}\,k_\alpha k_\beta\,G^{AB}(k)\,G^{CD}(-k)  \\
&=\int\frac{d^2k}{(2\pi)^2}\,\frac{k_\alpha k_\beta}{(k^2+m^2)^2}\,\Big(h^{AB}+\frac{i\,k_1k_2}{(k_1^2+m^2)}\,\eta^{AB}\Big)\Big(h^{CD}+\frac{i\,k_1k_2}{(k_1^2+m^2)}\,\eta^{CD}\Big)\;.
\end{split}    
\end{equation}
By considering the various components of $I_{\alpha\beta}^{ABCD}$ we see that, upon discarding the integrals that are zero by symmetry, the ones to be considered are
\begin{equation}
\begin{split}
J_1&=\int\frac{d^2k}{(2\pi)^2}\,\frac{k_1^2}{(k^2+m^2)^2}= \int\frac{d^2k}{(2\pi)^2}\,\frac{k_2^2}{(k^2+m^2)^2}\;,\quad J_2=\int\frac{d^2k}{(2\pi)^2}\,\frac{k_1^4k_2^2}{(k^2+m^2)^2(k_1+m^2)^2}\;,\\
J_3&=\int\frac{d^2k}{(2\pi)^2}\,\frac{k_1^2k_2^2}{(k^2+m^2)^2(k_1^2+m^2)}\;,\quad J_4=\int\frac{d^2k}{(2\pi)^2}\,\frac{k_1^2k_2^4}{(k^2+m^2)^2(k_1^2+m^2)^2}\;.
\end{split}    
\end{equation}
In particular, one can see that $J_4$ would be ill-defined if we removed the mass term from the $(k_1^2+m^2)^2$ denominator.

Apart from $J_1$, whose integrand is a tensor numerator times a Lorentz-invariant function, the other integrals cannot be treated by exploiting the usual $SO(2)$ symmetry. For this reason, we shall employ a split form of dimensional regularization: each integral $\int dk_1$ and $\int dk_2$ will be extended to $\nu=n/2=1+\epsilon/2$ dimensions. This allows to use only $SO(\nu)\times SO(\nu)$ symmetry, instead of $SO(n)$, but will suffice to recast all the above integrals in terms of
\begin{equation}
G(0)_{\rm reg}=\frac{1}{4\pi}\,\left(\frac{m^2}{4\pi\mu^2}\right)^{\frac{\epsilon}{2}}\Gamma(-\epsilon/2)=-\frac{1}{2\pi\epsilon}-\frac{1}{4\pi}\,\left(\gamma+\log\frac{m^2}{4\pi\mu^2}\right)+\cO(\epsilon)\;.
\end{equation}
In extending the $J_i$ integrals, one encounters the usual ambiguities in defining special tensor structures in arbitrary dimensions. For instance, one has
\begin{equation}
J_1^{\rm reg}=\frac{\mu^{2-2\nu}}{f(\nu)}\int\frac{d^\nu k}{(2\pi)^\nu}\frac{d^\nu q}{(2\pi)^\nu}\,\frac{k^2}{(k^2+q^2+m^2)^2}\;,\quad \nu=1+\frac{\epsilon}{2}\;.    
\end{equation}
Two natural choices are either $f(\nu)=1$ or $f(\nu)=\nu$, depending on whether one considers $k_1^2=k^2$ or $k_1^2=k_1k_1\to\frac{1}{\nu}\,\delta_{11}\,k^2=\frac{1}{\nu}\,k^2$ upon using $SO(\nu)$ invariance to substitute $k_\alpha k_\beta\to\frac{1}{\nu}\,\delta_{\alpha\beta}\,k^2$ in $\nu=1+\frac{\epsilon}{2}$ dimensions. 

With either choice, one can perform the integral by successive use of \eqref{general tadpole} (upon substituting $n\to\nu$)
finally obtaining
\begin{equation}
J_1^{\rm reg}=\frac{\nu}{2\,f(\nu)}\,G(0)_{\rm reg}=-\frac{1}{4\pi\epsilon}+\cO(\epsilon^0)  \;.  
\end{equation}
Analogously, we have
\begin{equation}
J_2^{\rm reg}=\frac{\mu^{2-2\nu}}{f(\nu)\,g(\nu)}\int\frac{d^\nu k}{(2\pi)^\nu}\frac{d^\nu q}{(2\pi)^\nu}\,\frac{k^4\,q^2}{(k^2+q^2+m^2)^2(k^2+m^2)^2}\;,    
\end{equation}
where $f(\nu)$ is the same as before, while $g(\nu)$ reflects a similar ambiguity in extending $k_1^4$ either to $k^4$ or to $k_1k_1k_1k_1\to \frac{3}{\nu(\nu+2)}\,k^4$ upon using $k_\alpha k_\beta k_\gamma k_\delta\to \frac{3}{\nu(\nu+2)}\,\delta_{(\alpha\beta}\delta_{\gamma\delta)}\,k^4$, that is $g(\nu)=1$ or $g(\nu)=\frac{\nu(\nu+2)}{3}$, respectively. In any case one obtains
\begin{equation}
J_2^{\rm reg}=\frac{\nu^2(\nu+2)}{2(4-\nu)(2-\nu)}\,\frac{1}{f(\nu)\,g(\nu)}\,G(0)_{\rm reg}=-\frac{1}{4\pi\epsilon}+\cO(\epsilon^0)\;.    
\end{equation}
The remaining integrals can be computed in the same way, leading to
\begin{equation}
\begin{split}
J_3^{\rm reg}&=\frac{\mu^{2-2\nu}}{f^2(\nu)}\int\frac{d^\nu k}{(2\pi)^\nu}\frac{d^\nu q}{(2\pi)^\nu}\,\frac{k^2\,q^2}{(k^2+q^2+m^2)^2(k^2+m^2)}\\
&=\frac{\nu^2}{2(2-\nu)}\,\frac{1}{f^2(\nu)}\,G(0)_{\rm reg}=-\frac{1}{4\pi\epsilon}+\cO(\epsilon^0)\;,\\[2mm]
J_4^{\rm reg}&=\frac{\mu^{2-2\nu}}{f(\nu)\,g(\nu)}\int\frac{d^\nu k}{(2\pi)^\nu}\frac{d^\nu q}{(2\pi)^\nu}\,\frac{k^2\,q^4}{(k^2+q^2+m^2)^2(k^2+m^2)^2}\\
&=-\frac{\nu(\nu+2)}{2(2-\nu)}\,\frac{1}{f(\nu)\,g(\nu)}\,G(0)_{\rm reg}=\frac{3}{4\pi\epsilon}+\cO(\epsilon^0)\;.
\end{split}    
\end{equation}
Let us notice that, for the present purpose of computing the one-loop $\beta-$functions, the different prescriptions for $f(\nu)$ and $g(\nu)$ do not matter, since the divergent part of the integrals is insensitive to this choice.

\bibliography{MS.bib}
\bibliographystyle{unsrt}

\end{document}